\documentclass{jpp}
\usepackage{graphicx}

\usepackage[utf8]{inputenc}
\usepackage[T1]{fontenc}
\usepackage[draft]{pgf}
\usepackage{amsmath}
\usepackage{graphicx}
\usepackage{dcolumn}
\usepackage{bm}
\usepackage{natbib}
\usepackage{soul}
\usepackage{xcolor}
\usepackage{amsmath}
\usepackage{amssymb}

\usepackage{color}
\usepackage{caption}

\usepackage{subfigure}
\usepackage{cmap}
\usepackage[T1]{fontenc}
\usepackage{enumerate}
\usepackage{makecell}
\usepackage{multirow}
\usepackage{hyperref}
\usepackage{array}

\shorttitle{Drift kinetic theory of alpha transport by tokamak perturbations}
\shortauthor{E. A. Tolman and P.J. Catto}

\title{Drift kinetic theory of alpha transport by tokamak perturbations}

\author{Elizabeth A. Tolman\aff{1,2}
  \corresp{\email{tolman@ias.edu}}
 \and Peter J.  Catto\aff{1}}

\affiliation{\aff{1}Plasma Science and Fusion Center, Massachusetts Institute of Technology,
Cambridge, MA 02139, USA \aff{2}Institute for Advanced Study,
Princeton, NJ 08540, USA}

\begin{document}

\maketitle

\begin{abstract}
Upcoming tokamak experiments fueled with deuterium and tritium are expected to have large alpha particle populations. Such experiments motivate new attention to the theory of alpha particle confinement and transport. A key topic is the interaction of alphas with perturbations to the tokamak fields, including those from ripple and  magnetohydrodynamic modes like Alfv\'{e}n eigenmodes. These perturbations can transport alphas, leading to changed localization of alpha heating, loss of alpha power, and damage to device walls. Alpha interaction with these perturbations is often studied with single particle theory.  In contrast, we derive a drift kinetic theory to calculate the alpha heat flux resulting from arbitrary perturbation frequency and periodicity (provided these can be studied drift kinetically).  Novel features of the theory include the retention of a large effective collision frequency resulting from the resonant alpha collisional boundary layer, correlated interactions over many poloidal transits, and finite orbit effects. Heat fluxes are considered for the example cases of ripple and the toroidal Alfv\'{e}n eigenmode (TAE).  The ripple heat flux is small.  The TAE heat flux is significant and scales with the square of the perturbation amplitude, allowing the derivation of constraints on mode amplitude for avoidance of significant alpha depletion. A simple saturation condition suggests that TAEs in one upcoming experiment will not cause significant alpha transport via the mechanisms in this theory. However, saturation above the level suggested by the simple condition, but within numerical and experimental experience, which could be accompanied by the onset of stochasticity, could cause significant transport.
\end{abstract}

\section{Introduction}
Next generation tokamak experiments fueled with deuterium and tritium, including ARC \citep{sorbom2015arc}, SPARC \citep{creely2020}, and ITER \citep{ikeda2007progress}, will have significant populations of highly energetic particles, including alpha particles and energetic particles resulting from external heating. These energetic particles will strongly affect overall experiment performance. Alpha particle behavior is particularly important because alphas are expected to dominate plasma heating in many cases \citep{heidbrink2002alpha}. Redistribution and loss of alpha particles could modify bulk plasma profiles due to changed localization of heating, could lead to loss of power needed to heat the plasma, and could cause damage to device walls \citep{heidbrink2002alpha,heidbrink2008basic}, making study of alpha confinement and transport critical to the design, operation, and analysis of next generation experiments. 

A major source of alpha particle transport is interactions with perturbations to the tokamak fields, which often come from ripple or magnetohydrodynamic (MHD) modes.  Ripple is a periodic, stationary magnetic perturbation of the plasma equilibrium due to the discrete nature of toroidal field coils.\footnote{Information on ripple in ITER can be found in \citet{kurki2009ascot}; information on ripple in SPARC in~\citet{scott2020}.} MHD modes are fluctuations of the plasma which include Alfv\'{e}n eigenmodes (AEs) \citep{cheng1986low,heidbrink2008basic}, driven by energetic particles, neoclassical tearing modes (NTMs) \citep{la2006neoclassical}, destabilized by tokamak bootstrap current, and fishbones \citep{betti1993destabilization}, also destabilized by energetic particles. A major mechanism of such transport is the anomalous radial motion introduced by the perturbation via drifts and changed magnetic field direction.  When an alpha particle's motion is in phase with the periodicity of the perturbation, this radial motion results in transport. Transport of this nature has sometimes been described as bounce harmonic resonant transport \citep{linsker1982banana,mynick1986generalized,park2009nonambipolar,kim2013numerical,logan2013neoclassical}.

Energetic particle transport is often analyzed numerically  [for examples, see \citet{snicker2013power} and \citet{collins2017phase}], using orbit following codes, including ASCOT \citep{hirvijoki2014ascot,varje2019high}, OFMC \citep{tani1981effect}, and ORBIT \citep{white1995rapid}.  These codes numerically evolve the trajectories of a large ensemble of energetic particles in a tokamak magnetic equilibrium modified by a perturbation, often including the effect of collisions on individual particle trajectories.  Understanding the physical mechanisms of transport can be challenging in numerical studies. In addition, codes may not have adequate phase space resolution to appropriately treat transport.  

Analytically, energetic particle transport is often studied by considering the behavior of a \textit{single particle}. Foundational works on transport identify the onset of significant energetic particle redistribution with individual particle trajectories becoming stochastic in phase space \citep{goldston1981confinement,sigmar1992alpha,collins2016observation}. For MHD modes, this stochasticization can be attributed to particles becoming trapped and de-trapped in overlapping phase space islands created by resonances with the mode \citep{white2010particle}, while for ripple transport, this stochastic diffusion can be caused by large radial jumps at the turning points of trapped particle orbits \citep{goldston1981confinement}. Other works use guiding center equations of motion to examine resonance and drifts that exist when a single particle moves in a mode \citep{poli2008observation}. 

Such theories do not reveal the behavior of the perturbed energetic particle distribution function and do not allow the calculation of the concomitant energetic particle flux. Thus, an analytic drift kinetic evaluation of energetic particle heat transport by perturbations is sorely needed. This theory should calculate the perturbed particle distribution caused by a field perturbation while retaining all phase differences between this distribution perturbation and the radial particle velocity perturbation that leads to transport. Past work \citep{nagaoka2008radial} as well as a recent review \citep{todo2019introduction} have highlighted this need. 

In this paper we develop such a theory for alpha particles. The theory as presently  formulated does not treat species of energetic particles other than alphas because of the importance alphas have in determining next generation tokamak behavior and because the standardized form of the alpha distribution simplifies the calculation. The theory could later be adapted to other types of energetic particles with more complex distribution functions. Our theory is a drift kinetic extension of single particle studies which use guiding center theory. It includes correlated interactions over many poloidal transits, finite orbit effects, and a large effective pitch angle scattering collision frequency resulting from the narrow boundary layers encompassing the resonant interaction of the perturbation with the alphas.  This large effective pitch angle scattering collision frequency resolves the alpha resonance with the perturbation and provides the dissipation and decorellation necessary for transport. [Traditional single particle theories of energetic particle transport often posit that collisions are not important because large energetic particle velocities reduce the particle collision frequency \citep{heidbrink19}. However, in a kinetic theory, the presence of sharply localized in phase space boundary layers enhances the importance of these collisions such that their consistent inclusion is crucial, as we will show.]

Throughout the work, we assume that the perturbations responsible for transport are below the threshold for stochasticity. Most tokamaks will experience a large number of magnetic and electric field perturbations localized at different flux surfaces and with different mode numbers. These distinct field perturbations can work together to cause transport of alphas across the device even in the absence of stochasticity. In this scenario, collisions provide radial steps between neighboring  mode surfaces, leading to the heat transport we calculate.\footnote{Such transport is analogous to the situation illustrated in figures 3 and 4 of \citet{berk1995line}.} The transport is akin to a quasilinear evaluation with a collisionally broadened overlap of the isolated mode surfaces spaced within a collisional step size of each other. Indeed, as our results will ultimately prove to be independent of collision frequency, it may be that they are even valid for mild resonance overlap, as long as the slowing down distribution is not significantly distorted. This "sea" of perturbations of various toroidal mode numbers is reduced to a tractable problem by focusing on a single toroidal mode centered at a flux surface. The transport associated with the first few bounce and transit harmonics and the most resonant poloidal mode is then evaluated, because these contributions dominate the overall transport. The transport due to other poloidal mode numbers and higher bounce and transit harmonics will be smaller
because of their phase mismatch. 

The techniques developed herein are kept in fully general form so that they can be used for tokamak perturbations of any frequency, toroidal periodicity, poloidal periodicity, and radial structure (provided that the perturbation is treatable with drift kinetics). However, two perturbations are given as examples. First, alpha particles are found to have many resonances with ripple, but these resonances drive negligible transport. Second, alpha particles resonate with toroidal Alfv\'{e}n eigenmodes (TAEs). These resonances drive significant alpha heat flux that increases with the square of mode amplitude.  The heat flux expression can be used to derive a constraint on mode amplitude for avoidance of significant alpha depletion. A simple model for TAE saturation is developed to estimate TAE amplitude. The saturation estimate is consistent with numerical results presented elsewhere \citep{slaby2018effects} and is comfortably below the stochastic threshold given in \citet{berk1992scenarios}. As an example, the constraints and the saturation amplitude are evaluated for SPARC-like parameters.  The estimate suggests that the TAE amplitude in SPARC will not cause significant depletion. However, a TAE saturation amplitude above that suggested by the simple model, but within experimental and modelling experience, could cause alpha particle depletion.

The structure of this paper follows. Section~\ref{sec:setup} introduces the mathematical description of the tokamak equilibrium and the drift kinetic equation which is used to describe alpha transport. It provides parameters for a SPARC-like tokamak which will be used as an example tokamak equilibrium throughout the paper. Then, the section and appendix~\ref{sec:slowingdown} derive the slowing down distribution, which describes the unperturbed alpha population. Section~\ref{sec:perts} introduces the form of the perturbations which we study in this paper, and describes the particular parameters that describe ripple and TAE, the perturbations we use as examples. Section~\ref{sec:response} discusses the plasma response to these perturbations, first giving the equation that governs how the alpha distribution function responds to perturbed fields, then solving this equation for the response as a function of the parameters describing the perturbation. Section~\ref{sec:flux} develops general formulae for the alpha particle heat flux from both trapped and passing alpha particle populations. These formulae are the key results of this paper. This section also considers the nature of a resonance function and phase factor that are key components of the formulae.  Section~\ref{sec:ripp} considers this heat flux for the case of ripple, and shows that it is very small. Section~\ref{sec:TAEflux} derives compact expressions for trapped and passing fluxes for the TAE. These fluxes are used to understand what amplitude of TAE causes significant alpha depletion. A simple saturation condition is developed and the amount of transport caused by this saturated amplitude is considered.  The conclusion, section~\ref{sec:concl}, summarizes the paper results, gives implications for future experiments, and presents avenues for future work.

\section{Equilibrium and governing equations}
\label{sec:setup} 
In this section, we first describe the tokamak magnetic equilibrium in which the alpha particle transport occurs and the parameters that describe alpha particles in this equilibrium.  Then, we introduce the drift kinetic equation which is used to study alpha behavior. Next, we calculate the equilibrium alpha particle slowing down distribution function. The section concludes by giving specific tokamak parameters which will be used as examples throughout the paper.
\subsection{Equilibrium and phase space coordinates}
The coordinates describing the tokamak equilibrium are $\psi$ (the poloidal flux function), $\vartheta$ (the poloidal angle)\footnote{At this point, $\vartheta$ is a fully general straightened field line poloidal coordinate which must only be chosen such that $q(\psi) = (\hat{b} \cdot \nabla \zeta)/(\hat{b} \cdot \nabla \vartheta)$.  This allows for shaping and finite aspect ratio. Later, in section~\ref{sec:flux}, a circular-cross-section,  high-aspect-ratio approximation is used to obtain analytic expressions for the flux.}, and $\alpha$, defined by
\begin{equation}
\label{eq:alphadef}
\alpha \equiv \zeta - q\left(\psi\right) \vartheta.
\end{equation}
Here, $\zeta$ is the toroidal angle. The safety factor is $q$, which characterizes the twist of the background magnetic field by $q \left(\psi \right) \equiv \hat{b} \cdot \nabla \zeta / \left( \hat{b} \cdot \nabla \vartheta \right)$ with $\hat{b}$ giving the direction of the equilibrium magnetic field.\footnote{Later in the paper, an approximate form of this definition,  $q \left(\psi \right) \approx  \left(R \hat{b} \cdot  \nabla \vartheta  \right)^{-1}$,  with $R$ the major radius coordinate, will be used to simplify expressions.}  
We study the behavior of alpha particles confined by a magnetic field which is axisymmetric and stationary except for the perturbation. Such a field is stated to zeroth order in the perturbation via the Clebsch representation as
\begin{equation}
\label{eq:equilibb}
\vec{B} = \nabla \alpha \times \nabla \psi =I \left( \psi \right) \nabla \zeta + \nabla \zeta \times \nabla \psi,
\end{equation}
with $I\left(\psi \right)$ characterizing the strength of the toroidal magnetic field by $B_\zeta = I\left(\psi\right)/R$, where $R$ is the major radius coordinate. We do not consider background electric fields because they do not affect alpha particle trajectories as strongly as the magnetic fields do.\footnote{This follows from the expression for the particle drifts, given in~\eqref{eq:vdtot}, where we can find that for alphas the tangential component of the $\vec{E}\times \vec{B}$ drift is given by $\vec{v}_{\vec{E}\times \vec{B}}\cdot \nabla \alpha = -c \partial\Phi_0/\partial \psi$, with $\Phi_0$ the background potential, while the $\nabla B$ drift is given by $\vec{v}_{\nabla B}\cdot \nabla \alpha  \sim -\rho_\alpha v_0 /\left(aR\right)$, with $\rho_\alpha$ the alpha gyroradius and $v_0$ the alpha birth speed.  Typically, $\partial \Phi_0/ \partial\psi \sim -\left(T_i/ en_i \right) \partial n_i/ \partial \psi $, with $T_i$ and $n_i$ the background ion temperature and density, respectively. Taking the scale length of the ion population to be the device minor radius $a$ gives $\vec{v}_{\vec{E}\times \vec{B}}\cdot \nabla \alpha \sim \rho_iv_i/a^2$, with $\rho_i$ the ion gyroradius.  The tangential component of the $\nabla B$ drift is then larger than the tangential component of the $\vec{E}\times \vec{B}$ drift by a factor of $\epsilon \rho_\alpha v_0 / \left( \rho_i v_i \right)$, which is large. } As noted previously, the unit vector corresponding to this field is
\begin{equation}
\frac{\vec{B}}{B} \equiv \hat{b}.
\end{equation}
The poloidal component of the field is denoted $B_p$, and can be found from $B_p \approx \epsilon B/q$, with the inverse aspect ratio $\epsilon \approx r /R $, where $r$ is local the minor radius. [The flux coordinate and the minor radius are related by $\partial/\partial \psi = \left(1/ RB_p \right) \partial /\partial r$.] The shear of the field is given by  
\begin{equation}
\label{eq:shear}
    s\equiv \left(r/q\right) \partial q /\partial r.
\end{equation}

The total magnetic and electric fields affecting the alpha particles will include both the equilibrium magnetic field and magnetic and electric fields resulting from the perturbations, which are discussed in section~\ref{sec:perts}. The total magnetic field from all of these sources is denoted $\vec{B}_{tot}$ (with unit vector $\hat{b}_{tot}$) and the total electric field is $\vec{E}_{tot}$. Several parameters can be defined in terms of the total field or the unperturbed field; the total quantities are in general only used before the perturbation analysis is carried out. 

 Alpha particles in these fields are characterized by their velocity, $v$ (or, equivalently, energy normalized to mass, $\mathcal{E} \equiv v^2/2$), the sign of $v_\parallel$, their velocity parallel to the equilibrium magnetic field, and their pitch angle, which may be defined relative to the total magnetic field,
\begin{equation}
\label{eq:lambdatot}
\lambda_{tot} = \frac{B_{0} v_{\perp,tot}^2}{B_{tot} v^2},
\end{equation}  
with $B_0$ the equilibrium on-axis magnetic field strength, or relative to the unperturbed field,
\begin{equation}
\label{eq:lambdal}
\lambda= \frac{B_{0} v_\perp^2}{B v^2}.
\end{equation} 
(Here,  $v_{\perp,tot}$ is the speed perpendicular to total field and $v_\perp$ is the speed perpendicular to the unperturbed field.)

The alpha particle mass and charge are given by $M_\alpha$  and $Z_\alpha $, respectively; $\Omega_{tot}  \equiv Z_\alpha eB_{tot}/M_\alpha c$ is the gyrofrequency in the total field and $\Omega \equiv Z_\alpha eB/M_\alpha c$ the gyrofrequency in the unperturbed field.  The alpha particle poloidal gyrofrequency is $\Omega_p \equiv Z_\alpha eB_p/M_\alpha c$. The alpha particle gyroradius is $\rho_\alpha \equiv v_\perp/ \Omega$, and the poloidal gyroradius $\rho_{p \alpha} \equiv v_\perp/ \Omega_p$.
\subsection{Drift kinetic equation and equilibrium alpha particle distribution}
\label{sec:dke}
We study the effect of perturbations on the energetic alpha particle distribution function $f$ using Hazeltine's drift kinetic equation \citep{hazeltine1973recursive}. This formalism is appropriate when $\rho_\alpha$ (which at the alpha particle birth speed is typically a couple of centimeters) is much less than the length scales relevant to the problem, including the perpendicular scale length of the perturbation and the scale length  of the alpha particle density,
\begin{equation}
\label{eq:aalpha}
    a_\alpha \equiv - \frac{n_\alpha}{R B_p \partial n_\alpha /\partial \psi},
\end{equation}
with $n_\alpha$ the alpha particle density. The drift kinetic equation reads:
\begin{multline}
\label{eq:dke}
\frac{\partial f}{\partial t} + \left(v_{\parallel , tot} \hat{b}_{tot} + \vec{v}_{d,tot} \right) \cdot \nabla f + \left[\frac{Z_\alpha e}{M_\alpha } \left(v_{\parallel,tot} \hat{b}_{tot} + \vec{v}_{d,tot} \right)\cdot \vec{E}_{tot} + \frac{\lambda_{tot} v^2}{2 B_0} \frac{\partial B_{tot}}{\partial t} \right] \frac{\partial f}{\partial \mathcal{E}} 
\\
= C\left\{f\right\} +\frac{S_{fus} \delta \left(v-v_0\right)}{4 \pi v^2}.
\end{multline}
Here, $S_{fus}$ is the source rate due to fusion, $v_0$ is the alpha particle birth speed (${1.3\times10^{9}}$ cm s$^{-1}$), and $\vec{E}_{tot}$ is the total electric field (equal to the contribution from the perturbations, since we neglect the background electric field). The velocity parallel to the total field is $v_{\parallel,tot}$. The energetic alpha collision operator represents collisions with the background plasma electrons and ions, and is given by \citep{cordey1976effects,catto2018ripple}
\begin{equation}
\label{eq:collop}
C\left\{f \right\} = \frac{1}{\tau_s v^2} \frac{\partial}{\partial v} \left[ \left(v^3 + v_c^3 \right) f \right] + \frac{2 v_\lambda^3 B_{0}}{\tau_s v^3 B}   \frac{v_\parallel}{v} \frac{\partial}{\partial \lambda} \left( \lambda \frac{v_\parallel}{v} \frac{\partial f}{\partial \lambda} \right).
\end{equation}
(Note that the distinction between the total and the unperturbed field is unimportant in the collision operator.)
The first term represents electron and ion drag while the second represents pitch angle scattering off of bulk ions. 
Here, the alpha slowing down time is given by
\begin{equation}
 \tau_s\left(\psi \right) = \frac{3M_\alpha T_e^{3/2}\left(\psi \right)  }{4 \left( 2 \pi m_e \right)^{1/2} Z_\alpha ^2 e^4 n_e\left(\psi \right) \ln \Lambda_c},
 \end{equation}
 with $\ln \Lambda_c$, $T_e$, $n_e$, and $m_e$, the Coulomb logarithm, the electron temperature, density, and mass, respectively.  The critical speed at which alpha particles switch from being mainly decelerated by electrons to being mainly decelerated by ions is found by summing over background ions,
 \begin{equation}
 v_c^3\left(\psi \right) = \frac{3 \pi^{1/2} T_e^{3/2} \left(\psi \right)}{\left(2m_e\right)^{1/2} n_e \left(\psi \right)} \sum_i \frac{Z_i^2 n_i\left(\psi \right)}{M_i},
 \end{equation}   
 with $Z_i$, $n_i$, and $M_i$ the charge, density, and mass of each of the background species. This is of similar size to $v_\lambda$, the speed at which pitch angle scattering is important to the behavior of the equilibrium energetic alpha population, which does not have the narrow boundary layers encountered later:
 \begin{equation}
 v_\lambda^3\left(\psi \right) \equiv \frac{3 \pi^{1/2} T_e^{3/2} \left(\psi \right)}{\left(2m_e\right)^{1/2} n_e \left(\psi \right) M_\alpha } \sum_i Z_i^2 n_i\left(\psi \right).
 \end{equation}
The drift velocity is given by
\begin{equation}
\label{eq:vdtot}
\vec{v}_{d,tot} =  \frac{c}{B_{tot}^2} \vec{E}_{tot}\times \vec{B}_{tot} + \frac{\lambda_{tot} v^2}{2 B_0\Omega_{tot}}\hat{b}_{tot} \times \nabla B_{tot} + \frac{v_\parallel^2}{\Omega_{tot}} \hat{b}_{tot} \times \left( \hat{b}_{tot} \cdot \nabla \hat{b}_{tot} \right),
\end{equation}
where the different terms represent, respectively, the $\vec{E} \times \vec{B}$ drift, the $\nabla B$ drift, and the curvature drift. 
We  use $\vec{v}_d$ to refer to the zeroth order drift in the presence of only the unperturbed field.

The alpha particle distribution, $f$, consists of an equilibrium component, $f_0$, and a response to any electromagnetic perturbations, $f_1$. Using~\eqref{eq:dke}, $f_0$ can be calculated to be the familiar slowing down distribution: 

 \begin{equation}
\label{eq:slowingdown}
f_0\left(\psi, v \right)= \frac{S_{fus}\left(\psi \right) \tau_s\left(\psi \right) H\left(v_0-v\right)}{4 \pi \left[v^3 + v_c^3\left(\psi\right)\right]}.
\end{equation}
The derivation of this expression for suprathermal alphas is given in Appendix~\ref{sec:slowingdown}. 

\subsection{Example tokamak parameters}
When numerical examples are needed in this paper, we use the equilibrium parameters given in table~\ref{tab:equilib}, which are similar to those planned for SPARC, a DT tokamak experiment currently under development \citep{creely2020,pablo2020}.

\begin{table}
  \begin{center}
\def~{\hphantom{0}}
  \begin{tabular}{lc}
Parameter & Value \\[3pt] 
$B_{0}$&$1.2 \times 10^{5}$ G \\ 
$R$  & $185$ cm\\
$n_e$, $n_i$  & $4 \times 10^{14}$ cm$^{-3}$\\
$T_e$  & $20$  keV\\
$v_0$  & $1.3 \times 10^{9}$  cm s$^{-1}$\\
  \end{tabular}
  \caption{Example tokamak parameters used in this paper, similar to those planned for SPARC \citep{creely2020,pablo2020}. The bulk plasma is assumed to be an equal mix of deuterium and tritium. For convenience, this table includes the alpha particle birth speed, $v_0$, even though this parameter is the same in any tokamak.}
  \label{tab:equilib}
  \end{center}
\end{table}

\section{Form of electromagnetic perturbations}
\label{sec:perts}
The aim of this paper is to calculate the response of the tokamak background described previously to a magnetic field perturbation, an electric field perturbation, or a combination thereof. In this section, we describe the form of these perturbations and discuss the example cases used in this paper. 
\subsection{General description of perturbations}
As discussed in~\ref{sec:dke}, the drift kinetic treatment bounds the scale length of the perturbation to be much longer than the alpha particle gyroradius, but the form of the perturbation is otherwise flexible. In general, perturbations to the tokamak background field can be represented by a perturbed vector potential,
\begin{equation}
\label{eq:adef}
\vec{A_1} = \vec{A}_\perp \left( \psi, \vartheta, \alpha,t \right) + A_\parallel \left(  \psi , \vartheta, \alpha, t\right) \hat{b},
\end{equation}
and a perturbed electric potential $\Phi\left(\psi, \vartheta, \alpha ,t \right)$.  These correspond to a perturbed electric field $\vec{E}_1 = -\nabla \Phi  - \left(1/c \right) \partial \vec{A}_1/\partial t$ and a perturbed magnetic field $\vec{B}_1 = \nabla \times \vec{A}_1$. We refer to the parallel perturbed magnetic field resulting from $\vec{A}_{\perp}$ as $B_{1 \parallel}$.
The overall magnetic field unit vector $\hat{b}_{tot}$ is related to the perturbation and the background field by
\begin{equation}
\hat{b}_{tot} \approx \hat{b} + B^{-1} \left(\nabla A_\parallel \times \hat{b} \right).
\end{equation}
A perturbation can be characterized by a frequency $\omega$, a toroidal mode number $n$, a poloidal mode number $m$, and an effective radial wave number $k_\psi$, or by a sum of components each characterized by these quantities. These fully-general forms are maintained in the paper until specific examples are computed in sections~\ref{sec:ripp} and~\ref{sec:TAEflux}. 
\subsection{Example perturbations: ripple and TAE}
\label{sec:chapex}
The first example perturbation considered in this paper is ripple, which results from the discrete nature of tokamak field coils. Ripple is composed of  a perpendicular vector potential perturbation and corresponding parallel magnetic field $B_{1\parallel}$ and is  time independent. It is characterized by a high $n \approx 18$ and low $m$ in most devices; its radial scale length is roughly determined by the device size or some fraction thereof, so $k_\psi \sim 1/a$ (note that $k_\psi$ depends on several other parameters, including $n$, but we require only knowledge of the rough size of  $k_\psi$ for ripple). These characteristics mean that the drift kinetic formalism is appropriate.  Ripple is typically of strongest magnitude on the low field side in regions of high safety factor $q$, high minor radius normalized to major radius $\epsilon$, and high shear $s$. [Information on ripple in ITER can be found in \citet{kurki2009ascot}; information on ripple in SPARC in~\citet{scott2020}.] A summary of these typical values, as well as the specific example values of key ripple parameters used to demonstrate the evaluation of alpha heat flux, is given in table~\ref{tab:perts}.

The second example perturbation is the TAE. The TAE is a special class of Alfv\'{e}n wave that exists in a tokamak; its structure is described by magnetohydrodynamics \citep{van2006radial}. The TAE is destabilized by energetic particles \citep{fu1989excitation}. TAEs, and many other MHD modes, are time dependent, with magnetic fields perpendicular to the background field, and result in no parallel electric field \citep{white2013theory}.\footnote{Inclusion of finite ion gyroradius effects introduces a parallel electric field, but these effects are small and are outside the scope of the MHD model often used to describe TAE structure.}  They are represented by a vector potential with $\vec{A}_\perp = 0$ and an electric potential that obeys
\begin{equation}
\label{eq:phidef}
-\frac{1}{c}\frac{\partial A_\parallel }{\partial t}- \hat{b} \cdot \nabla \Phi  = E_\parallel = 0.
\end{equation}
[This representation of the modes is used frequently in orbit-following codes \citep{hirvijoki2012alfven}.] The TAE frequency is given by $\omega \approx v_A/ \left(2q R\right)$, with $v_A$ the on-axis Alfv\'{e}n speed, $v_A \equiv B_0/\sqrt{4 \pi n_i m_i}$, with $m_i$ the average ion mass \citep{heidbrink2008basic}. TAEs mostly exist in the core and outer core of the tokamak, where  $q$, $\epsilon$, and $s$ have low to moderate values \citep{rodrigues2015systematic}.  A simple analytic estimate predicts that the most strongly driven TAEs are characterized by toroidal mode numbers $n$ ranging from $\sqrt{\epsilon} R \Omega_p /\left(q v_A\right)$ to $ R \Omega_p /\left(q v_A\right)$, depending on if TAEs are driven primarily by trapped or passing particles  \citep{fu1992excitation,breizman1995energetic,fulop1996finite,rodrigues2015systematic}.\footnote{This estimate results from balancing the width of the mode $1/k_\psi \sim \epsilon R/ \left(n q\right)$, with the width of the particle orbit of resonant alpha particles  (which scales with $v_A\sqrt{\epsilon}/\Omega_p$ for trapped particles and with $v_A/\Omega_p$ for passing particles). Various variations of this balance exist in the literature.}  Alpha-driven TAEs interact with both trapped and passing particle populations, such that the most destabilized values of $n$ will be intermediate between these values. However, for the specific value of $n$ we use to compute numerical examples in this paper, we select a value near the bottom of the range in order to strictly obey the drift kinetic ordering. (Higher $n$ modes are narrower and their width may approach the alpha gyroradius, such that drift kinetics is not a good description of their behavior.) The poloidal mode number $m$ is given by $nq-1/2$ \citep{heidbrink2008basic} and the radial wave number $k_\psi$ is given by $k_\psi = nq /\left(\epsilon R\right)$ \citep{breizman1995energetic,rodrigues2016sensitivity,tolman2019dependence}. The condition~\eqref{eq:phidef} and the TAE parameters can be used to find $\Phi = v_A A_\parallel /c$. A summary of the typical TAE characteristics and the specific values used as examples in this paper are given in table~\ref{tab:perts}.
\begin{table}
  \begin{center}
\def~{\hphantom{0}}
  \begin{tabular}{lcccc}
Parameter & Typical ripple val. & Example ripple val. &Typical TAE val. & Example TAE val. \\ [3pt] 
$q $&high&$3$& $\sim 1$ & $1.15$ \\
$\epsilon$ &high&$0.3$& medium & $0.2$ \\
$s\equiv \frac{r}{q} \frac{\partial q}{\partial r}  $&high&$1$& low & $0$ \\
$A_{\parallel}$&$0$&$0$& $A_{\parallel} $& N/A \\ 
$\Phi$  & $0$& $0$ & $v_A A_{\parallel}/c$ & N/A\\
$B_{1\parallel}$  & $B_{1\parallel}$ & N/A&  $0$ & $0$\\
$n$  & $\approx 18$ &$18$ & $\sqrt{\epsilon} R \Omega_p/\left(q v_A\right)$ to  $R \Omega_p/\left(q v_A\right)$  & $10$ \\
$m$  & $0$ &$0$ &  $nq-1/2$ & $11$ \\
$\omega$  & $ 0 $&$0$ & $v_A / \left(2qR\right)$ & $v_A/ \left(2qR\right)$\\
$k_\psi$  & $\frac{1}{a }$ &$\frac{1}{a}$ & $nq/ \left(\epsilon R \right)$ &$nq/\left(\epsilon R \right)$\\
  \end{tabular}
  \caption{Defining characteristics of this paper's two demonstration perturbations and typical equilibrium parameters where these perturbations exist. The value $n$ is, for ripple, a common ripple periodicity, and, for TAEs, the range of strongly-driven toroidal mode numbers. However, other values of $n$ are also sometimes of interest for these perturbations. The specific values of these quantities used for numerical examples in this paper are also included. "N/A" denotes parameters for which specific example values are not needed.  }
  \label{tab:perts}
  \end{center}
\end{table}

\section{Plasma response to perturbations}
\label{sec:response}
In this section, we calculate the plasma response to the perturbation described in the previous section. Doing so requires developing a phenomenological understanding of the transport and the role of the collision operator. Later, in section~\ref{sec:flux}, this plasma response and an expression for the radial velocity resulting from the perturbation will be used to calculate the alpha heat flux caused by the perturbation.
\subsection{Initial equation setup}
To first order in the perturbation, the drift kinetic equation~\eqref{eq:dke} is
\begin{multline}
\partial_t f_1 +\left(v_\parallel \hat{b} + \vec{v}_{d} \right)\cdot \nabla f_1+ \left[\left(v_\parallel \hat{b}_{tot} + \vec{v}_{d,tot} \right)\cdot \nabla \psi \right]_1\frac{\partial f_0}{ \partial \psi} + \left[ \frac{Z_\alpha e}{M_\alpha } \left(v_\parallel \hat{b} +\vec{v}_d \right)\cdot \vec{E}_1 \right] \frac{\partial f_0}{\partial \mathcal{E}}
\\
= C\left\{f_1\right\},
\end{multline}
where $f_1$ is the distribution function response to the perturbation and $\left[\left(v_\parallel \hat{b}_{tot} + \vec{v}_{d,tot} \right)\cdot \nabla \psi \right]_1$ is the component of $\left[\left(v_\parallel \hat{b}_{tot} + \vec{v}_{d,tot} \right)\cdot \nabla \psi \right]$ resulting from the perturbation.  
Further analysis of this equation is simplified by working in a coordinate that follows the alpha particle orbit, which has a finite departure from a flux surface. This variable is the following constant of the alpha particle's motion:\footnote{This drift kinetic angular momentum is a constant of the alpha particle's motion because the left-hand side of~\eqref{eq:dke0} operating on it is zero: $\left(v_\parallel \hat{b} + \vec{v}_d \right) \cdot \nabla \psi_\star = \vec{v}_d \cdot \nabla \psi - v_\parallel \hat{b} \cdot \nabla \left(I v_\parallel /\Omega \right) -\vec{v}_d \cdot \nabla \left( I v_\parallel/\Omega \right)$. It can be shown~\citep{helander2005collisional,parra2010transport} that $\vec{v}_d \cdot \nabla \psi = v_\parallel \hat{b} \cdot \nabla \left( I v_\parallel/\Omega \right)$ and $\vec{v}_d\cdot \nabla \left(I v_\parallel/ \Omega \right) = 0$, such that $\left(v_\parallel \hat{b} + \vec{v}_d \right) \cdot \nabla \psi_\star = 0$.}  
\begin{equation}
\label{eq:psistar}
    \psi_\star \equiv \psi - \frac{I v_\parallel}{\Omega}.
\end{equation}
The coordinate $\alpha$ is modified from~\eqref{eq:alphadef} in accordance:
\begin{equation}
    \alpha_\star \equiv \zeta - q_\star \vartheta,
\end{equation}
with 
\begin{equation}
\label{eq:qstar}
    q_\star = q \left(\psi_\star\right)
\end{equation} the safety factor evaluated at $\psi_\star$ instead of $\psi$.  Consistent with the approximation~\eqref{eq:gradapprox}, we can take that $\partial f_0/\partial \psi_\star \approx \partial f_0/ \partial \psi$.

The perturbed distribution $f_1$ has two parts,  such that $f_1 = h +  \left(Z_\alpha e\Phi/M_\alpha \right) \partial f_0/\partial \mathcal{E}$.  The second term is an adiabatic response to the perturbation, which does not participate in transport (it is exactly out of phase with the perturbed velocity that causes transport).  Evaluation of $\left[ \left(v_\parallel \hat{b}_{tot} + \vec{v}_{d,tot} \right)\cdot \nabla \psi \right]_1$ gives that the equation governing $h$ is 
\begin{multline}
\label{eq:govh}
\partial_t h+ \left( v_\parallel  \hat{b}+ \vec{v}_{d} \right) \cdot \nabla h -  C\left\{ h + \frac{Z_\alpha e\Phi}{M_\alpha } \frac{\partial f_0}{\partial \mathcal{E}} \right\}=
\\
 -\frac{Z_\alpha e}{M_\alpha } \left( \frac{\partial \Phi}{\partial t} - \frac{v_\parallel}{c} \frac{\partial A_\parallel}{\partial t} + \frac{\lambda v^2}{ 2 c \Omega} \frac{\partial B_{1\parallel}}{\partial t}\right)\frac{\partial f_0}{\partial \mathcal{E}} - c \left(\frac{\partial \Phi}{\partial \alpha_\star} - \frac{v_\parallel}{c} \frac{\partial A_\parallel}{\partial \alpha_\star } + \frac{\lambda v^2}{ 2c \Omega} \frac{\partial B_{1\parallel}}{\partial \alpha_\star}  \right) \frac{\partial f_0}{\partial \psi_\star}.
\end{multline}
Note that the collision operator acting on the adiabatic component is not zero, as it sometimes is in similar calculations. 

The perturbed fields $\Phi$, $A_\parallel$, and $B_{1\parallel}$ are functions of $\psi$, $\zeta$, and $\vartheta$ and can be Fourier analyzed in those coordinates, then rewritten in terms of $\psi$, $\alpha$, and $\vartheta$. For example, the electric potential is given by
\begin{multline}
    \Phi \left(\psi, \vartheta, \alpha \right) = \sum_{m,n,\omega} \Phi_{mn\omega} \cos{\left[n\alpha - \omega t + \left(nq -m \right) \vartheta + \int^\psi d\psi' \frac{k_\psi}{R B_p} \right]} 
    \\
    =
    \sum_{m,n, \omega}\Re{\left[\Phi_{mn\omega}e^{i n \alpha - i\omega t +i\left(n q-m \right)\vartheta + i \int^{\psi} d\psi' \frac{k_\psi}{R B_p} }\right]}.
\end{multline}
Moreover, for use in~\eqref{eq:govh}, the fields can also be written in terms of the starred coordinates, so that, for example,
\begin{equation}
    \Phi \left(\psi, \vartheta, \alpha \right) 
    =
    \sum_{m,n,\omega}\Re{\left[\Phi_{mn\omega}e^{i n \alpha_\star - i\omega t +i\left(n q_\star-m \right)\vartheta + i \int^{\psi_\star} d\psi' \frac{k_\psi}{R B_p} + \frac{ik_\psi I v_\parallel}{R B_p\Omega}}\right]}.
\end{equation}
 Due to the dominance of parallel streaming over drifts, we may take that $\vec{v}_d \cdot \nabla \vartheta \ll v_\parallel \hat{b}\cdot \nabla \vartheta$. Also, the form of the magnetic field introduced in~\eqref{eq:equilibb} can be used to show $v_\parallel \hat{b} \cdot \nabla h = v_\parallel \hat{b}  \cdot \nabla \vartheta \partial h / \partial \vartheta$. Then,~\eqref{eq:govh} becomes
\begin{multline}
\label{eq:tosolve}
\partial_t h + v_\parallel \hat{b} \cdot \nabla \vartheta \frac{\partial h}{\partial \vartheta} + \omega_{\alpha_\star} \frac{\partial h}{\partial \alpha_\star} 
 \\
  = C\left\{ h +\frac{Z_\alpha e\Phi}{M_\alpha } \frac{\partial f_0}{\partial \mathcal{E}} \right\} -i \sum_{m,n}D_{n \omega}S_{mn\omega}e^{i n \alpha_\star - i \omega t + i \left(nq_\star-m \right)\vartheta+ i\int^{\psi_\star} d\psi \frac{k_\psi}{RB_p} + \frac{i k_\psi I v_\parallel}{R B_p\Omega}} ,
\end{multline}
where\footnote{The quantity $\omega_{\alpha_\star}$ is found as follows:
$\left(v_\parallel \hat{b} + \vec{v}_d \right) \cdot \nabla \alpha_\star =  \left(v_\parallel \hat{b} + \vec{v}_d \right) \cdot \left(\nabla \zeta - q_\star \nabla \vartheta \right)$ $= \left(v_\parallel \hat{b} + \vec{v}_d \right)\cdot \left[ \nabla \zeta - q \nabla \vartheta + \left(Iv_\parallel /\Omega \right) \left( \partial q /\partial \psi \right) \nabla \vartheta \right]$ $= \left(v_\parallel \hat{b} + \vec{v}_d \right)\cdot \left[ \nabla \alpha + \vartheta \nabla q + \left(Iv_\parallel /\Omega \right) \left( \partial q /\partial \psi \right) \nabla \vartheta \right]$ $\approx \vec{v}_d\cdot \nabla \alpha + v_\parallel \hat{b} \cdot \nabla \left[\vartheta \left(I v_\parallel/\Omega \right) \left(\partial q / \partial \psi \right) \right]$. In the first equality, we have used that $\left(v_\parallel \hat{b} + \vec{v}_d \right)\cdot \nabla \psi_\star = 0$; in the last equality, we have used that $v_d\cdot \nabla \psi = v_\parallel \hat{b} \cdot \left(I v_\parallel/ \Omega \right)$. More information about the evaluation of $\vec{v}_d \cdot \nabla \alpha$ can be found in \citet{catto2019collisional}.}
\begin{equation}
\label{eq:omalphastar}
    \omega_{\alpha_\star} \equiv \vec{v}_d \cdot \nabla \alpha + v_\parallel \hat{b} \cdot \nabla \left( \vartheta \frac{I v_\parallel}{\Omega} \frac{\partial q}{\partial \psi} \right) = v_\parallel \hat{b} \cdot \nabla \vartheta \frac{\partial}{\partial \psi} \left( \frac{B v_\parallel}{\Omega \hat{b} \cdot \nabla \vartheta} \right)
\end{equation}
and  $S_{mn\omega}$ is given by 
\begin{equation}
 S_{mn\omega}\left(\vartheta, v, \lambda \right) =  \Phi_{mn\omega} - \frac{v_\parallel}{c} A_{\parallel mn\omega} + \frac{\lambda v^2}{2 c \Omega} B_{1 \parallel mn\omega}.
\end{equation}
(The $\vartheta$ dependence displayed in $S_{mn\omega}$ is due to the dependence of $v_\parallel$ on this quantity.)
Also, we define
\begin{equation}
\label{eq:Dnw}
    D_{n \omega}\left(v, \psi_\star\right) \equiv c n \frac{\partial f_0}{\partial \psi_\star}\left(1 - \frac{\omega}{n \omega_\star} \right)
\end{equation}
with 
\begin{equation}
\omega_\star \equiv \frac{c M_\alpha \partial f_0/ \partial \psi_\star}{Z_\alpha e \partial f_0/\partial \mathcal{E}}.
\end{equation}
Note that the drifts on the left side of~\eqref{eq:tosolve} are caused by the unperturbed fields. Thus, we can employ the following form of $h$:
\begin{equation}
\label{eq:postulate}
h\left(\psi_\star, \vartheta, \alpha_\star,t, v, \lambda, \sigma \right) =\sum_{n,\omega} h_{n\omega}\left(\vartheta, v, \lambda, \sigma \right)e^{i n \alpha_\star -i \omega t+ i \int^{\psi_\star} d\psi \frac{k_\psi}{RB_p}}.
\end{equation}
Here, we have introduced the variable $\sigma$, defined as follows:
\begin{equation}
\label{eq:sigmadef}
    \sigma = \left\{
     \begin{array}{@{}l@{\thinspace}l}
      0, \, \text{trapped particles}\\
      \frac{v_\parallel}{\left|v_\parallel\right|}, \, \text{passing particles} \\
     \end{array}
  \right. .
\end{equation}

\subsection{Role of collisionality, phenomenological description of transport, and form of collision operator}
At this point, a discussion of the role of collisionality in the transport reveals the appropriate treatment of the collision operator in~\eqref{eq:tosolve}. This role is directly analogous to that played by collisionality in superbanana plateau tokamak transport [see discussion in \citet{shaing2015superbanana,calvo2017effect,catto2019collisional2}]. It is also similar to the role played by collisionality in the  plateau regime of neoclassical transport [see \citet{helander2005collisional}], in the damping of plasma echos [see \citet{su1968collisional}], and in other wave-particle resonance processes \citep{duarte2019collisional,catto2020collisional}. 

\begin{figure}
\centering
\includegraphics[width=.6\columnwidth]{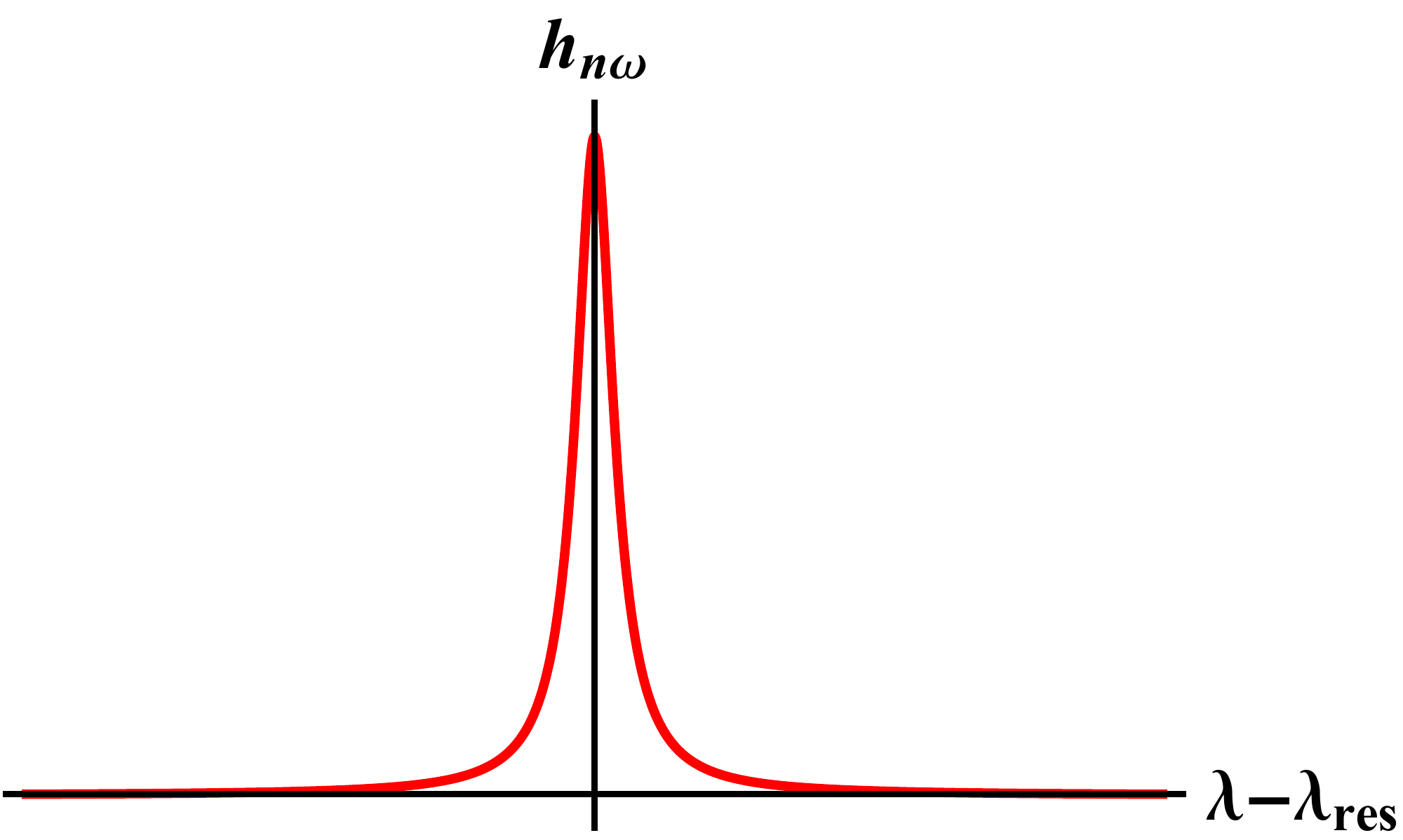}
\caption{A schematic representation of the boundary layer around the resonance. A specific value of pitch angle, $\lambda =\lambda_{res}$, is resonant with the mode, and the non-adiabatic alpha response $h_{n\omega}$ is largest for this value of $\lambda$. Pitch angle scattering collisions broaden $h_{n\omega}$ about this resonant value.  The sharp variation of $h$ near the resonant $\lambda_{res}$ enhances the importance of the pitch angle scattering operator [given by the second term in~\eqref{eq:collop}], which contains a second derivative with respect to $\lambda$.}
\label{fig:schemres}
\end{figure}

Our analysis of the transport equation~\eqref{eq:tosolve} will reveal that the effect of the perturbations on particle motion is highest for particles which are in resonance with the mode, that is, those for which all terms on the left side of~\eqref{eq:tosolve} vanish. If there were no collisionality, the effect of the perturbation drift [represented by the second term on the left side of~\eqref{eq:tosolve}], on such resonant particles would diverge.   So collisionality is necessary to resolve the singularity of the resonance.\footnote{Here, we refer only to behavior existing within the equation we solve: a linearized drift kinetic equation with a perturbation of zero or real frequency. Physically, it is possible for other effects to resolve the resonance. For instance, resonant particles will dephase from the perturbation as they are moved by it, resulting in nonlinear resolution of the resonance through the formation of phase space islands. The addition of a small imaginary component to the perturbation frequency is also able to resolve the resonance. However, it is not clear that phase space island formation (in the absence of resonance overlap) or an imaginary component of the frequency provide the decorellation and dissipation necessary for transport. The relationship between these mechanisms and pitch angle collisions in resolving the resonance and in causing transport is an interesting avenue for future work. One possibility is that resonance resolution by these mechanisms is similar enough to the collisional resolution that fluxes calculated in this paper are valid in cases where collisional resolution is less important than other mechanisms.} A small boundary layer of finite width in pitch angle space and velocity space will collisionally disrupt resonance with the mode, and the perturbed distribution function will vary sharply near this layer. Pitch angle scattering, represented by the second term in~\eqref{eq:collop}, rather than drag, is the important collisional behavior because  the second derivative of the pitch angle scattering operator is stronger near the resonance boundary layer than the first derivative of the drag. A schematic picture of $h$ about a resonant location in pitch-angle space is given in figure~\ref{fig:schemres}. A possible physical interpretation of this resonance structure is that particles drift while resonant with the wave until they pitch angle scatter, giving rise to gradual diffusion~\citep{helander2005collisional}.  

This understanding yields a phenomenological estimate of the heat flux which should result from the transport being considered. The heat flux will be given by
\begin{equation}
\label{eq:initest}
Q_\alpha\sim M_\alpha v_{res}^2 \delta \lambda  \left\{  \left[ \left(v_\parallel \hat{b}_{tot} + \vec{v}_{d,tot} \right)\cdot \nabla \psi \right]_1 \Delta t \right\}^2 \Delta t^{-1} \frac{\partial n_\alpha}{\partial \psi},
\end{equation}
with $M_\alpha$  the alpha particle mass, $v_{res}$ a characteristic alpha resonant velocity, $\delta \lambda$ the fraction of particles that are resonant with the mode, $  \left[ \left(v_\parallel \hat{b}_{tot} + \vec{v}_{d,tot} \right)\cdot \nabla \psi \right]_1$ the radial velocity caused by the perturbation, $\Delta t$ the time that a particle moves before becoming decorrelated by a collision (such that $  \left[ \left(v_\parallel \hat{b}_{tot} + \vec{v}_{d,tot} \right)\cdot \nabla \psi \right]_1 \Delta t$ is the step a particle takes before becoming decorrelated). The fraction of particles affected by the mode is estimated from the width of the  boundary layer representing the broadened resonance, given by balancing the left side of~\eqref{eq:tosolve} (we use $\omega_{\alpha_\star}$ to include pitch angle dependence) with the pitch angle scattering part of the collision operator,~\eqref{eq:collop},
\begin{equation}
n \omega_{\alpha_\star} \delta \lambda \sim  \frac{\nu_{pas}}{\delta \lambda^2}, 
\end{equation}
with $\nu_{pas}$ representing the coefficient of the second term of~\eqref{eq:collop}. [Note that the effect of the collision operator on the adiabatic component, $\left(Z_\alpha e\Phi/M_\alpha \right) \partial f_0 / \partial \mathcal{E} $, is negligible in comparison to the sharp variation of $h$ in the resonant boundary layer.] 
Thus, we have
\begin{equation}
\label{eq:deltlambda}
\delta \lambda \sim \left( \frac{\nu_{pas}}{n \omega_{\alpha_\star} }\right)^{1/3}.
\end{equation}
The decorrelation time $\Delta t$ is given by $\Delta t \sim \delta \lambda^2/\nu_{pas}$, such that estimated flux is
\begin{equation}
\label{eq:phenom}
Q_\alpha \sim \frac{M_\alpha v_{res}^2\left[ \left(v_\parallel \hat{b}_{tot} + \vec{v}_{d,tot} \right)\cdot \nabla \psi \right]_1^2}{n\omega_{\alpha_\star}} \frac{\partial n_\alpha}{\partial \psi}.
\end{equation}

Note that this flux is independent of the collision frequency $\nu_{pas}$ because $\delta \lambda$ times $\Delta t$ is independent of the collision frequency. The role of the collisions is crucial to the transport, but collisionality does not appear in the final expression for the flux. Thus, in the following steps we  use a Krook collision operator \citep{bhatnagar1954model,mynick1986generalized}, that is, an operator for which $C \left\{ h\right\} = - \nu h$. The Krook operator $\nu$ is related to $\nu_{pas}$ by $\nu \sim \nu_{pas}/ \delta  \lambda^2$ [recall that the pitch angle scattering operator~\eqref{eq:collop} depends on the second derivative of $h$ with respect to $\lambda$]. We also ignore the effect of the collision operator on the adiabatic response because the sharp variation of $h$ is more important. Whenever collisions play a role in the calculation, they allow the convergence of a sum or resolve a singularity, actions which are not qualitatively sensitive to the full form of the collision operator.
\subsection{Solution for  \texorpdfstring{$h$}{h}  with the Krook operator}
Inserting the Krook operator and the form~\eqref{eq:postulate} into~\eqref{eq:tosolve} reveals the equation to be solved for each $h_{n\omega}$:
\begin{equation}
\label{eq:initialeq}
v_\parallel \hat{b} \cdot \nabla \vartheta \frac{\partial h_{n\omega}}{\partial \vartheta} - i \left(\omega - n\omega_{\alpha_\star} + i\nu \right)h_{n\omega} = -iD_{n \omega} \sum_m S_{mn\omega}e^{i \left(nq_\star-m \right)\vartheta +\frac{i k_\psi I v_\parallel}{R B_p \Omega}}.
\end{equation}
Now, recall  the variable $\tau$, which characterizes the progression of particles along the magnetic field [see~\eqref{eq:deftau}]. Then, defining 
\begin{equation}
\label{eq:Lambda}
 \Lambda\left(\tau\right) \equiv \omega - n \omega_{\alpha_\star}-  v_\parallel \hat{b} \cdot \nabla \left[\left(nq_\star -m \right)\vartheta + \frac{k_\psi I v_\parallel}{R B_p\Omega} \right], 
\end{equation}  
\eqref{eq:initialeq} may be written
\begin{equation}
\frac{\partial}{\partial \tau} \left[h_{n\omega}  e^{-i\int_{\tau_0}^\tau d\tau ' \left( \omega - n \omega_{\alpha_\star} + i \nu \right)} \right] 
= -i D_{n\omega}\sum_mS_{mn\omega} e^{-i\int_{\tau_0}^\tau d\tau ' \left[ \Lambda \left(\tau' \right) + i \nu \right]}e^{\frac{ik_\psi I v_\parallel \left(\vartheta = 0 \right)}{R B_p\Omega}} .
\end{equation}
Here, we have chosen $\tau_0$ such that $\vartheta\left(\tau_0\right) = 0$. The perturbed distribution function $h_{n\omega}$ can then be found by integrating from $\tau = -\infty$, where $h_{n\omega}= 0$, forward to the "present" time $\tau=0$, with $\vartheta\left(\tau = 0\right) = \vartheta$,
\begin{equation}
\left. h_{n\omega}   e^{-i\int_{\tau_0}^\tau d\tau ' \left( \omega - n \omega_{\alpha_\star} + i \nu  \right)} \right|_{\tau = 0} 
= -iD_{n\omega}e^{\frac{ik_\psi I v_\parallel \left(\vartheta = 0 \right)}{R B_p\Omega}}\sum_m \int_{-\infty}^0 S_{mn\omega}   e^{-i\int_{\tau_0}^\tau d\tau ' \left[\Lambda\left(\tau' \right)+ i \nu  \right]  } d \tau .
\end{equation}
Moving the exponential on the left side to the right side gives
 gives
\begin{equation}
\label{eq:fullexpression}
h_{n\omega}  = -iD_{n \omega} \sum_m e^{i\left(nq_\star-m\right)\vartheta + \frac{i k_\psi I v_\parallel}{R B_p \Omega} }\int_{-\infty}^0 S_{mn\omega}   e^{-i\int_{0}^\tau d\tau ' \left[ \Lambda\left(\tau' \right)  + i \nu\right]} d \tau.
\end{equation}

The preceding expression represents the response of the plasma distribution to the presence of a perturbation of the tokamak fields. Next, we work to write this expression in a form which can be more easily evaluated and understood.
The expression~\eqref{eq:fullexpression} can be split into contributions from each complete bounce (for a trapped particle) or poloidal transit (for a passing particle) of duration given by
\begin{equation}
\label{eq:taub}
\tau_b\equiv \oint d\tau,
\end{equation}
such that
\begin{equation}
h_{n\omega} = -iD_{n\omega} \sum_m e^{i\left(nq_\star-m\right)\vartheta + \frac{i k_\psi I v_\parallel}{R B_p\Omega}}
\sum_{j=0}^{\infty} \left[ \int_{-\left(j+1\right)\tau_b }^{-j\tau_b}\right] S_{mn\omega}  e^{-i\int_{0}^\tau d\tau' \left[\Lambda\left(\tau' \right) + i \nu \right]} d \tau.
\end{equation} 
Note that at the trapped-passing boundary, the value of $\tau_b$ for a trapped particle is twice that for a passing particle. This complication will not affect the calculation that follows as velocity space integrations account for both signs of $v_\parallel$.
Each portion of the integral may be modified with the change of variables $\tau \rightarrow  \tau + j \tau_b$, such that the entire integral reduces to the geometric series
\begin{equation}
\label{eq:series}
h_{n\omega} = -iD_{n \omega} \sum_m e^{i\left(nq_\star-m\right)\vartheta + \frac{i k_\psi I v_\parallel}{RB_p\Omega}} \sum_{j=0}^{\infty}  e^{ij\oint d\tau \left[\Lambda \left(\tau \right)  + i \nu \right]} \int_{-\tau_b}^0 S_{mn\omega} e^{-i\int_{0}^\tau d\tau' \left[\Lambda\left(\tau' \right) + i \nu \right]} d \tau.
\end{equation}
This transformation requires that $\oint d \tau \Lambda \left(\tau \right)$ be the same (up to a factor of $2\pi$) for each bounce or transit, which corresponds to the existence of periodicity. Trapped particles, which traverse the same values of $\vartheta$ on each orbit, automatically fulfill this condition, because $\oint d \tau \Lambda \left(\tau \right)$ is trivially the same for each bounce. Passing particles do not reverse course and traverse values of $\vartheta$ ranging from negative infinity to infinity. For them, the split is only possible when the particle motion is resonant with the wave, such that the contributions to transport from each transit are the same.  When the particle is not in phase with the wave, however,~\eqref{eq:fullexpression} will approach zero integrated over the entire passing particle trajectory because of the rapid oscillation of the phase factor, unmitigated by periodicity. The following analysis is appropriate for all trapped particles and for resonant passing particles. The analysis will reveal in ~\eqref{eq:perturbed} onwards that transport comes from resonant particles only, such that the formulation is appropriate for all transport causing particles, passing and trapped.

Because of the small imaginary part provided by the collisionality, the series~\eqref{eq:series} is convergent and may be summed as 
\begin{equation}
h_{n\omega} = \sum_m\frac{-iD_{n\omega} e^{i\left(nq_\star-m\right)\vartheta + \frac{i k_\psi I v_\parallel}{RB_p\Omega}} \int_{-\tau_b}^0 S_{mn\omega}   e^{-i\int_{0}^\tau d\tau'   \left[\Lambda\left(\tau' \right)+ i \nu\right]} d \tau}{1- e^{i\oint d\tau \left[\Lambda \left(\tau \right)+ i \nu\right]} }.
\end{equation}
 Expanding the denominator about its zeros using $e^{i2 \pi l}=1$, with $l$ any integer, denoting the bounce or poloidal transit harmonic, gives 
\begin{equation}
\label{eq:toreduce}
h_{n\omega} =\sum_{m,l} \frac{ D_{n\omega} e^{i\left(nq_\star-m\right)\vartheta + \frac{i k_\psi I v_\parallel}{RB_p\Omega}} \int_{-\tau_b}^0 S_{mn\omega}  e^{-i\int_{0}^\tau d\tau ' \left[\Lambda\left(\tau' \right)+i \nu \right]}d \tau}{ \oint d\tau \left[\Lambda \left(\tau \right)  + i \nu \right]- 2 \pi l }.
\end{equation}
Alpha particles are in general very unlikely to pitch angle scatter because of their high energy.  Thus, for nearly all particles, $\nu \oint d\tau \ll \oint \Lambda d\tau - 2 \pi l $. In this limit, analogous to the limit that occurs in the plateau regime of neoclassical transport [see \citet{helander2005collisional}],~\eqref{eq:toreduce} approaches
 \begin{equation}
 \label{eq:perturbed}
h_{n\omega} =-i \pi D_{n \omega} \sum_{m,l}  \delta\left(  Q_l \right) e^{i\left(nq_\star-m\right)\vartheta + \frac{i k_\psi I v_\parallel}{RB_p\Omega}- i  \int_{0}^{\tau_0} d\tau ' \Lambda\left(\tau'\right) }
\oint S_{mn\omega}  e^{- i  \int_{\tau_0}^\tau d\tau' \Lambda\left(\tau'\right) } d \tau.
\end{equation}
Here, we have also split the integral in the exponential into two parts which both have limits at $\tau_0$, where $\vartheta\left(\tau_0\right)=0$. In addition, we have introduced a delta function $\delta \left(Q_l\right)$ with a resonance function as its argument,
\begin{equation}
\label{eq:res1}
Q_l\left(v,\lambda\right) \equiv\oint \Lambda d\tau -2 \pi l. 
\end{equation}
Only particles for which $Q_l$ is very close to zero participate in transport, and the strength of pitch angle scattering is accentuated for them from the normal weak level. The sharp variation in $h_{n\omega}$ between resonant particles and non-resonant particles (visible schematically in figure~\ref{fig:schemres}) leads to this accentuation. 

These resonances, discussed at greater length in section~\ref{sec:rescond}, are bounce harmonic resonances, which occur when the mode frequency, the particle bounce frequency, and the tangential precession frequency resonate. Such resonances are familiar from treatments of neoclassical toroidal viscosity of bulk tokamak plasmas~\citep{linsker1982banana,mynick1986generalized,park2009nonambipolar,kim2013numerical,logan2013neoclassical}  and also appear in treatments of Alfv\'{e}n eigenmode stability [see, for example, \citet{borba1999castor}] and in single particle analyses of energetic particle transport by neoclassical tearing modes, another type of tokamak perturbation \citep{poli2008observation}.

Finally, using~\eqref{eq:postulate} to obtain the full form of $h$ in terms of the unstarred coordinates gives the expression needed to form the heat flux,
\begin{multline}
    \label{eq:unstarred}
    h
    = -\sum_{m,n,\omega, l} \left\{ i \pi \left[e^{i n \zeta -i \omega t -im\vartheta + i \int^\psi d \psi' \frac{k_\psi}{RB_p} - i  \int_{0}^{\tau_0} d\tau ' \Lambda\left(\tau'\right)} \right]D_{n\omega}\delta \left(Q_l\right)\right.
    \\
    \left. 
\times \oint S_{mn\omega}  e^{- i  \int_{\tau_0}^\tau d\tau' \Lambda\left(\tau'\right) } d \tau \right\}.
\end{multline}

\section{Expression for flux} 
\label{sec:flux}
In this section, we show how to use the alpha distribution perturbation $h$ derived in the previous section to obtain the resulting alpha heat flux. This heat flux is the quantity of practical use in understanding and predicting tokamak discharges. In addition, we discuss key parts of the expression for this flux.
\subsection{Setup of flux expression}

To find the heat flux, the perturbation to the distribution function, $h$, found in~\eqref{eq:unstarred}, is multiplied by the perturbed radial velocity that also results from the electromagnetic perturbation, $ \left[ \left(v_\parallel \hat{b}_{tot} + \vec{v}_{d,tot} \right)\cdot \nabla \psi \right]_1$.\footnote{Some techniques for computing energetic particle transport, like the use of orbit following codes, focus only on the perturbed radial velocity. However, consistent consideration of both components is critical. An experimental demonstration of energetic particle transport caused by the product of the distribution function perturbation and the perturbed radial velocity is given in~\citet{nagaoka2008radial}.~\citet{todo2019introduction} discusses these results and the importance of developing energetic particle transport theories that self consistently treat both of these components.}  The result is weighted by energy and integrated over the distribution function.  Then, this quantity is averaged over the flux surface: 
\begin{equation}
\label{eq:gamma}
Q_\alpha =  \left<  \int d^3v \left(M_\alpha v^2/2 \right) h   \left[ \left(v_\parallel \hat{b}_{tot} + \vec{v}_{d,tot} \right)\cdot \nabla \psi \right]_1\right>.
\end{equation}
Here, the flux surface average is defined by
\begin{equation}
\left< A \right> \equiv  \frac{\oint \frac{A d\vartheta d \zeta}{\vec{B}\cdot \nabla \vartheta}}{ \oint \frac{d\vartheta d \zeta}{\vec{B}\cdot \nabla \vartheta}  } .
\end{equation}
To enable analytic evaluation of the flux, we now adopt the following approximate expression for the strength of the axisymmetric magnetic field:
\begin{equation}
\label{eq:bstrength}
B = B_{0} \left[1- \epsilon\left(\psi \right) \cos \vartheta \right],
\end{equation}
with $\epsilon \approx r /R $ (as introduced in section~\ref{sec:setup}) small. Then, 
\begin{equation}
\label{eq:fluxav}
\left< A \right>  \approx \frac{B_0}{4 \pi^2 q R } \oint \frac{A d\vartheta d \zeta}{\vec{B}\cdot \nabla \vartheta}.
\end{equation}
The perturbation to the distribution function $h$ may be found in~\eqref{eq:unstarred}. The perturbed radial velocity $\left[ \left(v_\parallel \hat{b}_{tot} + \vec{v}_{d,tot} \right)\cdot \nabla \psi \right]_1$ may be read off from~\eqref{eq:govh}, giving
\begin{equation}
\label{eq:perturbvel}
    \left[ \left(v_\parallel \hat{b}_{tot} + \vec{v}_{d,tot} \right)\cdot \nabla \psi \right]_1 = \sum_{m,n,\omega} inc S_{mn\omega}e^{in\zeta - i \omega t - i m\vartheta + i \int^\psi d\psi' \frac{k_\psi}{RB_p}}.
\end{equation}
The flux has contributions from both trapped and passing particles, and will be evaluated in slightly different ways for each of these populations. In general, we illustrate the trapped particle techniques first, then modify them for passing particles later. One of the key differences between the trapped and the passing calculations is the treatment of the particle pitch angle $\lambda$~\eqref{eq:lambdal}.  For trapped particles, integration in pitch angle is carried out in terms of the trapping variable $\kappa$, which is defined in terms of $\lambda$, 
\begin{equation}
\label{eq:kappadef}
\kappa^2 = \frac{1-\left(1-\epsilon\right)\lambda}{2 \epsilon \lambda}.
\end{equation}
This variable is defined such that, for a trapped particle, the angle of the turning point in the magnetic field~\eqref{eq:bstrength} is given by $2 \sin^{-1}\left(\kappa \right)$. Deeply trapped particles have $\kappa = 0$ and barely trapped particles have $\kappa = 1$. For passing particles, the analogous variable is 
\begin{equation}
\label{eq:kdef}
k \equiv \frac{1}{\kappa},
\end{equation}
where a value of $k= 1$ corresponds to a barely passing particle and a value of $k= 0$ corresponds to a fully passing particle.  [A pedagogical introduction to the trapping parameter can be found in~\citet{helander2005collisional}.] With the definitions of $\lambda$~\eqref{eq:lambdal} and $\kappa$~\eqref{eq:kappadef} we can write the differential in~\eqref{eq:gamma} as
\begin{equation}
\label{eq:differential}
d^3v = \frac{\pi B v^3 dv d \lambda}{B_{0} v_\parallel}; \, d \lambda = \frac{-4 \epsilon \kappa d \kappa }{\left(1 - \epsilon +2 \epsilon \kappa^2 \right)^2};
\end{equation}
an equivalent expression in terms of $k$ holds for passing particles.
\subsection{Discussion of synergistic transport from different  \texorpdfstring{$m$}{m}}
To evaluate the flux,~\eqref{eq:unstarred},~\eqref{eq:fluxav},~\eqref{eq:perturbvel}, and~\eqref{eq:differential} are inserted into~\eqref{eq:gamma}. Note that the resulting flux includes sums over two different sets of $m$ and $n$: one set for the perturbed distribution function~\eqref{eq:unstarred}, and one for the perturbed radial velocity~\eqref{eq:perturbvel}.   The complex quantities are written as their real equivalent and the $\zeta$ integral is evaluated using the identities $\oint d \zeta  \sin{\left(n\zeta -\phi_a\right)}\sin{\left(n'\zeta -\phi_b\right)} = \pi\delta_{n n'} \cos{\left(\phi_a - \phi_b\right)}$ and $\oint d \zeta  \cos{\left(n\zeta -\phi_a\right)}\sin{\left(n'\zeta -\phi_b\right)} = \pi\delta_{n n'} \sin{\left(\phi_a - \phi_b\right)}$, with $\phi_a$ and $\phi_b$ some phase and $\delta_{nn'}$ the Kronecker delta. 
Using these expressions, and stating the $\vartheta$ integral in terms of $\tau$~\eqref{eq:deftau}, we find for trapped particles
\begin{multline}
\label{eq:firstgamma}
Q_{\alpha,t} = \sum_{\substack{l,m,m'\\n,n'}}\frac{\delta_{nn'}cM_\alpha \epsilon \pi  }{2q R} \int_0^{v_0} dv \int_0^1 d \kappa \oint d\tau \frac{v^5 \delta\left(Q_l\right) \kappa n' S_{m'n'\omega'} D_{n\omega}}{\left(1-\epsilon + 2\epsilon \kappa ^2 \right)^2} 
\\
\times \left\{
\sin \left[ \left(m-m' \right) \vartheta + \left(\omega -\omega' \right)t - \int \frac{\left(k_\psi - k_\psi'\right)}{RB_p} d\psi \right] \left( \mathbb{S}_{mn\omega} \cos \int_{\tau_0}^\tau \Lambda d\tau'  - \mathbb{C}_{mn\omega} \sin \int_{\tau_0}^\tau \Lambda d\tau'  \right)
\right.
\\
\left.
-
\cos \left[ \left(m-m' \right) \vartheta + \left(\omega -\omega' \right)t - \int \frac{ \left(k_\psi - k_\psi'\right)}{RB_p} d\psi \right] \left( \mathbb{C}_{mn\omega} \cos \int_{\tau_0}^\tau \Lambda d\tau'  + \mathbb{S}_{mn\omega} \sin \int_{\tau_0}^\tau \Lambda d\tau'  \right)
\right\}
\end{multline}
(the subscript $t$ indicates this is a trapped particle expression). 
Here, we have defined
\begin{equation}
\label{eq:cmnw}
\mathbb{C}_{mn\omega}\left(v, \kappa  \right) \equiv \oint S_{mn\omega} \cos \left[  \int_{\tau_0}^\tau d\tau' \Lambda\left(\tau'\right) \right] d \tau
\end{equation}
and
\begin{equation}
\label{eq:smnw}
\mathbb{S}_{ mn\omega}\left(v, \kappa  \right) \equiv  \oint S_{mn\omega} \sin \left[ \int_{\tau_0}^\tau d\tau' \Lambda\left(\tau'\right) \right] d\tau.
\end{equation}
Equivalent expressions in terms of $k$ instead of $\kappa$ apply to passing particles.
Consideration of~\eqref{eq:firstgamma} makes clear that there is no coupling between perturbation components of different toroidal mode numbers: the perturbed distribution function of one $n$ will not interact with the perturbed radial velocity of another $n'$.  However, perturbations (or components of one perturbation) with the same toroidal mode number, but different poloidal mode numbers, can couple via the $\cos \left[ \left( m-m'\right) \vartheta \ldots \right]$ and $\sin \left[ \left( m-m'\right) \vartheta \ldots \right]$ terms. For perturbations of very different $m$, these terms are highly oscillatory, such that the integral over $\vartheta\left(\tau\right)$ will be small, as required by the Riemann-Lebesgue lemma.  The flux resulting from this coupling approaches zero. 

For perturbations of similar poloidal mode number, the coupling is significant and the perturbations cause synergistic transport in which the perturbed distribution function from one perturbation interacts with the perturbed drift velocity of the other to cause transport. Note that this synergistic transport will oscillate with time at a frequency equal to the difference between the two perturbations, as is typical for beat frequency phenomena.  Transport at such a beat frequency has been observed experimentally in studies of transport in the presence of NTMs and field perturbations caused by resonant magnetic perturbations \citep{snicker2018combined}. Evaluation of the flux for synergistic transport between poloidal harmonics requires numerical integration of~\eqref{eq:firstgamma}. Furthermore, for the case of interaction between TAE harmonics with the same $n$ and $\omega$, but different values of $m$, the relationship $nq-m = 1/2$ (see table~\ref{tab:perts}), ensures that the harmonics will have different values of $q$ and different radial profiles. The resulting rapid sinusoidal oscillations tend to reduce their contributions to transport. Thus, from here on, we evaluate the flux for a single perturbation, characterized by one value of $m$, $n$, and $\omega$, with the knowledge that if a perturbation of a similar poloidal mode number is also present in the device at the same radial location, transport may be altered in a synergistic manner. We denote the single $m$, $n$, and $\omega$ flux $Q_{\alpha,mn\omega}$.

\subsection{Assumption of \texorpdfstring{$q_\star \approx q$}{qs=q} and final statement of flux} 
\label{sec:shear}
The flux $Q_{\alpha,mn \omega}$ is a function of flux $\psi$; the perturbation is also a function of this variable. However, $\psi_\star$, defined in~\eqref{eq:psistar}, not $\psi$, is a constant of the alpha particle's motion, such that the alpha particle response is a function of $\psi_\star$. In the flux expression~\eqref{eq:firstgamma} $\psi_\star$ appears only in the variable $q_\star$ through the dependence of $Q_l$ and $\Lambda$ on this quantity. Recalling~\eqref{eq:qstar} and~\eqref{eq:psistar}, we have 
\begin{equation}
 q_\star \approx q\left(\psi\right) -  \frac{\partial q}{\partial \psi} \frac{I v_\parallel}{\Omega}  = q - \frac{qs  v_\parallel}{\epsilon R  \Omega_p}
\end{equation}
 [note that here we have provided $q_\star$ in terms of both $\partial q/\partial \psi$ and shear $s$,~\eqref{eq:shear}]. Because $v_\parallel$ is a function of $\vartheta$, this quantity varies poloidally in a complicated manner. Such complication introduces significant challenge in the evaluation of the flux, in particular making the term that enforces resonance, $\delta\left(Q_l\right)$, dependent on $\vartheta$. 

In this work, we thus choose to neglect the difference between $q$ and $q_\star$, leaving a detailed study of the effect of the difference between these quantities to future work. This simplification places constraints on the values of shear $s$ for which our work is valid. Because $q_\star$ appears in the combination $\left(n q_\star -m \right)$, we can compare $q_\star$ to the other terms in this combination to develop the following condition for the validity of setting $q_\star = q$:
\begin{equation}
\label{eq:qstarcond}
    s \ll \frac{\epsilon R \Omega_p \left(nq-m\right)}{nqv}.
\end{equation}
The degree of restriction that this condition places on the magnitude of the shear depends on the characteristics of the perturbation in question. For ripple, the condition is 
\begin{equation}
    s \ll \frac{\epsilon R}{\rho_{p\alpha}},
\end{equation}
a generous condition which allows the approximation to be valid even in the high shear regions near the tokamak edge where the ripple is important.  For the TAE, the equivalent expression is 
\begin{equation}
    s \ll \frac{\epsilon R}{nq \rho_{p \alpha}},
\end{equation}
which, given the fairly high values of $n$ characterizing TAEs, is a more restrictive condition, though it should still hold in much of the low-shear core plasma where TAE activity is most common.

With the replacement $q_{\star}\rightarrow q$, the outer $\tau$ integral in~\eqref{eq:firstgamma} can be commuted past $\delta\left(Q_l\right)$. In addition, using~\eqref{eq:slowingdown}, we can insert
\begin{equation}
\frac{\partial f_0}{\partial \psi} \approx \frac{\partial n_\alpha /\partial \psi }{4\pi \left(v^3 + v_c^3\right) \ln \left(v_0/v_c \right)  }   \approx  \frac{\partial n_\alpha /\partial \psi}{4\pi v^3 \ln \left(v_0/v_c \right)} 
\end{equation}
into the definition of $D_{n\omega}$~\eqref{eq:Dnw}. (Recall that we have also taken $\partial f_0 /\partial \psi_{\star} \approx \partial f_0 / \partial \psi$.)
This gives, for trapped particles,
\begin{equation}
\label{eq:flux,t}
Q_{\alpha,mn\omega ,t} =- \sum_l \frac{\epsilon M_\alpha c^2 n^2  \partial n_\alpha/ \partial \psi}{8qR \ln{\left(v_0/v_c\right)}}  \int_0^{v_0} dv \int_0^1 d\kappa \frac{\kappa v^2  \delta\left(Q_l\right) \left(1-\frac{\omega}{n\omega_\star} \right)\left( \mathbb{C}_{mn\omega}^2+ \mathbb{S}_{mn\omega}^2 \right)}{\left(1-\epsilon +2 \epsilon \kappa^2\right)^2} .
\end{equation}
The equivalent expression for passing particles is below, adding a subscript $p$ to make clear the expression refers to passing particles:
\begin{equation}
\label{eq:flux,p}
Q_{\alpha,mn \omega ,p} = -\sum_l \frac{\epsilon M_\alpha c^2 n^2  \partial n_\alpha/ \partial \psi}{8qR \ln{\left(v_0/v_c\right)}}  \int_0^{v_0} dv \int_0^1 dk \frac{k v^2  \delta \left( Q_l \right) \left(1-\frac{\omega}{n \omega_\star} \right)\left( \mathbb{C}_{mn\omega}^2+ \mathbb{S}_{mn\omega}^2 \right)}{\left[\left(1-\epsilon\right)k^2 +2 \epsilon \right]^2} . 
\end{equation}
These expressions are central results for this paper and can be used to evaluate the alpha flux caused by a wide range of tokamak perturbations. In later sections, they will be applied to the flux resulting from ripple and TAEs as examples. Already, the expressions display clear and expected trends.  That the flux is driven by the alpha density gradient is shown by the expression's dependence on $\partial n_\alpha/\partial \psi$. That the flux increases with the perturbed amplitude squared is also clear. The following subsections examine the elements of these fluxes that remain opaque: the term $\delta\left(Q_l\right)$, which enforces the resonance condition, and the term $\mathbb{C}_{mn\omega}^2+ \mathbb{S}_{mn\omega}^2$, which we refer to as a "phase factor."
\subsection{Resonance condition}
\label{sec:rescond}
The flux expressions~\eqref{eq:flux,t} and~\eqref{eq:flux,p}  include the delta function $\delta\left(Q_l\right)$. This delta function enforces the resonance given in equation~\eqref{eq:res1} and determines which particles are able to be transported by the mode and to contribute to the alpha flux. The expression for the resonance~\eqref{eq:res1} can be evaluated to read [setting $q_\star = q$, as discussed in section~\ref{sec:shear}]
\begin{equation}
\label{eq:res2}
Q_l\left(v,\kappa\right) = \omega \tau_b - n \overline{\omega_{\alpha_\star} } \tau_b - 2 \pi \sigma \left(nq-m\right) - 2 \pi l,
\end{equation}
where the bounce or transit time $\tau_b$ is defined in~\eqref{eq:taub}, $\sigma$ is defined in~\eqref{eq:sigmadef}, and the transit average drift is defined by 
\begin{equation}
\label{eq:transav}
\overline{\omega_{\alpha_\star} } \equiv \tau_b^{-1} \oint \omega_{\alpha_\star}  d \tau.
\end{equation}

\begin{table}
  \begin{center}
\def~{\hphantom{0}}
  \begin{tabular}{lcc}
Quantity & Trapped particles & Passing particles\\ 
$\tau_b$ & $\frac{8qRK\left(\kappa \right)}{v \sqrt{2\epsilon}}$  & $\frac{4qRk K\left(k\right)}{v\sqrt{2\epsilon \lambda}}$ \\ 
$\overline{\omega_{\alpha_\star} }$  & $\frac{v^2 \left\{2E\left(\kappa \right) - K\left(\kappa \right)+4s \left[E\left(\kappa \right) - \left(1-\kappa^2 \right) K\left(\kappa \right) \right] \right\}}{2\Omega_p R^2 K\left(\kappa \right) }$ 
& $\frac{\ v^2 \left[ 2 E\left(k\right) - \left(2-k^2\right) K\left(k\right)+4s E\left(k \right)\right]}{2\Omega_p R^2 \left[\left(1-\epsilon\right)k^2 + 2\epsilon \right] K\left(k\right)}$ \\
  \end{tabular}
  \caption{Quantities involved in resonance condition. The complete elliptic integral of the first kind is denoted by $K\left(\kappa \right)$; the complete integral of the second kind is $E\left(\kappa\right)$.}
  \label{tab:res}
  \end{center}
\end{table}

The terms in~\eqref{eq:res2} can be evaluated for trapped and passing particles to give the expressions in terms of $v$ and $\kappa$ or $k$ in Table~\ref{tab:res}.\footnote{This evaluation follows from using the  definitions of $\tau$~\eqref{eq:deftau} and $\omega_{\alpha_\star}$~\eqref{eq:omalphastar}.  In a magnetic field with strength described by~\eqref{eq:bstrength}, the parallel velocity appearing in $\omega_{\alpha_\star}$ and $d\tau$ is $v_\parallel = \pm v \sqrt{\left[1 - \left(1-\epsilon\right) \lambda \right] - 2 \epsilon \lambda \sin^2{\vartheta/2}}$. The changes of variable given later on in~\eqref{eq:xdeftrap} and~\eqref{eq:xdefpass} are needed.} The term $\delta \left(Q_l\right)$ in the flux expression thus relates the pitch angle and speed of particles that participate in transport and contribute to the flux.

\subsection{Evaluation of phase factor}
\label{sec:SmnCmn}
The term $\mathbb{C}_{mn\omega}^2 + \mathbb{S}_{mn\omega}^2$ in~\eqref{eq:flux,t} and~\eqref{eq:flux,p} depends on integrals defined in~\eqref{eq:cmnw} and~\eqref{eq:smnw}. These integrals are evaluated along particle trajectories, which advance with $\tau$~\eqref{eq:deftau}. This complex term, which we refer to as a "phase factor," determines how effective a given resonance is at creating a perturbation to the alpha distribution function. In this subsection, we write the phase factor for trapped and passing particles in a form that can be integrated numerically or approximated. 

\begin{figure}
\centering
\includegraphics[width=.8\columnwidth]{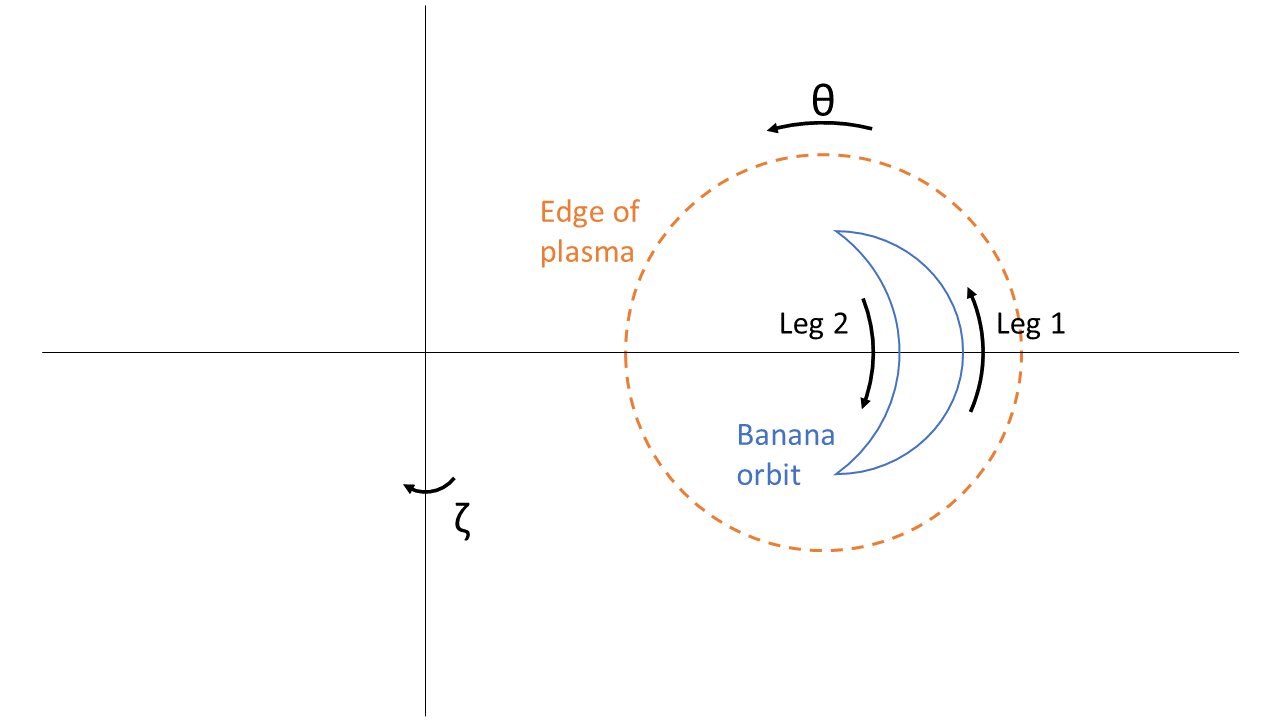}
\caption{A banana orbit, with legs 1 and  2 defined. Note that banana orbits always move in the sense indicated (i.e. with positive velocity on the outer side of the banana) for plasma current in the direction of positive $\zeta$ because of the conservation of $\psi_\star$~\eqref{eq:psistar}.}
\label{fig:bannanalegs}
\end{figure}

We begin with trapped particles. Trapped particles follow banana orbits, represented schematically in figure~\ref{fig:bannanalegs}, with two legs on which the particle velocity is in opposite directions.  Our evaluation assumes that the particle orbit is in resonance at a particular harmonic $l$ with the mode via~\eqref{eq:res2}. Only such resonant particles are able to contribute to the flux.  The integrals in the phase factor are evaluated along each of these legs, which contribute in distinct ways that depend on $l$ and finite orbit effects to the overall phase factor.

We begin by introducing the variable $x$, which is the following function of $\vartheta$:
\begin{equation}
\label{eq:xdeftrap}
    \kappa \sin x \equiv \sin \left(\vartheta/2 \right).
\end{equation}
The definition of $\kappa$~\eqref{eq:kappadef} can be used to show that the banana bounce points occur at $x = \pm \pi/2$, so that the use of $x$ standardizes the necessary integration. Now, to write $\mathbb{C}_{mn\omega}$ and $\mathbb{S}_{mn\omega}$ explicitly, we must write various quantities that appear in $\tau$~\eqref{eq:deftau} or $\Lambda$~\eqref{eq:Lambda} in terms of $x$.  The parallel velocity in the magnetic field~\eqref{eq:bstrength} is given by 
\begin{equation}
    v_\parallel = \pm \kappa v \sqrt{2\epsilon \lambda} \cos{x} \approx \pm \kappa v \sqrt{2\epsilon} \cos{x} ,
\end{equation}
where the last expression exploits that for trapped particles $\lambda$ differs from $1$ by at most $\pm \epsilon$.  The drift term appearing in $\Lambda$,~\eqref{eq:omalphastar}, is given by 
\begin{equation}
    \omega_{\alpha_\star} =  \frac{v^2}{2 \Omega_p R^2} \left(1 -2 \kappa^2 \sin^2 x + 4 s \kappa^2 \cos^2 x\right).
\end{equation}
We will also use $\hat{b} \cdot \nabla \vartheta  \approx 1/\left(q R\right)$ to simplify our expressions.

\begin{table}
  \begin{center}
\def~{\hphantom{0}}
  \begin{tabular}{lc}
Expression & Value \\ 
$a_t\left(x\right)$ & $ \frac{2q R}{v\sqrt{2\epsilon}} \int_0^x \frac{dx'\left[ \omega -\frac{nv^2}{2 \Omega_p R^2 } \left(1 -2 \kappa^2 \sin^2 x' + 4 s \kappa^2 \cos^2 x'\right)\right]}{\sqrt{1-\kappa^2 \sin^2 x'}}$ \\ 
$b_t$  &  $\frac{ k_\psi \sqrt{2\epsilon} v \kappa}{\Omega_p}$
 \\
$c_t\left(x\right)$  &  $b_t \cos{x}$
 \\
 $d_t\left(x\right)$  &  $2\left(nq -m \right)\sin^{-1}\left(\kappa \sin{x}\right)$
 \\
  \end{tabular}
  \caption{Definition of quantities used in the trapped particle phase factor.}
  \label{tab:trapphase}
  \end{center}
\end{table}

To make our presentation of the phase factor more compact, we define a set of expressions [$a_t\left(x\right)$,$b_t$, $c_t\left(x\right)$, and $d_t\left(x\right)$]  given in table~\ref{tab:trapphase}. Then, the contribution of the outer leg $1$, which  has velocity in the direction of positive $\vartheta$,  to $\mathbb{C}_{mn\omega}$~\eqref{eq:cmnw}  can be shown to equal (where we have again taken $\lambda \approx 1$)
\begin{multline}
        \mathbb{C}_{mn \omega, outer} = \frac{2q R \left( \Phi_{m n \omega} + \frac{v^2}{2c \Omega}B_{1 \parallel m n\omega} \right)}{v\sqrt{2\epsilon}} \int_{-\pi/2}^{\pi/2}dx \frac{\cos{\left[ a_t\left(x\right) +b_t -c_t\left(x\right) -d_t\left(x\right) \right]}}{\sqrt{1-\kappa^2 \sin^2 x}}
        \\
        - \frac{ A_{\parallel mn \omega} q R}{c} \int_{-\pi/2}^{\pi/2}dx \frac{2\kappa \cos{x} \cos{\left[a_t\left(x\right) +b_t -c_t\left(x\right) -d_t\left(x\right) \right]} }{\sqrt{1-\kappa^2 \sin^2 x}}.
\end{multline}
Meanwhile, the inner leg is given by 
\begin{multline}
        \mathbb{C}_{mn \omega, inner} = \frac{2q R \left( \Phi_{m n \omega} + \frac{v^2}{2c \Omega}B_{1 \parallel m n\omega} \right)}{v\sqrt{2\epsilon}} \int_{\pi/2}^{-\pi/2} dx \frac{- \cos{\left[ \pi l - a_t\left(x\right) +b_t +c_t\left(x\right) -d_t\left(x\right) \right]}}{\sqrt{1-\kappa^2 \sin^2 x}}
        \\
        - \frac{ A_{\parallel mn \omega} q R}{c} \int_{\pi/2}^{-\pi/2} dx \frac{ 2\kappa \cos{x} \cos{\left[ \pi l - a_t\left(x\right) +b_t +c_t\left(x\right) -d_t\left(x\right) \right]}}{\sqrt{1-\kappa^2 \sin^2 x}}.
\end{multline}
Here, our use of $l$ follows from the recognition that only particles which are in resonance according to~\eqref{eq:res2} contribute to the flux, and that the quantity $2\pi l$ accumulates from $\omega$ and $n \omega_{\alpha_\star}$ integrated over the entire orbit starting from $x =0$. The inner leg value can be restated by applying the relationship $\cos\left(x+ \pi l \right) = (-1)^l \cos x$ and flipping the limits so they agree with those given for the outer leg:
\begin{multline}
        \mathbb{C}_{mn\omega l, inner} = \frac{2q R \left(-1\right)^l \left( \Phi_{m n \omega} + \frac{ v^2}{2c \Omega}B_{1 \parallel m n\omega} \right)}{v\sqrt{2\epsilon}} \int_{-\pi/2}^{\pi/2} dx \frac{ \cos{\left[ - a_t\left(x\right) +b_t +c_t\left(x\right) -d_t\left(x\right) \right]}}{\sqrt{1-\kappa^2 \sin^2 x}}
        \\
        + \frac{ A_{\parallel mn \omega} q R \left(-1\right)^l}{c} \int_{-\pi/2}^{\pi/2} dx \frac{ 2\kappa \cos{x} \cos{\left[ - a_t\left(x\right) +b_t +c_t\left(x\right) -d_t\left(x\right) \right]}}{\sqrt{1-\kappa^2 \sin^2 x}}.
\end{multline}
Equivalent expressions hold for $\mathbb{S}_{mn\omega}$.  Forming and evaluating the quantity $\mathbb{C}_{mn\omega}^2 + \mathbb{S}_{mn\omega}^2$ reveals that the dependence on $b_t$ vanishes, as is to be expected given that this parameter depends on the parallel velocity evaluated at $\vartheta\left(\tau_0\right)= 0$, an arbitrary location. Suppressing the $x$ dependence of $a_t$, $c_t$, and $d_t$ for compactness, the full phase factor reads
\begin{multline}
\label{eq:trappedphase}
  \frac{\mathbb{C}_{mn\omega}^2 + \mathbb{S}_{mn\omega}^2}{4q^2 R^2} = 
    \\
  \left\{\frac{  \Phi_{m n \omega} + \frac{ v^2}{2c \Omega}B_{1 \parallel m n\omega}}{v\sqrt{2\epsilon}} \int_{-\pi/2}^{\pi/2} dx \frac{ \cos{\left( a_t -c_t -d_t \right)} + \left(-1\right)^l \cos{\left(- a_t +c_t -d_t \right) }}{\sqrt{1-\kappa^2 \sin^2 x}}\right.
    \\
    \left.
    - \frac{A_{\parallel mn\omega} \kappa}{c} \int_{-\pi/2}^{\pi/2}dx \frac{\cos{x} \left[  \cos{\left( a_t -c_t -d_t \right)} - \left(-1\right)^l \cos{\left(- a_t +c_t -d_t \right)}\right] }{\sqrt{1-\kappa^2\sin^2{x}}}\right\}^2
    \\
    + 
     \left\{\frac{  \Phi_{m n \omega} + \frac{ v^2}{2c \Omega}B_{1 \parallel m n\omega}}{v\sqrt{2\epsilon}} \int_{-\pi/2}^{\pi/2} dx \frac{ \sin{\left( a_t -c_t -d_t \right)} + \left(-1\right)^l \sin{\left(- a_t +c_t -d_t \right) }}{\sqrt{1-\kappa^2 \sin^2 x}}\right.
    \\
    \left.
    - \frac{A_{\parallel mn\omega}  \kappa}{c} \int_{-\pi/2}^{\pi/2}dx \frac{\cos{x} \left[  \sin{\left( a_t -c_t -d_t \right)} - \left(-1\right)^l \sin{\left(- a_t +c_t -d_t \right)}\right] }{\sqrt{1-\kappa^2\sin^2{x}}}\right\}^2.
\end{multline}
Though this expression is daunting, in practice many simplifications are possible. First, most modes have only a subset of $\Phi_{mn\omega}$, $B_{1 \parallel m n \omega}$, and $A_{\parallel mn\omega}$, and these quantities may be proportional to each other. In addition, it is sometimes possible to neglect some of the terms in the arguments of the cosine. The integrals can also be approximated in a variety of ways. 

Passing particles are simpler to treat than trapped particles. For passing particles the variable $x$ is defined as:
\begin{equation}
\label{eq:xdefpass}
     x \equiv \frac{\vartheta}{2}.
\end{equation}
Recalling the variable $k$ defined in~\eqref{eq:kdef}, the parallel velocity in the magnetic field~\eqref{eq:bstrength} for passing particles is given by 
\begin{equation}
    v_\parallel = \pm \frac{v \sqrt{2\epsilon \lambda}}{k} \sqrt{1-k^2 \sin^2{x}}. 
\end{equation}
Note that $\lambda$ can be very small for passing particles and thus cannot be set to 1 as it was for the trapped particles. For passing particles, the drift~\eqref{eq:omalphastar} is: 
\begin{equation}
    \omega_{\alpha_\star} =  \frac{\lambda v^2}{2 \Omega_p R^2 k^2} \left[k^2 -2 k^2 \sin^2 x + 4 s\left(1-k^2\sin^2x\right)\right].
\end{equation}
\begin{table}
  \begin{center}
\def~{\hphantom{0}}
  \begin{tabular}{lc}
Expression & Value \\ 
$a_p\left(x\right)$ & $ \frac{2q Rk \sigma}{v\sqrt{2\epsilon\lambda}} \int_0^x \frac{dx'  \left[ \omega - \frac{nv^2\lambda \left(k^2 -2 k^2 \sin^2x'+ 4 s \left(1-k^2\sin^2 x'\right)  \right)}{2 k^2\Omega_p R^2 }\right]}{\sqrt{1-k^2 \sin^2 x'}} $ \\ 
$b_p$  &  $ \frac{\sigma k_\psi  \sqrt{2\epsilon\lambda} v }{k\Omega_p}$
 \\
$c_p\left(x\right)$  &  $b_p \sqrt{1-k^2\sin^2 x}$
 \\
 $d_p\left(x\right)$  &  $2\left(nq -m \right)x$
 \\
  \end{tabular}
  \caption{Definition of quantities used in the passing particle phase factor. Note that $\sigma$ is the parameter introduced~\eqref{eq:sigmadef}, which characterizes the direction of a particle's motion in $\vartheta$.}
  \label{tab:passphase}
  \end{center}
\end{table}We again define certain expressions to make the presentation of the phase factor more compact in Table~\ref{tab:passphase}. A passing particle orbit goes around the entire device, from $x = -\pi/2$ to $x= \pi/2$. Its $\mathbb{C}_{mn\omega}$ is given by
\begin{multline}
    \mathbb{C}_{mn\omega} = \frac{2kq R \left(\Phi_{m n \omega} + \frac{\lambda v^2}{2c \Omega}B_{1 \parallel m n\omega} \right)}{v \sqrt{2\epsilon \lambda}}\int_{-\pi/2}^{\pi/2} \frac{dx}{\sqrt{1-k^2\sin^2{x}}}  \cos{\left(a_p + b_p -c_p -d_p\right)}
    \\
    - \frac{2\sigma q R A_{\parallel m n \omega}}{c} \int_{-\pi/2}^{\pi/2} dx \cos{\left(a_p +b_p -c_p -d_p \right)}.
\end{multline}
An analogous expression holds for $\mathbb{S}_{mn\omega}$. Forming the sum of the two terms gives
\begin{multline}
\label{eq:phasepass}
  \frac{\mathbb{C}_{mn\omega}^2 + \mathbb{S}_{mn\omega}^2}{4q^2 R^2} = 
    \\
  \left\{\frac{ k\left( \Phi_{m n \omega} + \frac{\lambda v^2}{2c \Omega}B_{1 \parallel m n\omega}\right)}{v\sqrt{2\epsilon \lambda}} \int_{-\pi/2}^{\pi/2} dx \frac{ \cos{\left( a_p -c_p -d_p \right)} }{\sqrt{1-k^2 \sin^2 x}}
    - \frac{\sigma A_{\parallel mn\omega}}{c} \int_{-\pi/2}^{\pi/2}dx  \cos{\left( a_p -c_p -d_p \right)}  \right\}^2
    \\
    + 
     \left\{\frac{ k\left( \Phi_{m n \omega} + \frac{\lambda v^2}{2c \Omega}B_{1 \parallel m n\omega}\right)}{v\sqrt{2\epsilon \lambda}} \int_{-\pi/2}^{\pi/2} dx \frac{ \sin{\left( a_p -c_p -d_p \right)} }{\sqrt{1-k^2 \sin^2 x}}
    - \frac{\sigma A_{\parallel mn\omega} }{c} \int_{-\pi/2}^{\pi/2}dx  \sin{\left( a_p -c_p -d_p \right)} \right\}^2.
\end{multline}
Again, in practice this complicated expression can be simplified in a variety of ways. 
\section{Ripple flux}
\label{sec:ripp}
In this section, we consider the trapped and passing fluxes~\eqref{eq:flux,t} and~\eqref{eq:flux,p} for our first example perturbation: ripple, which has characteristic parameters described in table~\ref{tab:perts}. We will find that the contribution of bounce harmonic resonances to the transport for this perturbation is small because $nq$ is very high. Because of this realization, we do not provide an analytic expression for the flux. 

\begin{figure}
  \centering
  \includegraphics[width=7.4cm]{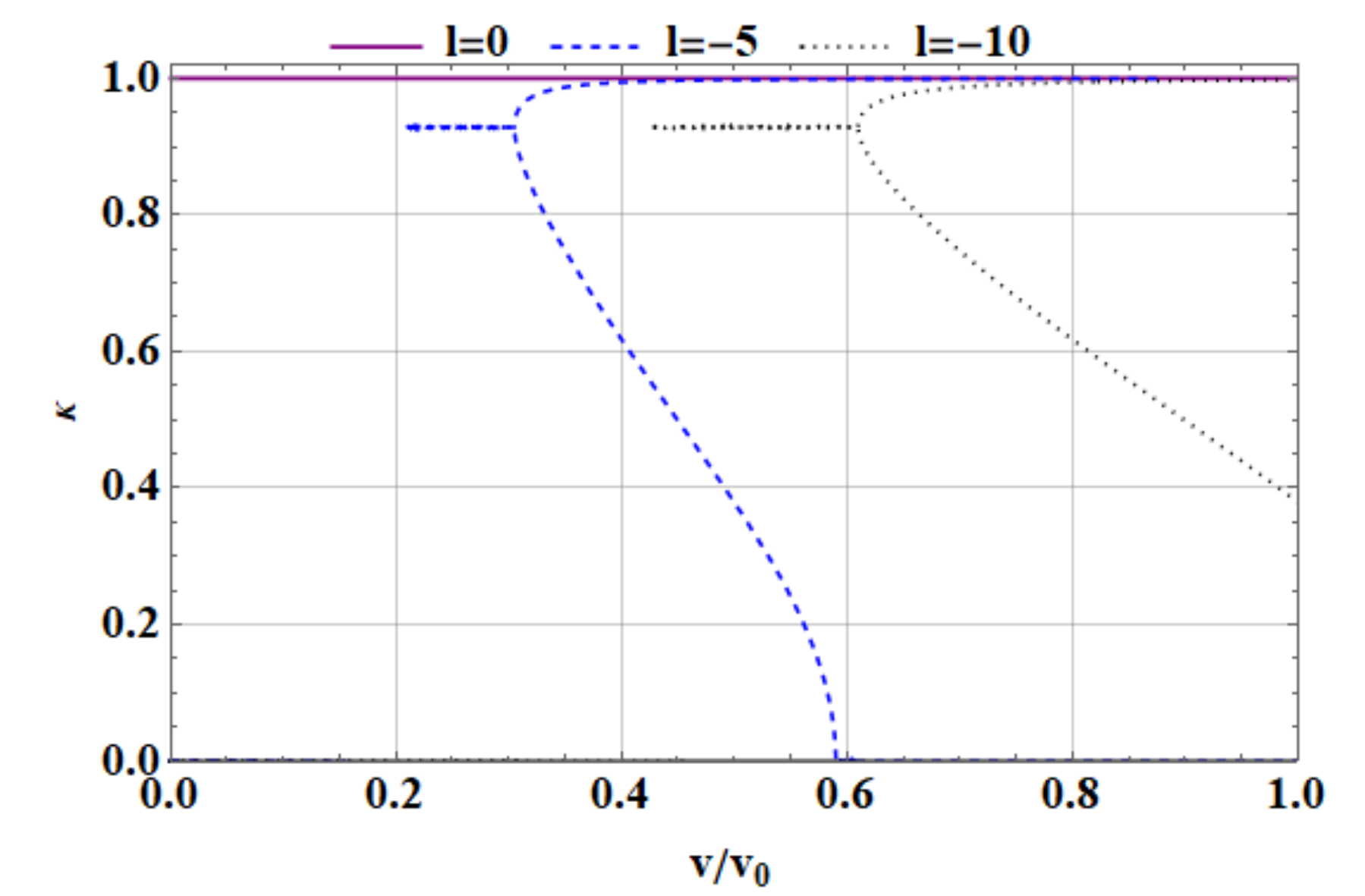}
  \caption{Trapped particle ($\sigma = 0$) resonance structure [i.e., where $Q_l\left(v, \kappa\right) = 0$] for ripple with parameters given in table~\ref{tab:perts} in a tokamak described by the values in table~\ref{tab:equilib}.}
\label{fig:resstructrip0}
\end{figure}

\begin{figure}
  \centering
  \includegraphics[width=7.4cm]{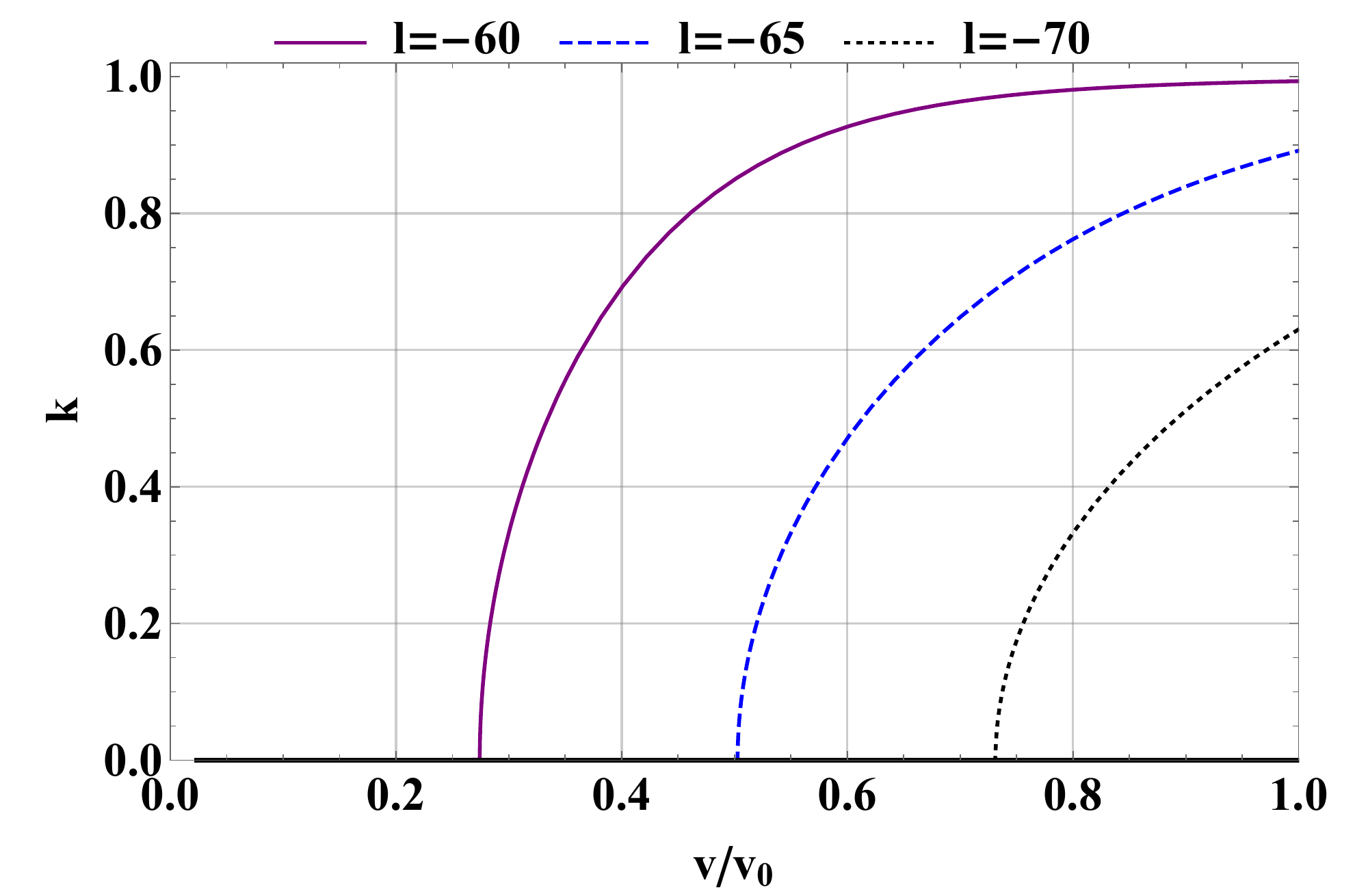}
  \caption{Passing, $\sigma = 1$, particle resonance structure [i.e., where $Q_l\left(v, \kappa\right) = 0$] for ripple with parameters given in table~\ref{tab:perts} in a tokamak described by the values in table~\ref{tab:equilib}.}
\label{fig:resstructrip1}
\end{figure}

\begin{figure}
  \centering
  \includegraphics[width=7.4cm]{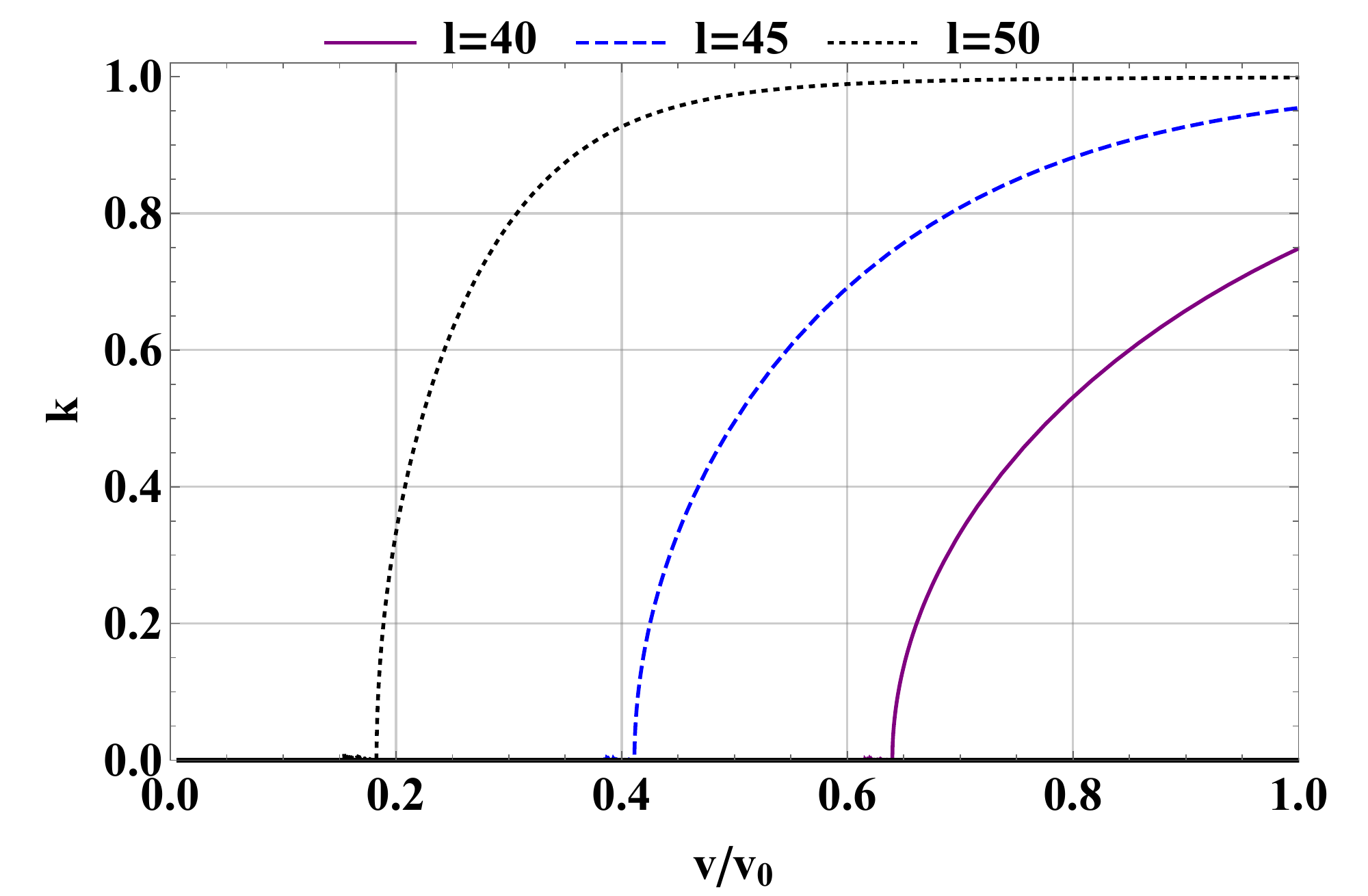}
  \caption{Passing, $\sigma = -1$, particle resonance structure [i.e., where $Q_l\left(v, \kappa\right) = 0$] for ripple with parameters given in table~\ref{tab:perts} in a tokamak described by the values in table~\ref{tab:equilib}.}
\label{fig:resstructrip-1}
\end{figure}

First, we consider the structure of the resonances between alphas and ripple in a tokamak with parameters given in table~\ref{tab:equilib}. These are plotted in figures~\ref{fig:resstructrip0},~\ref{fig:resstructrip1}, and~\ref{fig:resstructrip-1}, for trapped particles with $\sigma=0$, for passing particles with $\sigma = 1$, and for passing particles with $\sigma=-1$. Specifically, these figures show, for typical values of $l$, the values of $v$ and $\kappa$ (or $k$) for which $Q_l\left(v,\kappa\right) = 0$ and the delta function in~\eqref{eq:flux,t} or~\eqref{eq:flux,p} has support. The lines in the plot identify the areas of phase space which could contribute to transport. The resonance structure is only shown up to the alpha birth velocity $v_0$ because alpha particles are unable to interact with perturbations through resonances above this velocity. 

Trapped ripple resonances are shown in figure~\ref{fig:resstructrip0}.  Trapped resonances can occur at most values of $\kappa$ for negative values of $l$ of intermediate magnitude ($l=-5$ and $l=-10$ are shown in the plot). An alpha particle is born at $v_0$ and a specific value of $\kappa$. As the particle slows down, it will move left along the $x$-axis while maintaining the same value of $\kappa$. During this process, the alpha interacts with the perturbation when it crosses a resonance line. Note that in this case, the lines representing some harmonics are multivalued in $\kappa$ at some velocities.  This simply means that at that velocity, multiple pitch angles are able to interact with the mode. 

A series of resonances near the trapped-passing boundary ($\kappa = 1$) also exists.  These include parts of harmonics $l<0$ and all of $l= 0$,\footnote{Note that the $l=0$ resonance is so close to $\kappa = 1$ because we selected as an example parameter $s=1$ (see table~\ref{tab:perts}). For lower values of shear, the $l=0$ resonance occurs at lower values of $\kappa$ and does not suffer from the resonance crowding described in this paragraph. However, ripple does tend to be strongest in regions of high shear, so $s = 1$ is an appropriate example value.} shown in the plot.  There are also several more resonances, corresponding to $l>0$, at $\kappa\approx 1$, which are not displayed in the plot for clarity. Because the resonances at $\kappa \approx 1$  are very close to each other in pitch angle and very near to the trapped-passing boundary layer where other boundary layer physics enters \citep{catto2018ripple}, the use of the Krook operator to represent pitch angle scattering is not well justified, making the treatment in this paper unsatisfactory for this region. A more elaborate boundary layer theory is necessary to satisfactorily treat these resonances, so in this paper we focus on the lower $\kappa$ parts of the trapped resonances with $l<0$.  

For passing particles, shown in figures~\ref{fig:resstructrip1} and~\ref{fig:resstructrip-1}, which have $\sigma \neq 0 $ in~\eqref{eq:res2}, the large value of $nq$ characterizing ripple means that resonances occur at values of $l$ of large magnitude. At alpha birth (the right side of the plot), particles near trapped-passing boundary ($k = 1$) resonate with ripple, while at lower speeds, freely passing particles (small $k$) resonate with ripple.

\begin{figure}
\centering
\includegraphics[width=\columnwidth]{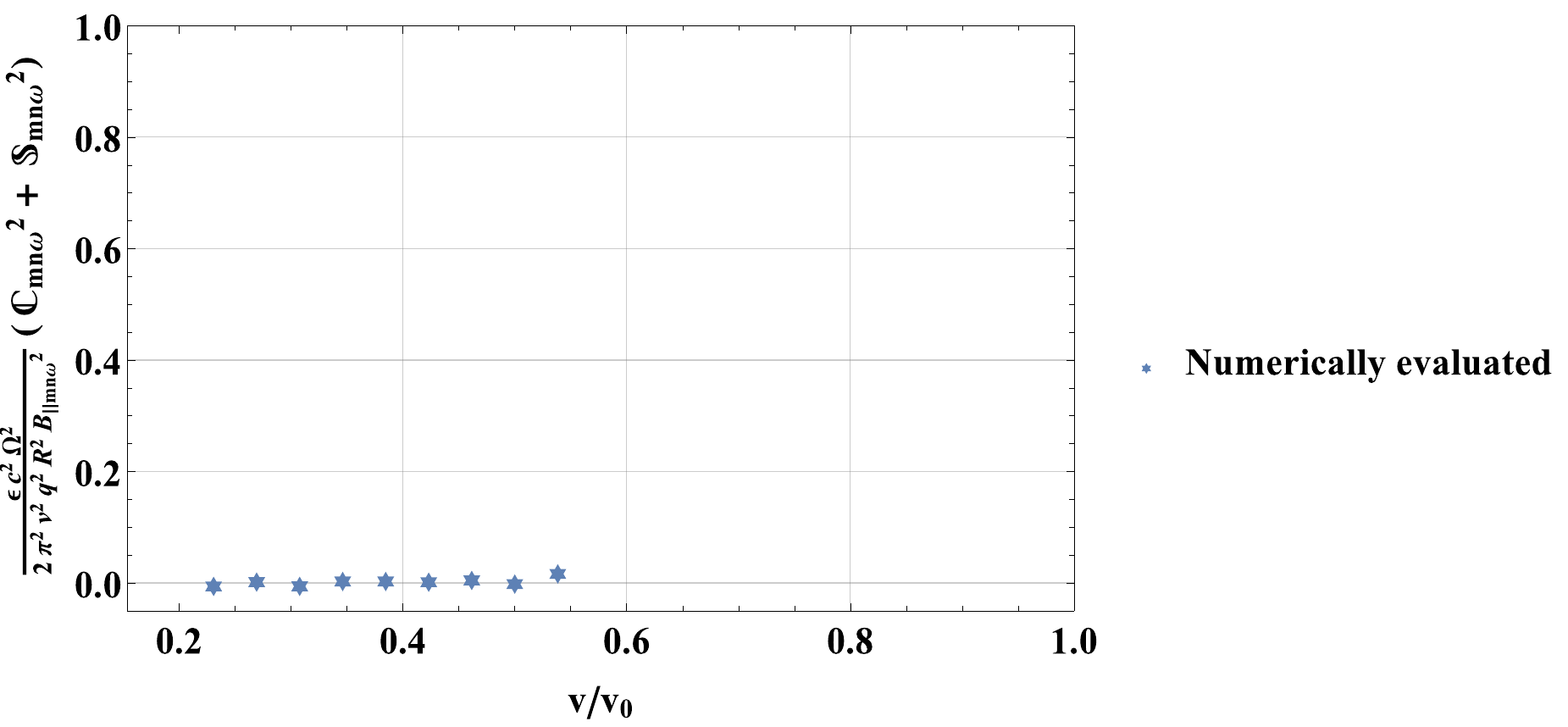}
\caption{Exact, numerically evaluated trapped phase factor for the $l= -5$ resonance of ripple. (Note that the phase factor is evaluated for the lower $\kappa$ branch of the resonance only, whenever two branches exist. Our approach is not capable of treating the higher branch, where resonances are bunched very closely together.) The normalization is determined  using the coefficients of $B_{1 \parallel m n \omega}$ in~\eqref{eq:trappedphase}; the sum of integrals $\left( \int_{-\pi/2}^{\pi/2} dx \left[\cos{} \ldots\right]/\sqrt{\left[1-\kappa^2 \sin^2 x \right]}  \right)^2 + \left( \int_{-\pi/2}^{\pi/2} dx \left[\sin{} \ldots\right]/\sqrt{\left[1-\kappa^2 \sin^2 x \right]}  \right)^2$ is normalized to $4 \pi^2$.}
\label{fig:ripplephase}
\end{figure}

After understanding the resonance structure, the next step in the evaluation of flux is to consider the phase factor $\mathbb{C}_{ mn\omega}^2 + \mathbb{S}_{ mn\omega}^2$, discussed in section~\ref{sec:SmnCmn}. However, this phase factor is very small for ripple resonances. As an example, figure~\ref{fig:ripplephase} shows the numerically evaluated phase factor for the $l=-5$ trapped harmonic displayed in figure~\ref{fig:resstructrip0}. Specifically, this plot shows the quantity~\eqref{eq:trappedphase} as a function of $v$, where~\eqref{eq:trappedphase} is numerically evaluated at the value of $\kappa$ that is resonant at that value of $v$. The small values of the phase factor evident in figure~\ref{fig:ripplephase} are typical of those for all ripple resonances.

These small values result from the highly oscillatory nature of ripple phase factor, which we demonstrate using the original definition of $\mathbb{C}_{ mn\omega}$,~\eqref{eq:cmnw}, and of $\mathbb{S}_{ mn\omega}$,~\eqref{eq:smnw}. In particular, the argument of the sinusoids in~\eqref{eq:cmnw} and~\eqref{eq:smnw} can be roughly approximated by 
\begin{equation}
\label{eq:linearphase}
    \int_{\tau_0}^\tau \Lambda\left( \tau', v, \kappa \right) d\tau' \approx  -n \overline{\omega_{\alpha_\star}} \left(\tau -\tau_0\right) - n q  \vartheta\left(\tau\right) 
\end{equation}
for values of $v$ and $\kappa$ satisfying the resonance condition. [Here, $\overline{\omega_{\alpha_\star}} $  is defined in~\eqref{eq:transav}.] To obtain this expression, we have considered the definition of $\Lambda$,~\eqref{eq:Lambda}, and inserted $m = 0$, $q_\star \approx q$, and $\omega = 0$, as appropriate for ripple. Then, we have treated $\omega_{\alpha_\star}$ as though it were constant and independent of $\tau$ [we note that a similar approximation was used to evaluate related phase factors in previous studies of the effect of non-axisymmetries on bulk plasma \citep{park2009nonambipolar}].  Then, the resonance condition~\eqref{eq:res2} can be applied to~\eqref{eq:linearphase} to reveal
\begin{equation}
\label{eq:phasescaling}
    \int_{\tau_0}^\tau \Lambda\left( \tau', v, \kappa \right) d\tau' \approx
    \begin{cases}
      - \frac{2\pi l \left(\tau - \tau_0\right)}{\tau_b}-n q \vartheta\left(\tau \right), & \textrm{trapped particles}  \\
      -\frac{2\pi l\left(\tau-\tau_0\right)}{\tau_b}, & \textrm{passing particles}.
    \end{cases}
\end{equation}
(For the passing expression, we have taken the additional approximation $\sigma\left(\tau - \tau_0\right)/\tau_b \approx \vartheta /2 \pi $.)
These expressions indicate that the arguments of the sinusoids in~\eqref{eq:cmnw} and~\eqref{eq:smnw} change very quickly if, for trapped particles, $\left|l+nq\right|$ is large or, for passing particles, $\left|l\right|$ is large. The corresponding phase integrands in the phase factor are highly oscillatory. The Riemann-Lebesgue lemma states that such integrals must approach zero as $\left|l + nq \right|^{-1}$ for trapped particles and $\left|l\right|^{-1}$ for passing particles.

Ripple is characterized by $nq >> 1$.  Then, consideration of the harmonics $l$ characterizing resonances that exist in a typical tokamak, illustrated in figures~\ref{fig:resstructrip0},~\ref{fig:resstructrip1}, and~\ref{fig:resstructrip-1}, reveals that for trapped particles $\left|l+nq\right|$ is always large and for passing particles $\left|l\right|$ is large.\footnote{The reader might ask if it is possible to construct any tokamak in which $\left|l+nq\right|$ is small for trapped particle ripple resonances and $\left|l \right|$ is small for passing particle ripple resonances. For trapped particles, this would require $nq\sim -l$ in~\eqref{eq:res2}. This reduces to the requirement $v\sim \Omega_p R$, which requires unrealistically small machine size or magnetic field. A similar condition is found for passing particles.} Based on Eq.~\eqref{eq:phasescaling} and the discussion in the previous pragraph, this means the phase factors for ripple perturbations are extremely small, as exemplified in figure~\ref{fig:ripplephase}.

With such a small phase factor, the flux described by~\eqref{eq:flux,t} and~\eqref{eq:flux,p} is negligible. Therefore, bounce harmonic resonant transport is very weak for ripple perturbations and will not create noticeable alpha heat flux.  Alpha transport from ripple may, however, occur via other mechanisms, including stochastic ripple diffusion and ripple trapping \citep{goldston1981confinement,white1995toroidal,catto2018ripple}.\footnote{Some of these mechanisms are collisionless. One might at first expect collisionless processes to be unimportant in steady state, where refilling of loss regions is necessary for sustained transport. However, alpha particles are continuously introduced to the tokamak via fusion, which serves to refill loss regions. Therefore, consideration of collisionless processes in tokamak design is critical to achieving satisfactory steady state behavior.}  In addition, further work should consider the impact of externally-applied 3D magnetic field perturbations, such as those used for ELM control, which could cause significant alpha transport \citep{sanchis2018characterisation}.

\section{TAE flux}
\label{sec:TAEflux}
In this section, we evaluate the trapped and passing fluxes~\eqref{eq:flux,t} and~\eqref{eq:flux,p} for the example case of the TAE, which has characteristic parameters described in table~\ref{tab:perts}. We find that the TAE can cause significant alpha heat flux, and develop a constraint the TAE amplitude must obey to prevent significant alpha depletion. At the end of the section, we develop a simple saturation model for TAE. This model suggests that TAEs in SPARC-like tokamaks will saturate below the level at which bounce harmonic resonant transport can cause significant alpha depletion. However, saturation amplitudes above those suggested by our model, but within  computational and experimental experience, could cause depletion.

\subsection{Evaluation of flux}
\begin{figure}
  \centering
  \includegraphics[width=7.4cm]{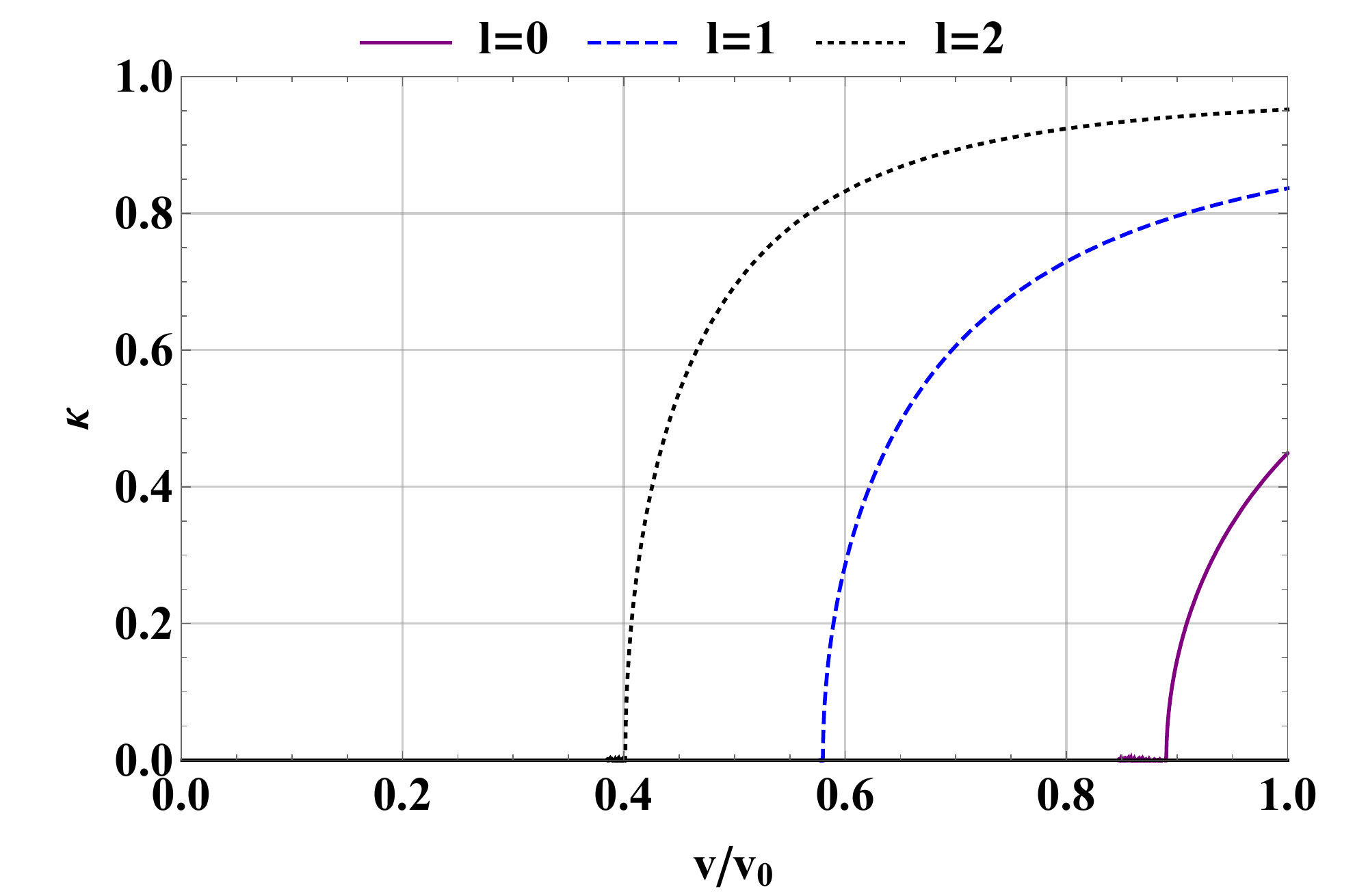}
  \caption{Trapped particle ($\sigma = 0$) resonance structure [i.e., where $Q_l\left(v, \kappa\right) = 0$] for a TAE with parameters given in table~\ref{tab:perts} in a tokamak described by the values in table~\ref{tab:equilib}.}
\label{fig:resstructtae0}
\end{figure}
\begin{figure}
  \centering
  \includegraphics[width=7.4cm]{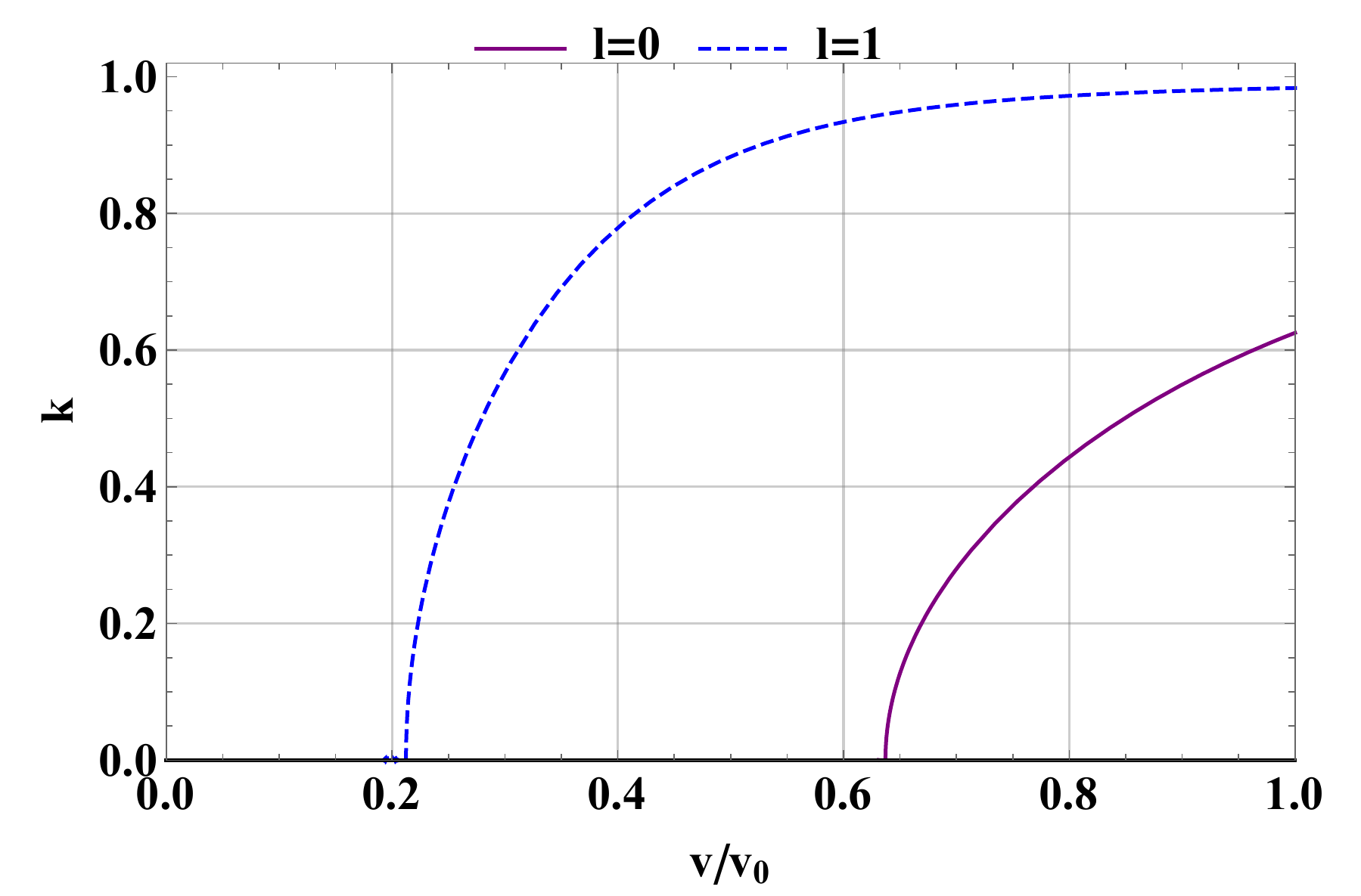}
  \caption{Passing, $\sigma = 1$, particle resonance structure [i.e., where $Q_l\left(v, \kappa\right) = 0$] for a TAE with parameters given in table~\ref{tab:perts} in a tokamak described by the values in table~\ref{tab:equilib}.}
\label{fig:resstructtae1}
\end{figure}
\begin{figure}
  \centering
  \includegraphics[width=7.4cm]{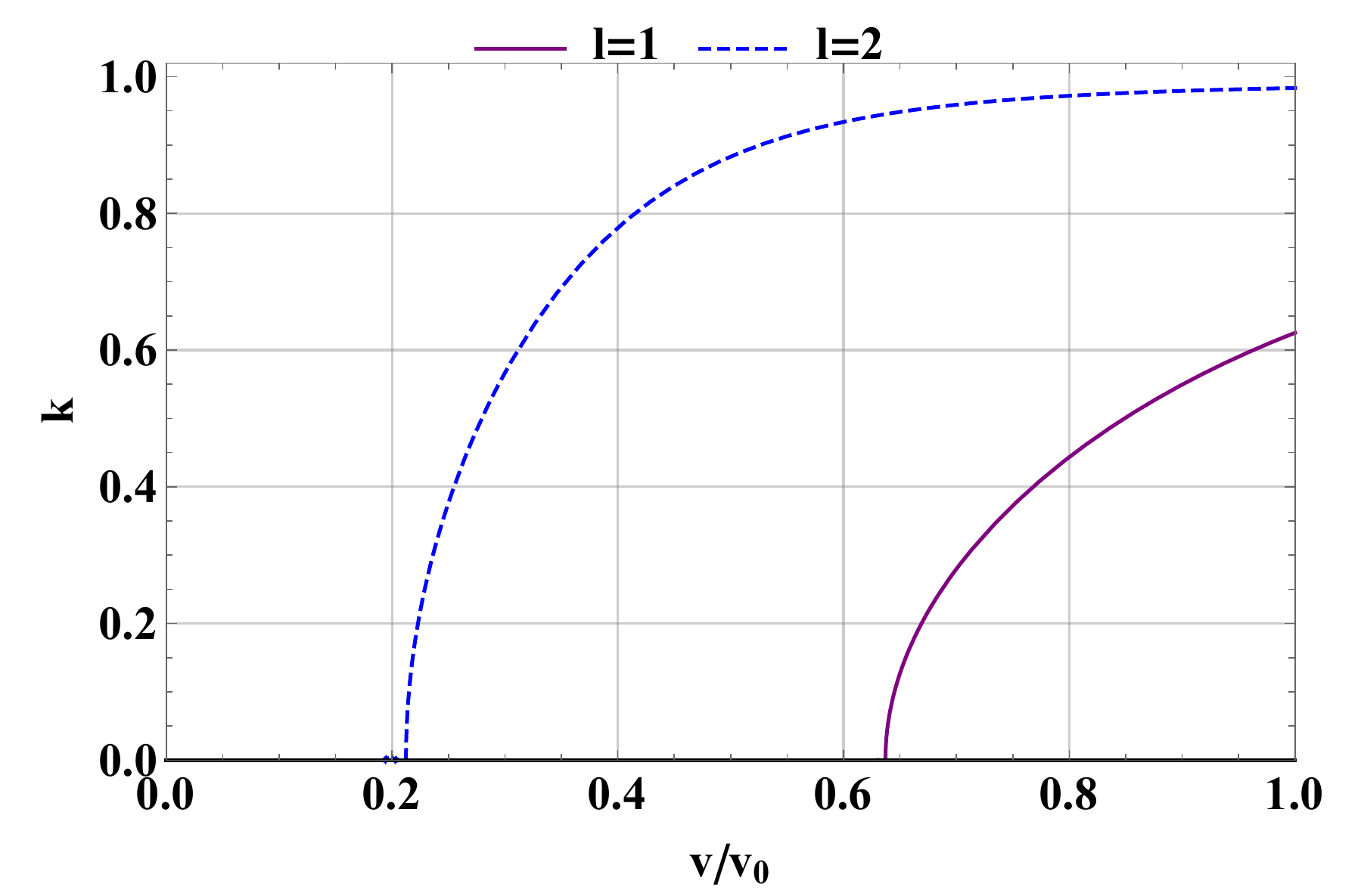}
  \caption{Passing, $\sigma = -1$, particle resonance structure [i.e., where $Q_l\left(v, \kappa\right) = 0$] for a TAE with parameters given in table~\ref{tab:perts} in a tokamak described by the values in table~\ref{tab:equilib}.}
\label{fig:resstructtae-1}
\end{figure}

The resonance structures for the TAE are shown in figures~\ref{fig:resstructtae0},~\ref{fig:resstructtae1}, and~\ref{fig:resstructtae-1}. Trapped particles are able to resonate with the TAE at low values of the harmonic $l$, as seen in figure~\ref{fig:resstructtae0}. Near the alpha particle birth speed (the right edge of the plot) the mode resonates with a small subset of pitch angles nearer the trapped-passing boundary ($\kappa = 1$). As the alpha particles slow down (moving left along the x-axis), a wider range of pitch angles can resonate with the mode. For a given value of $\kappa$, lower values of $l$ are able to resonate with higher speed particles.

Passing particle resonances (with both $\sigma = 1$ and $\sigma = -1$) with the TAE are similar, with alpha particles near birth resonating with barely passing particles and those that have slowed down resonating at a wider variety of pitch angles. These resonance structures can be seen in figures~\ref{fig:resstructtae1} and~\ref{fig:resstructtae-1}. Again, lower values of $l$ tend to resonate with higher velocity particles for a given value of $k$. As the drift $\overline{\omega_{\alpha_\star}}$ is negative, an $l=0$ resonance is only possible for $\sigma = 1$, but we will soon find the transport associated with this sign of $\sigma$ is very small. Note that fully passing particles (with $k=0$) resonate at the Alfv\'{e}n speed and one third of the Alfv\'{e}n speed. This aligns with the resonance structure described in simplified treatments of TAE resonance which focus on freely passing particles~\citep{heidbrink2008basic,betti1992stability}.

\begin{figure*}
  \centering
  \subfigure[ $l=0$]{\label{fig:l2} \includegraphics[width=.9\columnwidth]{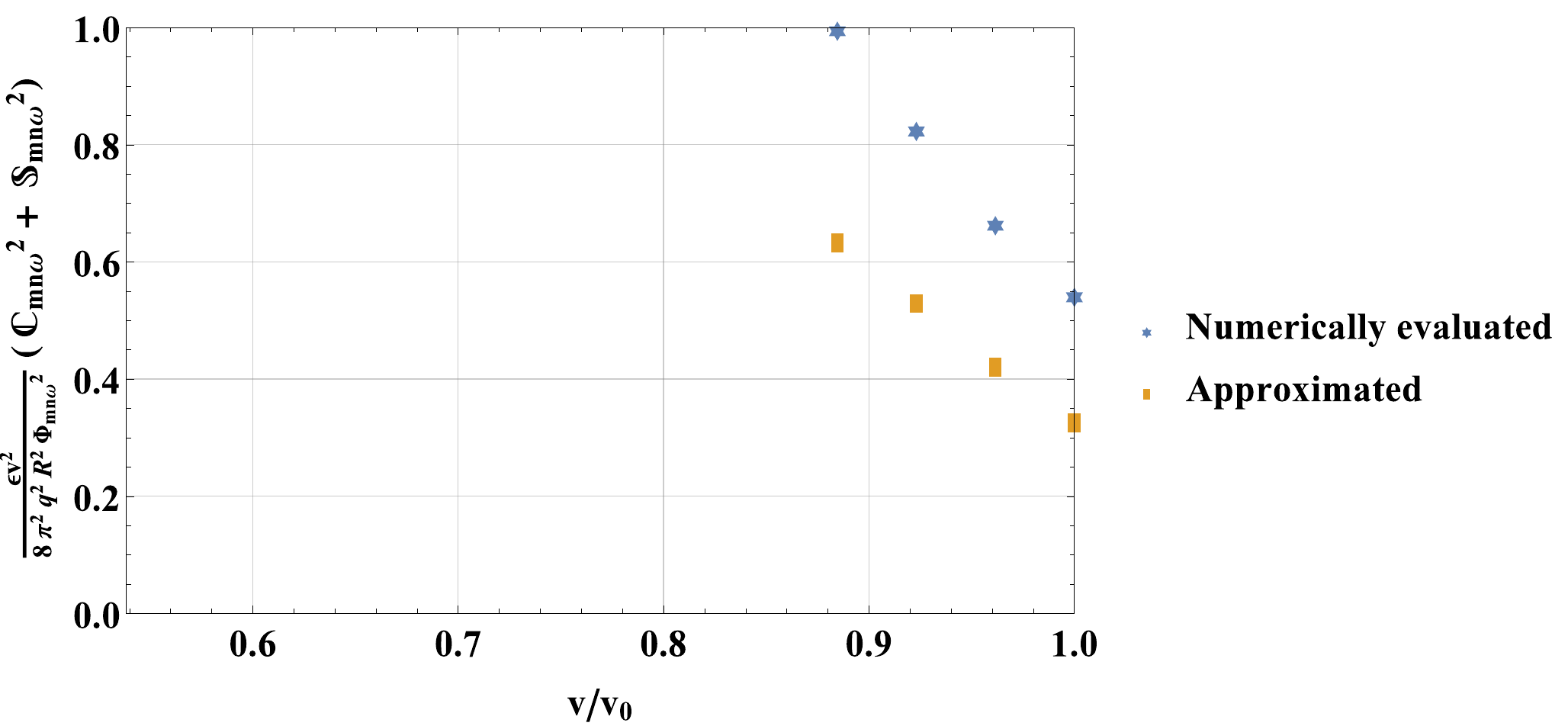}} \hfill
\subfigure[ $l=1$]{\label{fig:l1}\includegraphics[width=.9\columnwidth]{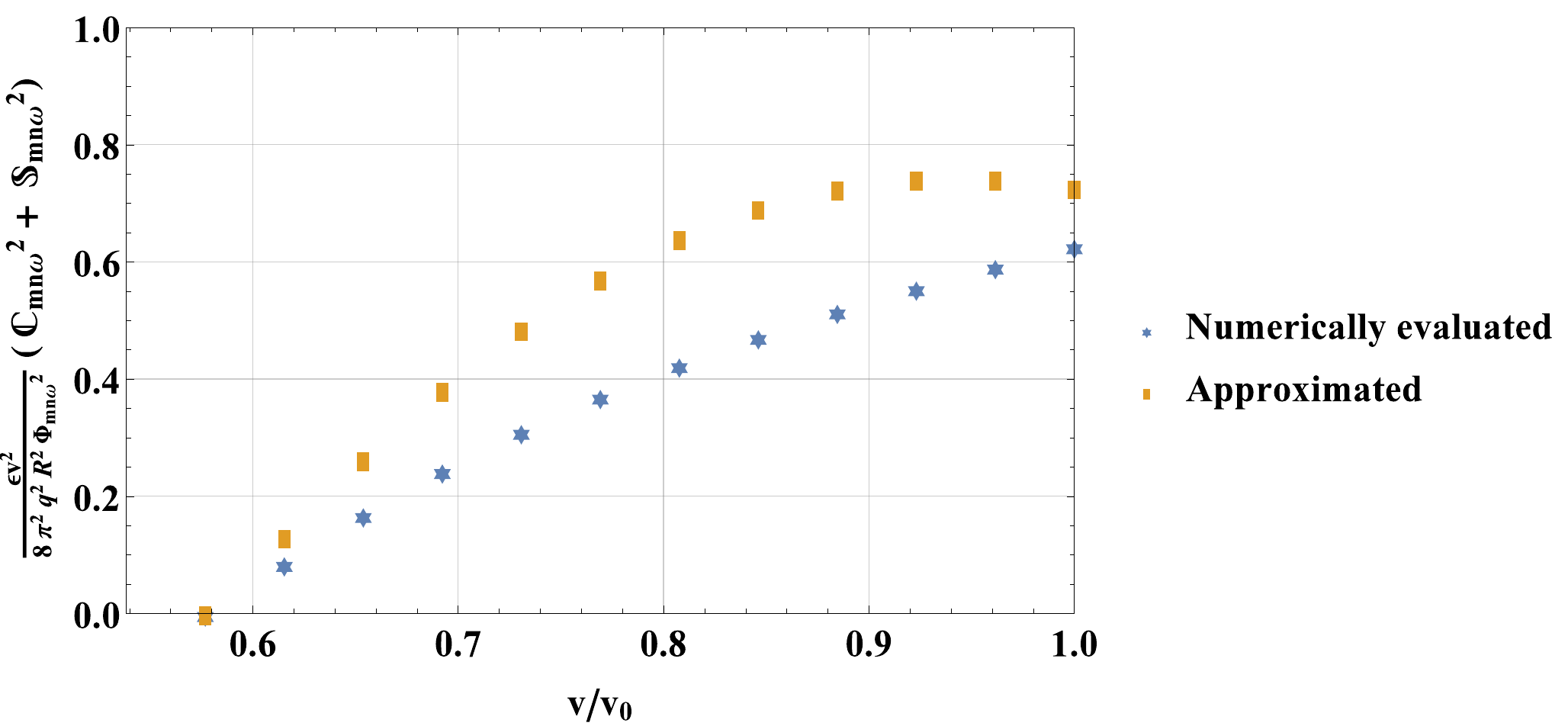}}
\subfigure[ $l=2$]{\label{fig:l2}\includegraphics[width=.9\columnwidth]{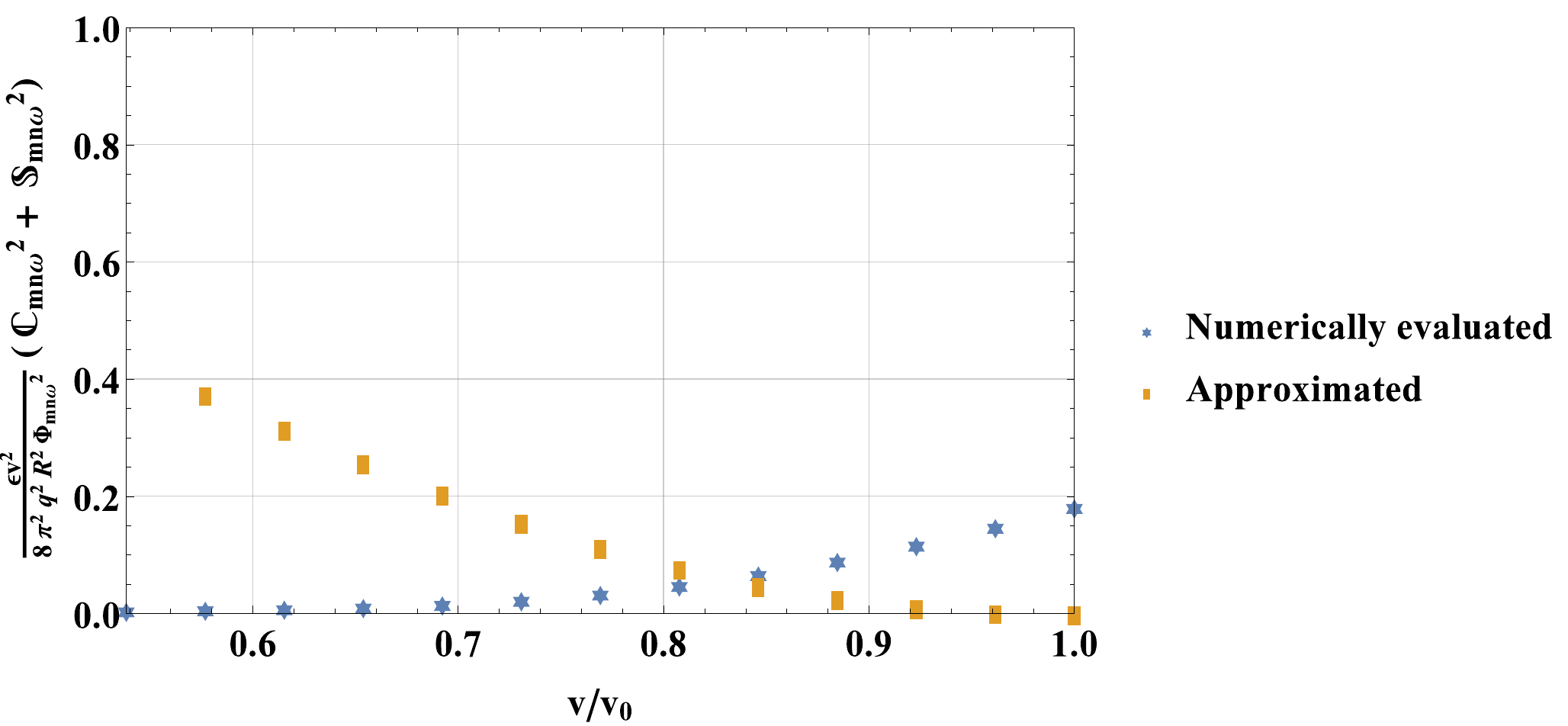}}
  \caption{Exact, numerically evaluated phase factor for trapped particles interacting with the TAE compared to the approximate form given in~\eqref{eq:trappedtaeapprox2}, for $l=0,1,2$. The approximation reproduces overall trends well for low values of $l$. For higher values of $l$, the neglect of $a_t$ in the approximation is inappropriate; however, higher values of $l$ have very small contributions to the flux. The normalizations in these plots are found in the same way as those in figure~\ref{fig:ripplephase}.}
\label{fig:phaseint}
\end{figure*}

Now, we move to the evaluation of the TAE flux, starting with trapped particles. Again, this procedure begins with consideration of the phase factor~\eqref{eq:trappedphase}. Full evaluation of this quantity must be done numerically. However, a tolerable approximation of its value can be made for low values of $l$.  At low $l$ for typical TAE parameters, the dominant term in the sinusoids of the phase factor is the finite orbit width term $c_t$. This can be verified numerically. After neglecting $a_t$ and $d_t$, the terms in the phase factor which are proportional to $A_\parallel$ can be neglected as higher order in $\epsilon$. Then, the terms $\cos{\left(c_t\right)} = \cos{\left(b_t \cos{x} \right)}$ (for even $l$) and $\sin{\left(c_t\right)} = \sin{\left(b_t \cos{x} \right)}$ (for odd $l$) can be written in terms of Bessel functions using the Bessel generating function $e^{i b_t \cos{x}} = \sum_{n = - \infty}^{n= \infty} i^n  e^{inx} J_n\left(b_t \right)$. Here, $J_n$ represents the Bessel function of the first kind of order $n$. Higher values of $n$ in the sum are more oscillatory and do not contribute much to the phase factor integral. So, only the lowest $n$ are maintained.   This process gives:
\begin{equation}
\label{eq:trappedtaeapprox2}
      \frac{\mathbb{C}_{mn\omega}^2 + \mathbb{S}_{mn\omega}^2}{4q^2 R^2}\approx
    \begin{cases}
      \frac{8 \Phi_{ m n \omega}^2  J_0\left(b_t\right)^2 K\left(\kappa\right)^2}{\epsilon  v^2} , &l  \text{ even}\  \\
      \frac{32 \Phi_{mn\omega}^2 J_1\left(b_t\right)^2 \left(\sin^{-1}{\kappa}\right)^2}{\epsilon \kappa^2 v^2}, & l \text{ odd}.
    \end{cases}
\end{equation}
  The approximation in~\eqref{eq:trappedtaeapprox2} is compared to a full numerical evaluation of~\eqref{eq:trappedphase} for $l=0$, $l=1$, and $l=2$ in figure~\ref{fig:phaseint}. This comparison shows that, although the approximation is not exact, it captures key trends in the phase factor for low values of $l$. The approximation is less good for higher values of $l$, where $a_t$ becomes more important and cannot be neglected.

However, higher values of $l$ will have minimal contribution to the flux for three reasons.  First, the resonance structures in figures~\ref{fig:resstructtae0},~\ref{fig:resstructtae1}, and~\ref{fig:resstructtae-1} make clear that, for a given value of $\kappa$, higher values of $l$ correspond to lower values of $v$.  These values of $l$ contribute less heat flux, which depends on $v^2$.  Second, the second term in $1 - \omega/\left(n\omega_\star \right)$, which appears in the flux expression~\eqref{eq:flux,t}, scales with velocity like $1/v^2$ [as we will see later on in~\eqref{eq:omnom}], further reducing the contribution of lower velocities to the flux.  Finally, similar arguments to those used in the consideration of the ripple phase factor show that the sinusoids in the phase factor will become highly oscillatory for high values of $l$, such that the value of the phase factor must  shrink by the  Riemann-Lebesgue lemma. This can be verified computationally. In figure~\ref{fig:phaseint}, the phase factor for $l=2$ is already smaller than for $l=0$ and $l=1$, which is an example of this effect.  Because higher $l$ harmonics have such small contributions to the flux, in this section we compute the contributions from $l=0,1,2$ only.

We now proceed with the evaluation of~\eqref{eq:flux,t}. To lowest order in $\epsilon$, the denominator of~\eqref{eq:flux,t} is 1.  The delta function enforcing the resonance is $\delta\left(Q_l\right)$, with $Q_l$ as evaluated in~\eqref{eq:res2} with $\sigma = 0 $ for trapped particles. This delta function is used to evaluate the velocity integral, recalling the general property that 
\begin{equation}
\label{eq:delprop}
\delta \left[f \left(x \right)\right] = \sum_i \frac{\delta \left(x - a_i \right)}{\left| \frac{df}{dx}\left(a_i \right) \right|}
\end{equation}
for any function $f\left(x \right)$ with roots $a_i$.
Furthermore, we can write that 
\begin{equation}
\label{eq:omnom}
    1 - \frac{\omega}{n \omega_\star} = 1 -\frac{3 \Omega_p \omega R a_\alpha}{n v^2},
\end{equation}
with $a_\alpha$ the alpha scale length defined in~\eqref{eq:aalpha}. For typical values of $a_\alpha$, $a_\alpha/R\sim \epsilon$; inserting this scaling and the resonant value of $v$ where $Q_l\left(v,\kappa\right) = 0$ gives that
\begin{equation}
    1 - \frac{\omega}{n \omega_\star} = 1 - \mathcal{O}\left(\epsilon \right),
\end{equation}
such that $1 - \omega/\left(n\omega_\star\right)$ may be set to $1$ to lowest order in $\epsilon$  (at higher values of $l$ than those considered here, where the resonant $v$ becomes very small for many values of $\kappa$, this replacement would not be appropriate). Then, the contribution to the flux from each $l$ reduces to a fairly compact integral over $\kappa$. This integral is different for $l$ even and $l$ odd:
\begin{equation}
\label{eq:trappedflux}
    Q_{\alpha,mn\omega,t} = -\frac{ 4M_\alpha n^2 c^2  \Phi_{ m n \omega}^2 qR}{\ln{\left(v_0/v_c\right)} } \frac{\partial n_\alpha }{\partial \psi } \sum_l \left\{
     \begin{array}{@{}l@{\thinspace}l}
      \int_0^{\kappa_0} d \kappa  \frac{\kappa J_0 \left(\frac{\sqrt{2\epsilon } k_\psi  v_{res} \kappa}{ \Omega_p}\right)^2 K\left(\kappa\right)^2}{\left|\frac{\partial Q_l}{\partial v} \left(v_{res}\right) \right| }, \, l\text{ even}\\
      \int_0^{\kappa_0} d \kappa  \frac{4  J_1 \left(\frac{\sqrt{2\epsilon } k_\psi  v_{res} \kappa}{ \Omega_p}\right)^2 \left(\sin^{-1}{\kappa} \right)^2}{\kappa\left|\frac{\partial Q_l}{\partial v} \left(v_{res}\right) \right| }, \, l\text{ odd} \\
     \end{array}
  \right. .
\end{equation}
Here, $v_{res}$ is the resonant velocity at which $Q_l \left(v_{res}, \kappa \right) = 0$. The value of $\kappa$ in resonance at the birth velocity $v_0$ is called $\kappa_0$, i.e., using~\eqref{eq:res2} and table~\ref{tab:res},
\begin{equation}
\label{eq:kappamaxdef}
    0 = \frac{8qR K\left(\kappa_0 \right) \omega}{v_0 \sqrt{2\epsilon}} - \frac{4nqv_0 \left[2 E\left(\kappa_0\right) - K\left(\kappa_0\right) \right]}{\Omega_p R \sqrt{2\epsilon}} - 2 \pi l.
\end{equation}
The integrals in~\eqref{eq:trappedflux} can easily be evaluated numerically for specific values of the relevant parameters. 
 
To obtain a closed expression, we now make another set of approximations for the values of $\kappa_0$.    For $l= 0$, $\kappa_0$ is low, and the resonance condition~\eqref{eq:kappamaxdef} is expanded about $\kappa_0 = 0$ to $\mathcal{O}\left( \kappa_0^2 \right)$ and then solved for $\kappa_0$.  For higher values of $l$, with higher values of $\kappa_0$, different approximations are used. The quantity $2E\left(\kappa_0 \right) - K\left(\kappa_0\right)$ vanishes at the point $\kappa_0 \approx 0.91$ and is small near that value.  Resonances with $l>0$ have values of $\kappa_0$ in this region. For $l=1$, we can simply approximate $\kappa_0$ with this value. For resonances of higher $l$, we can neglect the term proportional to $2E\left(\kappa_0 \right) - K\left(\kappa_0\right)$ in the resonance condition~\eqref{eq:kappamaxdef}. Also, we make the replacement $K\left(\kappa_0\right)\rightarrow \ln{\left(4/\sqrt{1-\kappa_0^2}\right)}$ in the remaining terms [this is the limit of $K\left(\kappa_0\right)$ as $\kappa_0 \rightarrow 1$].  This process gives, where we have inserted $\omega \approx v_A/\left(2 q R\right)$ to simplify the expressions,\footnote{The $l=0$ approximation, and the fluxes that follow from it, only provide physical values when $v_A < v_0^2 q n /\left(\Omega_p R\right)$.  When device Alfv\'{e}n speed violates this inequality, the $l=0$ resonance vanishes, and its contribution should be neglected.}
\begin{equation}
\label{eq:kappamax}
    \kappa_0 \approx  \left\{
     \begin{array}{@{}l@{\thinspace}l}
     \sqrt{1 - \frac{v_A \Omega_p R}{nqv_0^2}}, \, l=0\\
     0.91, \, l=1\\
      \sqrt{1-16 e^{-\frac{\pi l v_0 \sqrt{2\epsilon}}{v_A}}}, \, l > 1 \\
     \end{array}
  \right. .
\end{equation}

 The integrands can be expanded about $\kappa = 0$ to lowest non-trivial order in $\kappa$ [$\mathcal{O}\left(\kappa\right)$ for $l$ even and $\mathcal{O}\left(\kappa^3\right)$ for $l$ odd]. Prior to making the expansion for odd $l$, the velocity in the argument of $J_1$ is replaced with the birth velocity. Comparison of the resulting approximate integrands to the numerically evaluated integrands shows that the approximations are good throughout the integration interval.  Then, we calculate 
\begin{equation}
\label{eq:trappedfinal}
    Q_{\alpha,mn\omega,t} \approx -\frac{\sqrt{\epsilon} M_\alpha n\pi c^2  \Phi_{m n \omega}^2 \Omega_p R^2}{4\sqrt{2} \ln{\left(v_0/v_c\right)} }\frac{\partial n_\alpha }{\partial \psi } \sum_l C_{l,t},
\end{equation}
where $C_{l,t}$ is a parameter given in table~\ref{tab:Clt} for the most important low $l$ harmonics. 
\begin{table}
  \begin{center}
\def~{\hphantom{0}}
  \begin{tabular}{lc}
Expression & Value \\ 
$C_{0,t}$ & $ 1- \frac{\Omega_p R v_A}{n q v_0^2}$ \\ 
$C_{1,t}$  &  $\frac{0.28 n^2 q^2 v_0^2}{\epsilon R^2\Omega_p^2} \left(1- \sqrt{\frac{1}{1 +\frac{2 n q v_A }{\epsilon R \Omega_p}}} \right)$ 
 \\
$C_{2,t}$  &  $ \left(  1-16 e^{-\frac{2\pi  v_0 \sqrt{2\epsilon}}{v_A}} \right) \left(1 - \sqrt{\frac{1}{1 +\frac{ n q v_A }{2\epsilon R \Omega_p}}} \right)$
 \\
  \end{tabular}
  \caption{Coefficients of harmonics of contributions to flux used in Eq.~\eqref{eq:trappedfinal}. Only values for $l=0,1,2$ are given because higher values of $l$ have negligible contributions to the flux. We have inserted $\omega \approx v_A/\left(2qR \right)$ to simplify the coefficients.}
  \label{tab:Clt}
  \end{center}
\end{table}

This expression, giving the trapped particle alpha heat flux from a TAE of a given magnitude, is a central result of this paper.   For the example parameters in this paper, $\sum_l C_{l,t} \approx 1.07$.

Now, we consider passing particles, with flux given by the expression in~\eqref{eq:flux,p}. The first step is to approximate the phase factor~\eqref{eq:phasepass}. The dominant term in the phase factor sinusoids is again $c_p$. However, it is not possible to use the Bessel generating function for the passing particles, so instead we simply take $\cos{c_p} \approx \cos{b_p}$ and $\sin{c_p} \approx \sin{b_p}$. Then the phase factor becomes, for $l$ even or odd,
\begin{equation}
\label{eq:passingtaeapprox}
    \frac{\mathbb{C}^2_{mn\omega} + \mathbb{S}_{mn\omega}^2 }{4q^2 R^2} = \frac{\pi^2 k^2  \Phi_{m n \omega}^2 }{2\epsilon \lambda v^2} \left(\frac{2K\left(k\right)}{\pi} - \frac{\sigma v \sqrt{2\epsilon \lambda} }{kv_A} \right)^2.
\end{equation}
This approximation is compared to the full numerical evaluation of the phase factor in table~\ref{tab:passingphaseint}.
 \begin{table}
\centering
\centerline{
\begin{tabular}{m{.8cm}| >{\centering\arraybackslash}m{6cm} >{\centering\arraybackslash}m{6cm}}
 &$\sigma = 1$ &$\sigma = -1$\\
  & & \\
  \hline
$l=0$ &\includegraphics[width=6cm]{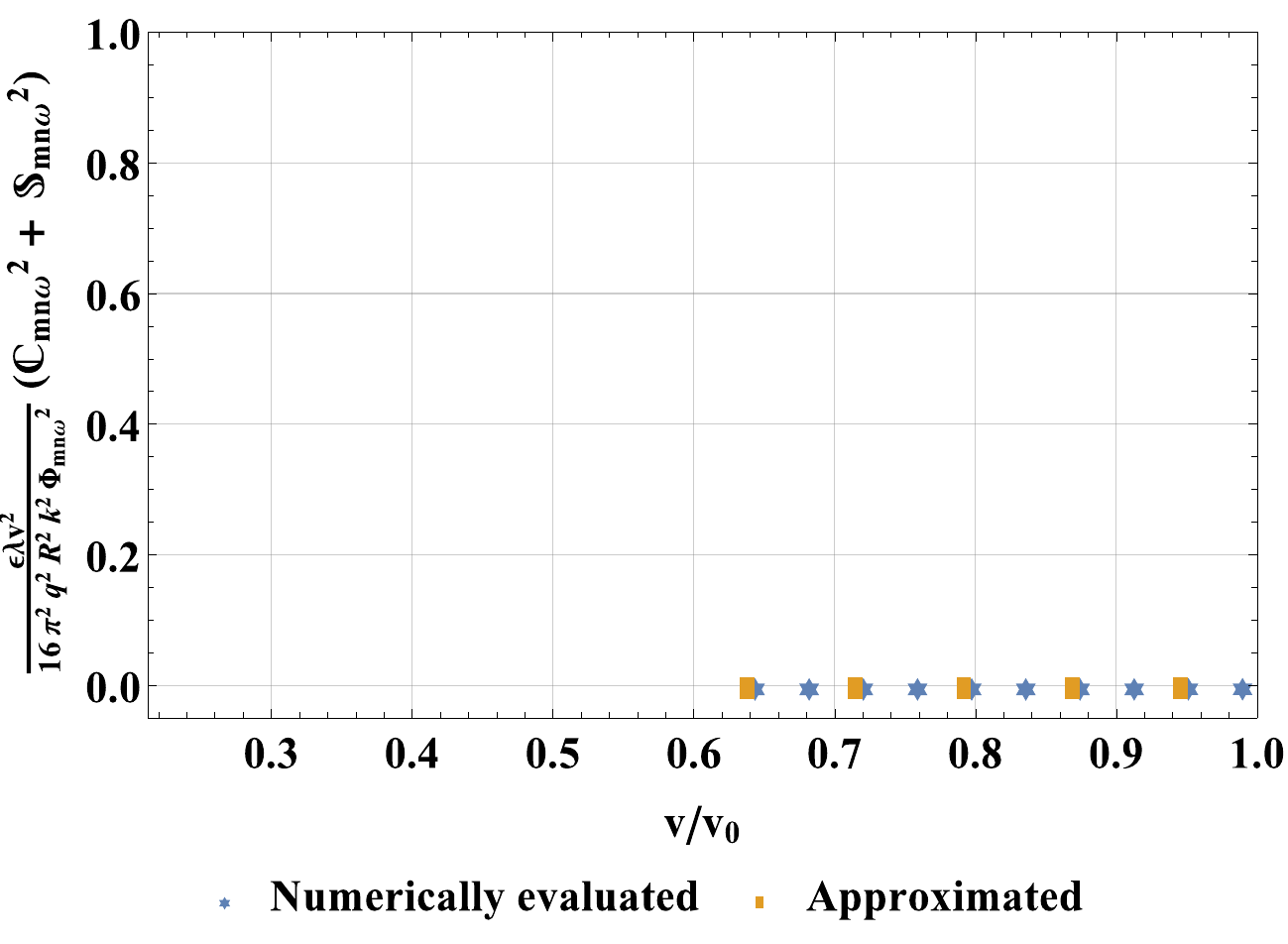} & \textit{No resonance}\\
$l=1$ &\includegraphics[width=6cm]{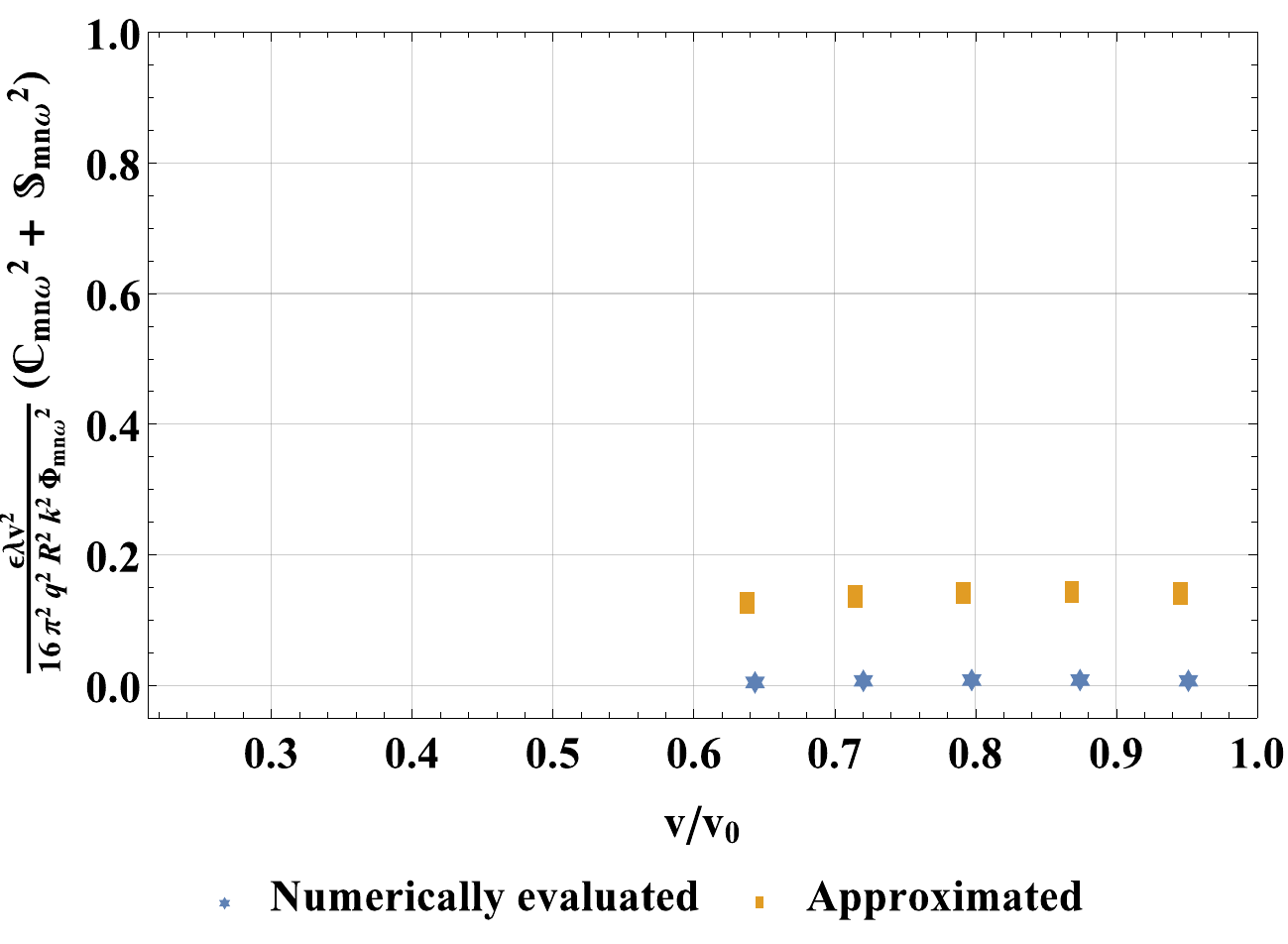} &\includegraphics[width=6cm]{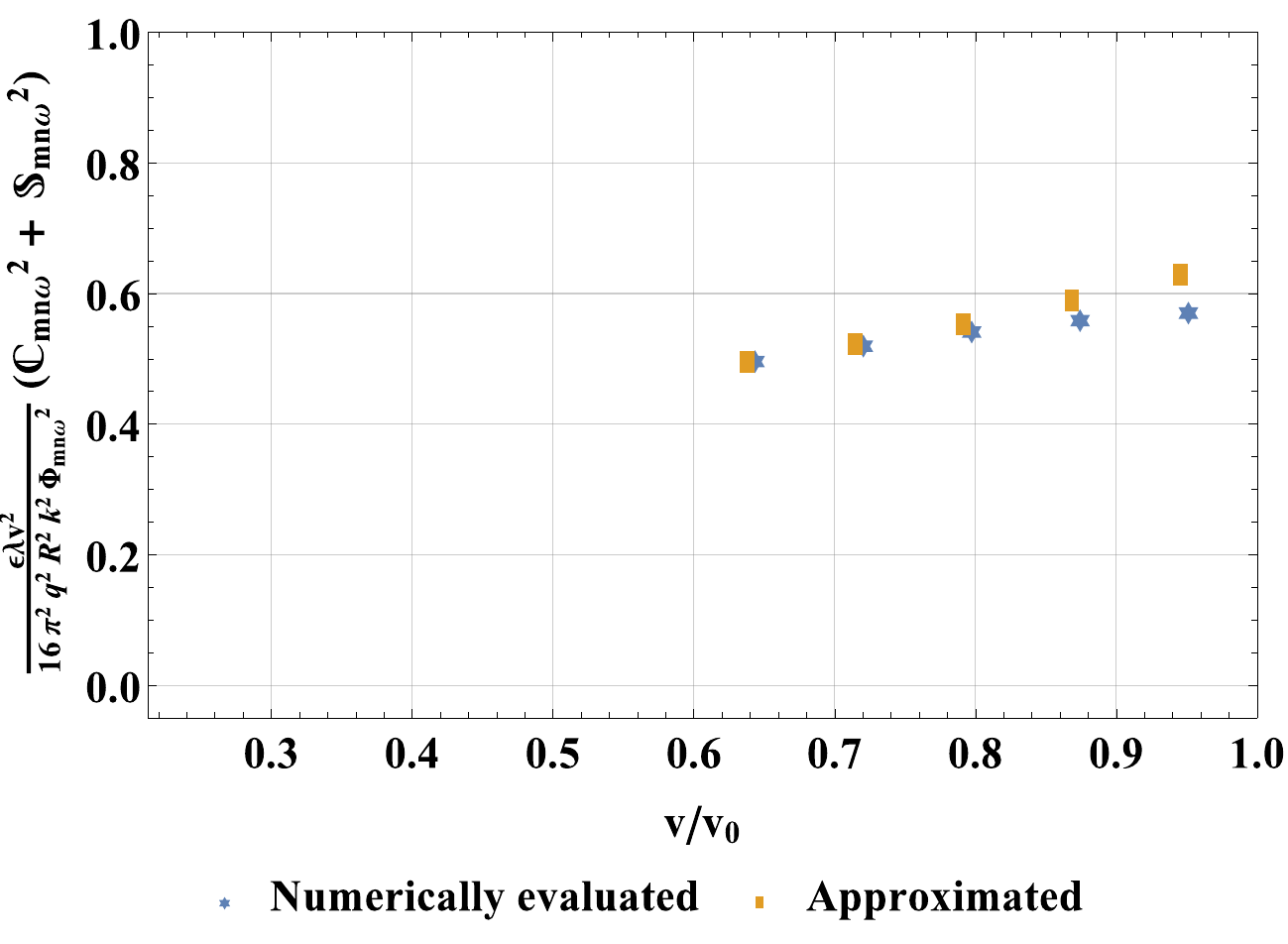}\\
$l=2$ &\textit{Very low contribution to flux} &\includegraphics[width=6cm]{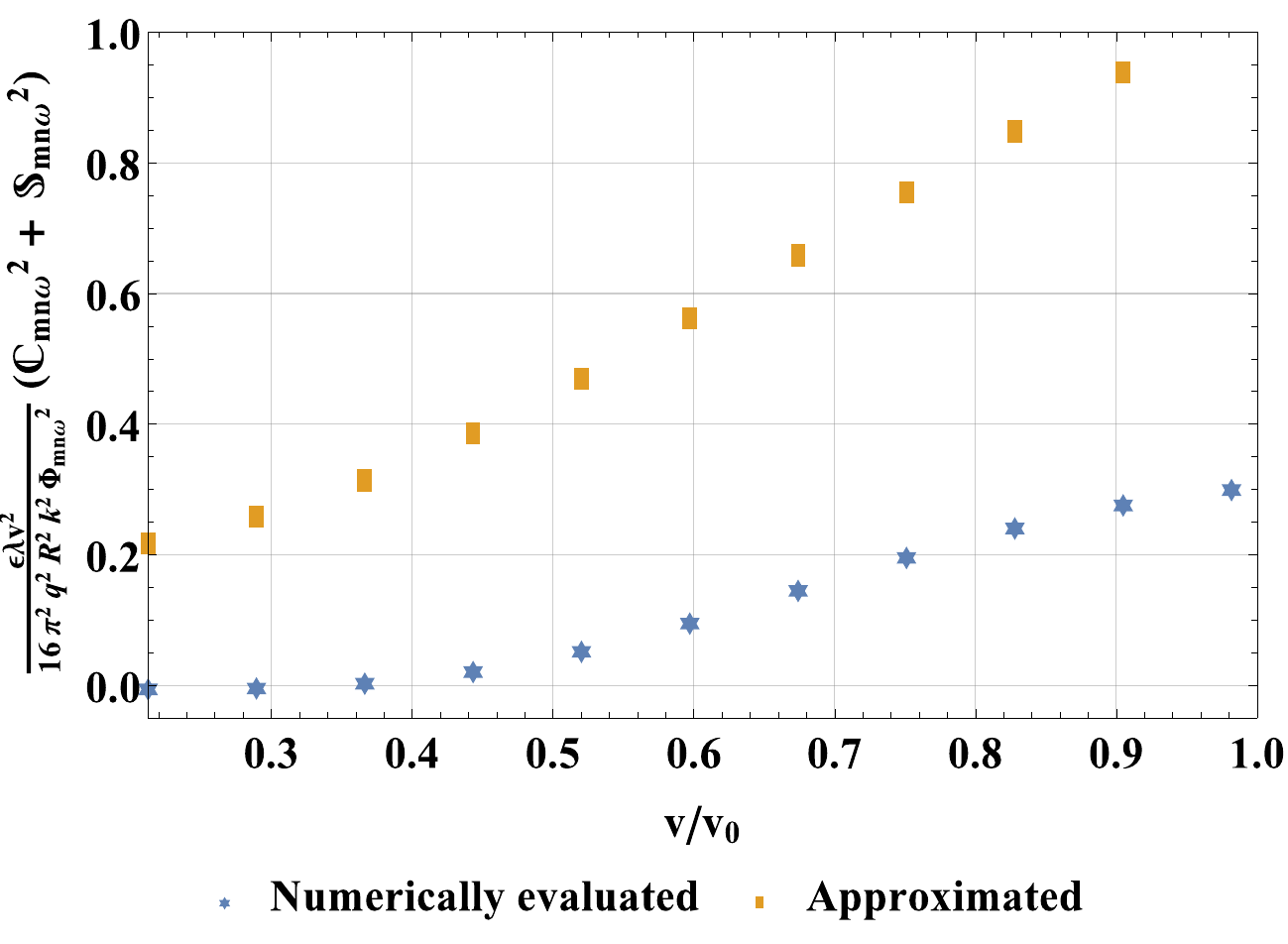}\\
\end{tabular}}
\caption{Exact, numerically evaluated phase factor for passing particles interacting with the TAE compared to the approximate form given in~\eqref{eq:passingtaeapprox}. The approximation reproduces overall trends well for low values of $l$. Note that the phase factor is much smaller for $\sigma=1$ than for $\sigma=-1$.  The normalizations in the plots in this figure are different from those used in the analogous plots for trapped particles, found in figure~\ref{fig:ripplephase} and figure~\ref{fig:phaseint}. In addition to being in terms of $k$ and $\lambda$, the normalization is a factor of $2$ higher, which reflects that passing phase factors are intrinsically larger.}
\label{tab:passingphaseint}
\end{table}

Notably, the phase factor is very small for $\sigma = 1$ in both the numerically evaluated and the approximated forms. This results from the cancellation between the contribution to the phase factor from the electric potential and that from the parallel vector potential.  Thus, moving forward we evaluate only the contribution from $\sigma = -1$. 

The denominator in~\eqref{eq:flux,p} cannot be set to $1$ as it was for the evaluation of the trapped flux~\eqref{eq:flux,t}; however, the delta function is applied in the same way as for the trapped particles, using~\eqref{eq:res2}, and $1- \frac{\omega}{n \omega_\star}$ is again $1$ to lowest order in $\epsilon$. We set $\sigma = -1$ because, as discussed previously, this is the only significant contribution to the flux.   Then the integral to be evaluated is
\begin{equation}
    Q_{\alpha,mn\omega, p} = -\frac{M_\alpha n^2 \pi^2 c^2 \Phi_{ m n \omega}^2 qR}{4 \ln{\left(v_0/v_c\right)}} \frac{\partial n_\alpha}{\partial \psi} \sum_l\int_0^{k_0}dk \frac{k \left(\frac{2 K\left(k\right)}{\pi} + \frac{v_{res} \sqrt{2\epsilon \lambda}}{kv_A} \right)^2 }{\left[\left(1-\epsilon \right)k^2 + 2 \epsilon  \right] \left| \frac{\partial Q_l}{\partial v}  \left(v_{res}\right)\right|},
\end{equation}
where $v_{res}$ is the resonant velocity at which $Q_l \left(v_{res}, k \right) = 0$. The upper limit of integration is the value of $k$ in resonance at the birth velocity $v_0$, which can be found for each value of $l$ using~\eqref{eq:res2} and the passing quantities in table~\ref{tab:res}. 

For $l=1$, a good approximation for $k_0$ is found from expanding the resonance condition about $k_0 = 0$ to $\mathcal{O} \left( k_0^2 \right)$. In contrast, for $l=2$, a good approximation is found by expanding about $k_0 = 1$ to $\mathcal{O} \left[ \ln\left(1-k_0\right) \right]$. This process gives, with $\omega \approx v_A/\left(2 qR\right)$ used to simplify the expressions:

\begin{equation}
\label{eq:kmax}
    k_0 =  \left\{
     \begin{array}{@{}l@{\thinspace}l}
     \sqrt{2 \epsilon} \sqrt{\frac{v_0^2}{v_A^2} -1}, \, l=1\\
      1 - 8 e^{\frac{-v_0 \left(3 \sqrt{2\epsilon} \Omega_p \pi R + 4 n q v_0 \right)}{nqv_0^2 + \Omega_p  R v_A}}, \, l > 1 \\
     \end{array}
  \right. .
\end{equation}
The integrand  can again be expanded about $k=0$ to lowest non-trivial order in $k$.  Then, we find, where we have inserted $\omega \approx v_A/\left(2qR\right)$
\begin{equation}
\label{eq:fluxpassing}
    Q_{\alpha,mn\omega, p} \approx  -\frac{M_\alpha n^2 \pi v_0 c^2\Phi_{m n \omega}^2 qR }{ \ln{\left(v_0/v_c\right)} } \frac{\partial n_\alpha}{\partial \psi}\sum_l C_{l,p} ,
\end{equation}
where $C_{l,p}$ are parameters given in table~\ref{tab:Clp}. Again, we include only harmonics $l\leq 2$. Higher harmonics have small contribution to the flux.  This expression, giving the passing particle alpha heat flux from a TAE of a given magnitude, is another central result of this paper. For the example parameters in this paper, $\sum_l C_{l,p} \approx 0.41 $. The passing flux is larger than the trapped flux.  This results from multiple factors. For instance, there are more passing particles than trapped particles in the isotropic alpha population. Also, the passing phase factor tends to be larger than the trapped phase factor.

\begin{table}
  \begin{center}
\def~{\hphantom{0}}
  \begin{tabular}{lc}
Expression & Value \\ 
$C_{1,p}$ &  $1-\frac{v_A}{v_0} $\\ 
$C_{2,p}$  &  $\frac{4v_A}{8 1 \sqrt{2 \epsilon} v_0} \left[ 1 - 8 e^{\frac{-v_0 \left(3 \sqrt{2\epsilon} \Omega_p \pi R + 4 n q v_0 \right)}{nqv_0^2 +  \Omega_p R v_A}} \right] $ 
 \\
  \end{tabular}
  \caption{Coefficients of harmonics of contributions to the flux used in Eq.~\eqref{eq:fluxpassing}.}
  \label{tab:Clp}
  \end{center}
\end{table}

The expressions for the trapped,~\eqref{eq:trappedfinal}, and passing,~\eqref{eq:fluxpassing}, fluxes can be compared to the phenomenological flux,~\eqref{eq:phenom}. In particular, for the trapped flux,~\eqref{eq:trappedfinal}, setting $n\omega_{\alpha_\star}\sim nv_{res}^2/\left(\Omega_p R^2 \right)$ and ${\left(\vec{v}_\parallel \hat{b}_{tot} + \vec{v}_{d,tot} \right)_1\cdot\nabla\psi\sim n c \Phi_{m n \omega}}$  reproduces the dependencies seen in~\eqref{eq:trappedfinal}. For the passing flux, setting $n\omega_{\alpha_\star}\sim\omega \approx v_A/\left(2qR\right)$, $v_{res} \sim v_A\sim v_0$ and making the same replacement as for the trapped for the radial velocity reproduces the dependencies seen in~\eqref{eq:fluxpassing}. As intuition suggests, the heat flux scales with the square of the perturbation amplitude, $\Phi_{m n \omega}$, and increases with the periodicity of the mode, $n$.
\subsection{Interpretation of flux and development of saturation condition}
\begin{figure}
\centering
\includegraphics[width=.9\columnwidth]{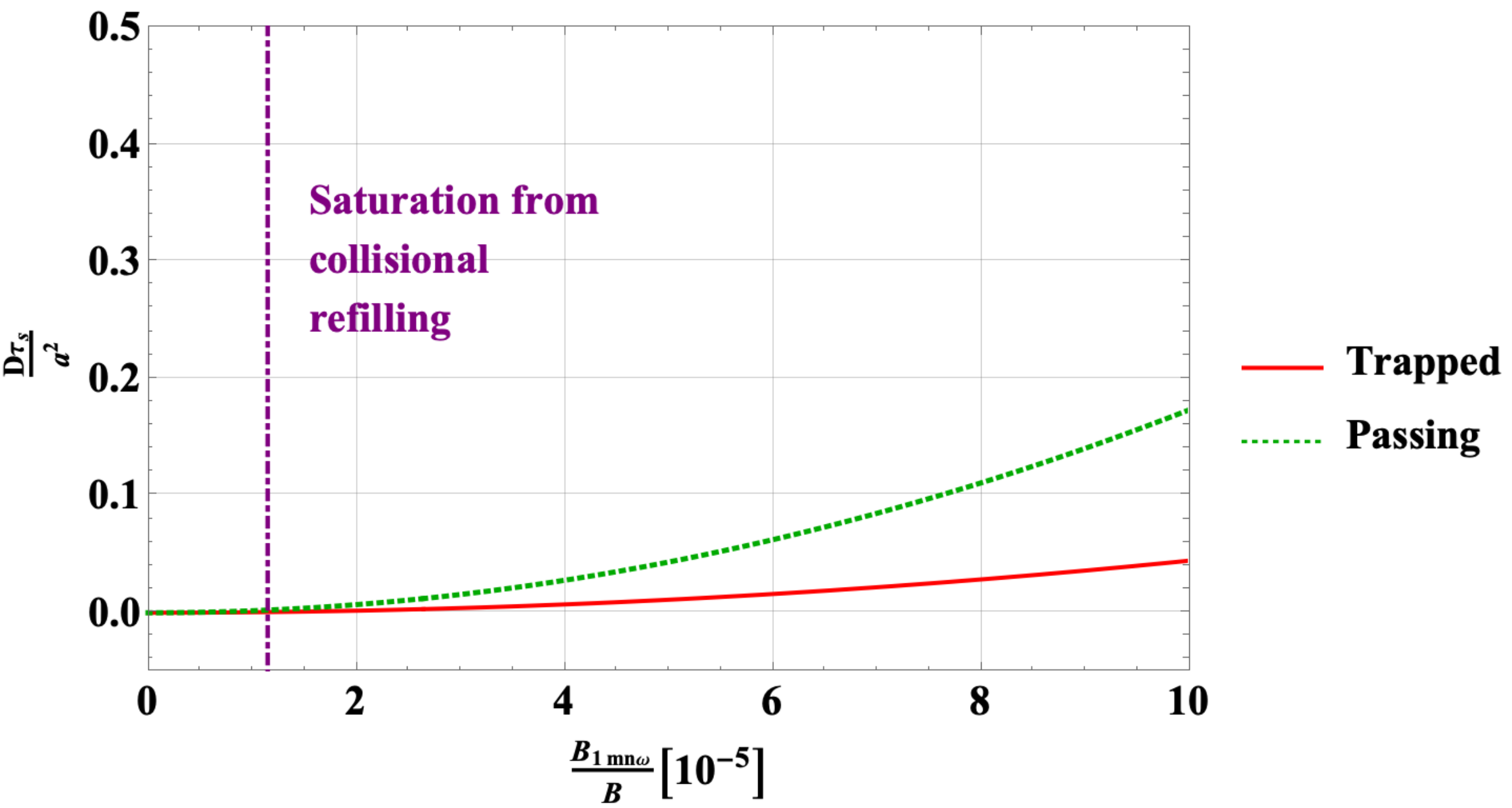}
\caption{The strength of the diffusion caused by a TAE [i.e., the left side of~\eqref{eq:trappeddtaua} and~\eqref{eq:passingdtaua}] as a function of amplitude for trapped and passing particles. This quantity must remain well below 1 to avoid significant alpha depletion. The saturated TAE amplitude suggested by a simple model presented in this paper, Eq.~\eqref{eq:satamp}, is indicated by the purple dash-dotted line. However, experiments and other models and numerical treatments of saturation often find amplitudes ranging in magnitude from $1 \times 10^{-5}$ to $100 \times 10^{-5}$.}
\label{fig:diffusion}
\end{figure}

The magnitude of the alpha flux caused by the TAE is understood by comparing the rate of the diffusion it causes to the rate of slowing down. Slowing down (described in appendix~\ref{sec:slowingdown}) removes alpha particles from the energetic population and transfers their energy to the local bulk plasma, leaving the alpha particles as helium ash. So, with $D$ the diffusion coefficient describing the TAE flux and $a$ the device minor radius, if 
\begin{equation}
\label{eq:weak} 
    \frac{D\tau_s}{a^2} \ll 1, 
\end{equation}
TAE diffusion is weak and does not remove alpha particles  before they can give their energy to the background plasma and transition to ash. When 
\begin{equation}
\label{eq:strong} 
    \frac{D\tau_s}{a^2} \sim 1, 
\end{equation}
TAE diffusion is very strong. It would be capable of diffusing alpha particles across the whole poloidal tokamak cross section (with an area that scales as $a^2$) in a slowing down time if mode activity of similar strength existed across that area. (Note that the width of an individual mode is significantly smaller than $a$.) At this level, alpha particles diffuse sufficiently fast to cause significant modification to the alpha heat deposition profile. (At this point the perturbative approach used in this paper becomes inappropriate. For losses at this level, a quasilinear treatment which retains the radial flux losses in~\eqref{eq:dke0} for $f_0$ is needed.)
The diffusion coefficient is related to the flux by 
\begin{equation}
    D \equiv \frac{-2Q_\alpha}{R^2 B_p^2 \frac{\partial n_\alpha }{\partial \psi} M_\alpha v_0^2},
\end{equation}
such that the condition for avoiding strong TAE diffusion,~\eqref{eq:weak}, becomes
\begin{equation}
\label{eq:trappeddtaua}
    \frac{\sqrt{2\epsilon} \Omega_p \pi v_A^2 R^2  \sum_l C_{l,t} }{4 nv_0^2} \left(\frac{B_{1 m n \omega} }{B} \right)^2 \frac{\tau_s}{a^2} \ll 1
\end{equation}
 for trapped particles and 
\begin{equation}
\label{eq:passingdtaua}
    \frac{2 \pi  q  v_A^2  R \sum_l C_{l,p} }{v_0}  \left(\frac{B_{1 m n \omega} }{B} \right)^2 \frac{\tau_s}{a^2}\ll 1
\end{equation}
for passing particles. Here, we have stated the flux in terms of the strength of the perturbed magnetic field, $B_{1mn\omega}/B \sim nc \Phi_{mn\omega} / \left( v_A R B_p \right)$, which is more commonly seen in the literature than $\Phi_{mn\omega}$. In addition, we have used $\ln{\left(v_0/v_c\right) }\approx 1$.
For fixed tokamak equilibrium parameters and fixed radial location $\epsilon$, these conditions constrain the maximum amplitude for the mode to avoid serious alpha depletion.   We plot the value of $D \tau_s/a^2$ as a function of the normalized mode amplitude, $B_{1mn\omega}/B$, in figure~\ref{fig:diffusion}.

Understanding how this constraint compares to typical TAE amplitudes requires a rough estimate for mode saturation. We now develop such an estimate.  TAE saturation has been estimated using a variety of methods. Some works balance the linear growth rate of the instability with the rate at which particles get trapped by the wave and flatten the gradient driving the instability, sometimes including collisions which restore the original gradient  \citep{wu1994self,fu1995nonlinear,wang2016saturation,zhou2016collisional,todo2019introduction}. Other works consider the effects of zonal flows and nonlinear mode couplings \citep{spong1994nonlinear,chen2012nonlinear}. Our technique for estimating saturation is similar to the first of these approaches in that it also considers TAE drive reduction due to the flattening of the alpha gradient and collisional refilling of this gradient. However, we consider the removal of the drive to be due to flattening, rather than due to particle trapping.  

\begin{figure}
\centering
\includegraphics[width=.5\columnwidth]{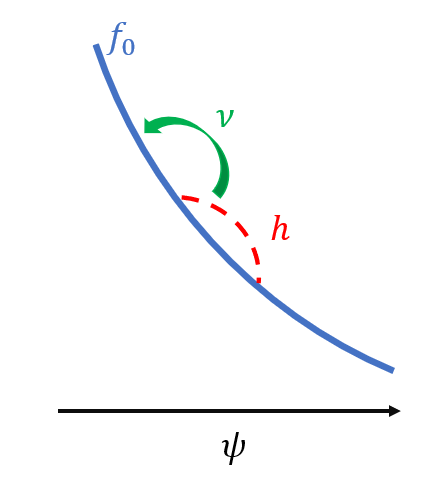}
\caption{Schematic representation of TAE saturation. The non-adiabatic particle response $h$ reduces the alpha gradient, $\partial f_0/\partial \psi$ and the drive to the TAE. Collisions (represented by $\nu$) restore the original particle gradient. The balance of these processes, as expressed in~\eqref{eq:satbal}, results in mode saturation. }
\label{fig:saturation}
\end{figure}

The mechanism of saturation is represented schematically in figure~\ref{fig:saturation}.
As the TAE grows, the non-adiabatic particle response $h$ reduces the alpha gradient responsible for mode drive. However, collisions concurrently act to restore the original particle gradient by counteracting this flattening. The balance of these processes results in mode saturation. (If the original drive were not restored fast enough by collisions, $h$ and the mode amplitude would shrink, while if it were restored faster than it was reduced, $h$ and mode amplitude would grow.) Mathematically, this balance is stated  
\begin{equation}
\label{eq:satbal}
    cnS_{mn\omega}\frac{\partial h_{mn\omega}}{\partial \psi} \sim \nu_{pas} \frac{\partial^2 h_{mn\omega}} {\partial \lambda^2}.
\end{equation}
The left side is the rate at which the drive for $h_{mn\omega}$, $cn S_{mn\omega} \partial f_0/\partial \psi$, is reduced by $h_{mn\omega}$. The right side, $\nu_{pas} \partial^2 h_{mn\omega}/\partial \lambda^2$, is the rate at which the original drive is restored by collisions.  

This saturation condition can be evaluated in two different limits.  The first, low collisionality, limit occurs when $\partial h_{mn\omega} /\partial \psi \sim  \left(  \partial h_{mn\omega}/\partial \kappa \right) \left( \partial \kappa/ \partial \psi \right)\sim h_{mn\omega}/\left(  R^2 B_p \delta \lambda \right)$, with $\delta \lambda$ given by~\eqref{eq:deltlambda}. (This expression is for trapped particles; the equivalent passing expression is in terms of $k$.) The second, high collisionality, limit occurs when $\partial h_{mn\omega} /\partial \psi \sim k_\psi h_{mn\omega}/\left(R B_p \right)$, with $k_\psi$ given by the value for the TAE given in table~\ref{tab:perts}. The transition between the two limits occurs when $k_\psi \sim 1/\left(R \delta \lambda \right)$. 
Inserting $\partial^2 h_{mn\omega} /\partial \lambda^2 \sim h_{mn\omega}/ \left(\delta \lambda \right)^2$, with $\delta \lambda$ given by~\eqref{eq:deltlambda} with $n\omega_{\alpha_\star} \sim \omega$, and $cn S_{mn\omega} \sim c n \Phi_{m n \omega} \sim v_A R B_p B_1/ B $ reveals the saturated amplitude of the TAE:
\begin{equation}
\label{eq:satamp}
    \frac{B_1}{B} = \begin{cases}
\frac{R \nu_{pas}^{2/3} \omega^{1/3} }{v_A},& \frac{\nu_{pas}}{\omega} < \left( \frac{\epsilon}{nq} \right)^3\\
\frac{\epsilon R \nu_{pas}^{1/3} \omega^{2/3} }{nq v_A} &\frac{\nu_{pas}}{\omega} > \left( \frac{\epsilon}{nq} \right)^3.
\end{cases}
\end{equation}
Here, $\nu_{pas}\sim v_\lambda^3/\left(v_0^3 \tau_s\right)$ [from the second term in the collision operator~\eqref{eq:collop}]. Similar scalings of saturated amplitude with collisionality are found elsewhere in the literature \citep{berk1992scenarios,slaby2018effects,white2019collisional}.
For the parameters used in tables~\ref{tab:equilib} and~\ref{tab:perts}, the low collisionality limit is appropriate, and gives an amplitude of roughly $B_1/B \approx 1.1 \times 10^{-5}$. This saturation estimate is comfortably below the stochastic threshold given in \citet{berk1992scenarios}, making it consistent with our assumption throughout the paper that the TAE amplitude is below the threshold for stochasticity.  

Saturated AE amplitudes observed experimentally \citep{nazikian1997alpha,van2006radial,fasoli1997alfven} and predicted computationally \citep{fitzgerald2016predictive} or analytically \citep{zonca1995nonlinear} using other models have found amplitudes ranging from $10^{-5}$ to $10^{-3}$ across a wide range of machine parameters and types of energetic particle populations; in extreme cases, amplitudes of $10^{-2}$ have even been found \citep{todo2003simulation}. Thus, our estimated amplitude is at the bottom of the range typically found in the literature. 

The expected saturated amplitude at the example parameters used in this paper, $1.1 \times 10^{-5}$, is indicated on the plot of diffusion strength, figure~\ref{fig:diffusion}, with a purple dash-dotted line, revealing that at this amplitude, $D\tau_s/a^2$ is small for both the trapped and the passing populations. At this amplitude, significant redistribution of alphas due to this TAE is not expected, though some localized alpha gradient flattening may occur.  (In addition, synergistic action with overlapping modes of different $m$ and $n$ could change the transport; since TAEs consist of coupled poloidal harmonics, this change is likely to occur.) An increase in saturated amplitude from the level predicted in our simplified model, but still within the $10^{-5} - 10^{-3}$ range commonly seen in the literature, could lead to significant redistribution. This significant activity is especially likely for the passing population.

\section{Conclusion}
\label{sec:concl}

This work has developed the drift kinetic theory of transport that results from resonance between alpha particle motion and a perturbation to the tokamak magnetic and electric fields. The theory allows the calculation of the alpha particle heat flux of trapped and passing particles that results from these perturbations,~\eqref{eq:flux,t} and~\eqref{eq:flux,p}.  These expressions are examined for two specific perturbations: ripple and TAE.  For ripple, this mechanism causes negligible alpha particle transport, as discussed in section~\ref{sec:ripp}.  For TAE, the mechanism causes significant heat flux [given by~\eqref{eq:trappedfinal} for trapped particles and~\eqref{eq:fluxpassing} for passing particles] which allows the development of constraints on TAE amplitude to avoid significant depletion of the alpha population [given by~\eqref{eq:trappeddtaua} for trapped particles and~\eqref{eq:passingdtaua} for passing particles]. A simple saturation condition suggests that the TAE amplitude in a tokamak scenario similar to that which might occur in the next generation tokamak SPARC is below this limit. This is a reassuring conclusion because significant alpha transport due to TAEs has the potential to seriously impair tokamak operation. However, the presence of other mechanisms for TAE transport [such as the removal of alphas due to their being born in stochastic regions as considered further in~\citet{sigmar1992alpha} and~\citet{collins2016observation}] and the precedent for larger saturation amplitude in the experimental and analytical literature suggest that caution is still important.

Our expressions for the radial heat transport are independent of the collision frequency even though each wave-particle resonance path in velocity space is at the center of a narrow collisional boundary layer. Indeed, our resonant plateau model is reminiscent of quasilinear treatments in the sense that the narrow collisional boundary layer results in and gives meaning to the delta function behavior that removes the collision frequency when the alpha heat flux is evaluated [see the diffusive limit in \citet{duarte2019collisional}  and the pitch angle scattering treatment of \cite{catto2020collisional}]. Our expressions for the alpha heat fluxes are most appropriate when the phase space islands about the wave-particle resonance curves do not overlap. However, the expressions may remain reasonable even as the islands begin to overlap provided the TAE amplitude remains below the threshold for full stochasticity.

Avenues for future work based on this paper fall into three categories: analytical, numerical, and experimental. On the analytical front, future work should adapt and improve the formalism developed in this paper so that it can be used to study other species of energetic particle, which have more complicated distribution functions than the alpha distribution function. This will enable comparison of this model's predictions to current experiments without alpha particles. In addition, analytical work could focus on the calculation of the flux for perturbations for which the approximation $q \approx q_\star$, discussed in section~\ref{sec:shear}, is not possible. This includes the neoclassical tearing mode \citep{la2006neoclassical}, for which $nq \sim m$, making it necessary to avoid the condition in~\eqref{eq:qstarcond}. Evaluation of the flux from such perturbations requires modified techniques. Finally, additional analytical work should consider adaptation of the techniques developed in this paper for use with a more precise collision operator than the Krook operator. This would allow the study of resonances, like some of the ripple resonances discussed in section~\ref{sec:ripp}, which are spaced too closely for the Krook operator to be appropriate. 

On the numerical front, effort should be made to determine the relative importance of the transport mechanisms studied in this paper and those that are the focus of other works on energetic particle transport.  Other mechanisms include overlapping phase space islands from different resonances, the transfer of alpha particles across topological boundaries (for example, the trapped-passing boundary), and convective transport which occurs when an alpha particle near the edge stays in phase with a perturbation without scattering until it leaves the device \citep{heidbrink2008basic}.  Simulations of alpha transport due to tokamak perturbations using the orbit following codes discussed in the introduction can be examined to distinguish transport resulting from these mechanisms. Such simulations should take care to properly resolve the boundary layer around the resonant velocity and pitch angles. We note that collisional boundary layer behavior was recently observed in~\citet{white2019collisional}, but was attributed to an alternate mechanism. 

On the experimental front, the predictions in this paper should be validated against estimates of alpha particle flux from measurements of actual experiments. Such validation will be possible in next generation experiments which are dominantly alpha heated, like SPARC and ITER, and may be possible in "afterglow" scenarios of experiments with some alpha heating, like the forthcoming JET DT campaign. [Afterglow scenarios, discussed in \citet{sharapov1999stability}, occur when external heating is suddenly turned off in order to transiently observe alpha particle effects.]  If the formalism in this paper is adapted to distributions typical of energetic particles resulting from ICRF or beams, which are common in current day experiments, precise comparison to current experiments is possible. Already, current experimental measurements can be examined for evidence of general trends suggested by our work. For instance, part of the energetic particle flux resulting from energetic particle mode perturbations has been nicely demonstrated to scale with mode amplitude squared in \citet{nagaoka2008radial}. This agrees with our predictions.

The authors thank Steven Scott, Ryan Sweeney, Abhay Ram, Ian Abel, Ian Hutchinson, Eero Hirvijoki, Antti Snicker, Konsta S\"{a}rkim\"{a}ki, Paulo Rodrigues, Earl Marmar, Felix Parra, Iv\'{a}n Calvo, Vin\'{i}cius Duarte, Lucio Milanese, and Martin Greenwald for helpful conversations. The authors are especially grateful to Nuno Loureiro for suggesting P.J.C. become involved in the problem and for several insightful suggestions. The authors acknowledge support from the National Science Foundation Graduate Research Fellowship under Grant No. DGE-1122374, support from US Department of Energy award DE-FG02-91ER54109, and support from the Bezos Membership at the Institute for Advanced Study.

\appendix 
\section{Derivation of equilibrium alpha particle slowing down distribution}
\label{sec:slowingdown} 
The equilibrium energetic alpha distribution, $f_0$, is determined by~\eqref{eq:dke} in the unperturbed magnetic field~\eqref{eq:equilibb}:
\begin{equation}
\label{eq:dke0}
 \left(v_\parallel \hat{b} + \vec{v}_{d} \right) \cdot \nabla f_0 = C\left\{f_0\right\} +\frac{S_{fus} \delta \left(v-v_0\right)}{4 \pi v^2}.
\end{equation}
This expression is solved by separating its terms by size and solving for the corresponding size terms in the distribution function, with $f_0 = f_0^0+ f_0^1 + ...$.
The largest term in the unperturbed drift kinetic equation~\eqref{eq:dke0} is the streaming term, so that 
\begin{equation}
v_\parallel \hat{b} \cdot \nabla f_0^0 = 0. 
\end{equation}
This equation shows $f_0^0$ is a flux function, i.e.,
\begin{equation}
\label{eq:f00}
f_0^0 = f_0^0 \left( \psi, v ,\lambda \right).
\end{equation}
The next order expression includes the drift term (which can be shown to be small compared to the streaming term by the poloidal alpha gyroradius over the perpendicular scale length of $f_0$), the collision operator (which is small by the connection length $qR$ over the alpha mean free path), and the source term (which must be the same size as the collision operator for alpha particles to slow down before they are lost due to other processes). This expression reads
\begin{equation}
\label{eq:nextorder}
v_\parallel \hat{b} \cdot \nabla f_0^1 + \vec{v}_d \cdot \nabla f_0^0 = C \left\{f_0^0\right\} +\frac{S_{fus} \delta \left(v-v_0\right)}{4 \pi v^2}. 
\end{equation}
Because $f_0^0$ spatially depends only on $\psi$, we have that
\begin{equation}
\vec{v}_d \cdot \nabla f_0^0 =\vec{v}_d \cdot \nabla \psi \frac{\partial f_0^0}{\partial \psi} = v_\parallel \hat{b} \cdot \nabla \left( \frac{I v_\parallel}{\Omega} \right) \frac{\partial f_0}{\partial \psi},
\end{equation}
where the final expression can be derived as explained in equation (8.14) of \citet{helander2005collisional}.

We define the variable $\tau$ characterizing the progression of particles along the magnetic field,
\begin{equation}
\label{eq:deftau}
d \tau \equiv \frac{d l}{v_\parallel} = \frac{d \vartheta}{v_\parallel \hat{b} \cdot \nabla \vartheta\left(\tau \right)},
\end{equation} 
with $dl$ a unit of distance along the magnetic field.
Integration over $\tau$ over a closed orbit annihilates the left side of~\eqref{eq:nextorder}, giving
\begin{equation}
\label{eq:annihil}
\oint \frac{d \vartheta}{v_\parallel \hat{b} \cdot \nabla \vartheta} v_\parallel \hat{b} \cdot \nabla \left( f_0^1 + \frac{Iv_\parallel}{\Omega} \frac{\partial f_0^0}{\partial \psi} \right) = 0 = \oint \frac{d \vartheta}{v_\parallel \hat{b} \cdot \nabla \vartheta}\left[ C\left\{f_0^0 \right\} + \frac{S_{fus} \delta \left(v-v_0\right)}{4 \pi v^2} \right]. 
\end{equation}
Because $f_0^0$ is not a function of $\vartheta$ [recall~\eqref{eq:f00}], the integrand on the right side must be identically zero. Enforcing this condition using~\eqref{eq:collop} (note that alphas are born isotropically, so only the first term in $C$ is important here) gives the slowing down distribution,
\begin{equation}
f_0^0 \left(\psi, v\right) = \frac{S_{fus}\left(\psi \right) \tau_s\left(\psi \right) H\left(v_0-v\right)}{4 \pi \left[v^3 + v_c^3\left(\psi\right)\right]}.
\end{equation}
 Insertion of the slowing down distribution into~\eqref{eq:nextorder} gives that 
 \begin{equation}
 v_\parallel\hat{b} \cdot \nabla \left(f_0^1 + \frac{Iv_\parallel}{\Omega} \frac{\partial f_0^0}{\partial \psi} \right) = 0,
 \end{equation} 
 which shows that $f_0^1 + \left(Iv_\parallel/ \Omega \right) \partial f_0 / \partial \psi$ is a flux function. Evaluation of the drift kinetic equation to the next order allows the derivation of neoclassical alpha transport driven by the term $\left(Iv_\parallel / \Omega\right) \partial f_0^0/\partial \psi$, as discussed in \citet{catto2018ripple}. However, in this paper we focus on transport caused by perturbations and take that $f_0 = f_0^0$. This means that we assume
 \begin{equation}
 \label{eq:gradapprox}
 \frac{Iv_\parallel}{\Omega} \frac{1}{f_0}\frac{\partial f_0^0}{\partial \psi} \ll 1,
 \end{equation}
 which is equivalent to stating $\rho_{p\alpha} /a_\alpha \ll 1$, with $a_\alpha$ the alpha particle scale length defined in~\eqref{eq:aalpha}.

\bibliographystyle{jpp}

\bibliography{jpp-instructions}

\end{document}